\documentclass[twocolumns]{aastex631}
\usepackage{color}
\usepackage{graphicx}
\usepackage[normalem]{ulem}
\usepackage{rotating}

\usepackage{lineno}
\usepackage{braket}
\usepackage{gensymb}

\shorttitle{GCs Structures in Fornax cluster}
\shortauthors{}

\date{\today}

\begin{document}

\title{Spatial Structures in the Globular Cluster Distribution of Fornax Cluster Galaxies}

\correspondingauthor{Raffaele D'Abrusco}
\email{rdabrusc@cfa.harvard.edu}

\author[0000-0003-3073-0605]{Raffaele D'Abrusco}
\affil{Center for Astrophysics \textbar\ Harvard \& Smithsonian, 60 Garden Street, Cambridge, MA 20138, USA}

\author{David Zegeye}
\affil{Department of Astronomy and Astrophysics, University of Chicago, 5640 South Ellis Avenue, Chicago, IL 60637, USA}

\author{Giuseppina Fabbiano}
\affil{Center for Astrophysics \textbar\ Harvard \& Smithsonian, 60 Garden Street, Cambridge, MA 20138, USA}

\author{Michele Cantiello}
\affil{INAF Osservatorio Astronomico d’Abruzzo, Via Maggini, 64100 Teramo, Italy}

\author{Maurizio Paolillo}
\affil{Dipartimento di Fisica ``Ettore Pancini'', Universit\`a di Napoli ``Federico II'', via Cinthia 9, 80126 Napoli, Italy}
\affil{INFN - Sezione di Napoli, via Cinthia 9, 80126 Napoli, Italy}
\affil{INAF - Osservatorio Astronomico di Capodimonte, Salita Moiariello 16, I-80131, Napoli, Italy}

\author{Andreas Zezas}
\affil{Center for Astrophysics \textbar\ Harvard \& Smithsonian, 60 Garden Street, Cambridge, MA 20138, USA}
\affil{Physics Department and Institute of Theoretical and Computational Physics, University of Crete,  
71003 Heraklion, Crete, Greece}
\affil{Institute of Astrophysics, Foundation for Research and Technology, Hellas}

\begin{abstract}

We report the discovery of statistically significant spatial structures in the projected two-dimensional 
distributions of the Globular Cluster (GC) systems of 10 among the brightest galaxies in the Fornax Cluster.
We use a catalog of GCs extracted from the Hubble Space Telescope (HST) Advanced Camera for Surveys (ACS) 
Fornax Cluster Survey (ACSFCS) imaging data. We characterize the size, shape and location relative 
to the host galaxies of the GC structures and suggest a classification based on their morphology
and location that is suggestive of different formation mechanisms. We also investigate the GC 
structures in the context of the positions of their host galaxies relative to the general spatial 
distributions of galaxies and intra-cluster globular clusters in the Fornax Cluster. We finally 
estimate the dynamical masses of the progenitors of some GC structures, under the assumption 
that they are the relics of past accretion events of satellite galaxies by their hosts. 

\end{abstract}

\section{Introduction}
\label{sec:introduction}

The study of Globular Cluster systems of massive elliptical galaxies has been spurred
by the growing evidence that these systems can be used as fossil records of the formation 
history of their host galaxies~\citep{harris1979,brodie2006} and can constrain the virial mass
of the halo of the hosts~\citep{blakeslee1997,spitler2009,burkert2020}. This opportunity has led, 
in turn, to a quick growth of our knowledge of the global 
properties of the GC systems. 
The large-scale spatial distribution of Globular Clusters (GCs) has been investigated in 
depth: GCs extend radially to larger radii than the diffuse stellar 
light of the galaxies~\citep{harris1986} and the most luminous galaxies have more centrally 
concentrated GC distributions than fainter hosts~\citep{ashman1998}. The radial profiles 
of metal-rich (red) and metal-poor (blue) GCs differ: red GCs follow more closely the 
light profile of the stellar bulge, with the exception of central flattened cores in the GC 
distribution, and blue GCs show a flatter, more extended 
distribution~\citep{rhode2001,dirsch2003,bassino2006,mineo2014}. 

Several studies of the full two-dimensional spatial distributions of GC systems have indicated 
that their ellipticity and position angles are consistent, on average, with those of the stellar 
spheroids of the host galaxy~(e.g. NGC4471 in~\citealt{rhode2001}, 
NGC1399 in~\citealt{dirsch2003,bassino2006}, NGC3379, NGC4406 and NGC4594 in~\citealt{rhode2004}, 
NGC4636 in~\citealt{dirsch2005}, multiple galaxies in~\citealt{hargis2012}, NGC3585 and NGC5812 
in~\citealt{lane2013}, NGC4621 in~\citealt{bonfini2012} and~\citealt{dabrusco2013}). More 
recent works that took advantage of Hubble Space Telescope imaging data collected for 
several galaxies in the Advanced Camera for Surveys (ACS) Virgo Cluster Survey 
(ACSVCS)~\citep{cote2004}, have uncovered azimuthal anisotropy in the GC distributions~\citep{wang2013}, 
with GCs preferentially aligned with the major axis in host galaxies that have significant elongation and 
intermediate to high luminosity. By combining HST and large-field, ground-based observations, the 
geometry of the GC color subclasses in ETGs have also been found to differ: red GCs correlate with the 
ellipticity and size of the host's stellar halo, while the blue GCs populations show smaller ellipticities and 
extend to larger radii~\citep{park2013,kartha2014}. These results suggest two different formation 
mechanisms for red and blue GCs and the potential presence of two distinct halo components, with red GCs 
forming at the same time of the stars of the galaxy and blue GCs possibly formed in low-mass dwarf galaxies 
subsequently accreted to their current host~\citep{forbes1997,cote1998,cantiello2018}.

The average radial distributions and metallicity properties of the red and blue GCs are
reproduced by N-body smoothed particle hydrodynamical simulations~\citep{bekki2002}
under the assumption that the elliptical galaxy was the result of early major dissipative merging. 
Tidal stripping of GCs and accretion of satellites in elliptical galaxies in high density regions have been
suggested as the main mechanisms to reconcile the simulations with the data~\citep{bekki2003}. The gap 
between observations of the spatial distribution of GCs and galaxy evolution models has been further 
narrowed by recent cosmological simulations combined with semi-analytic models for the formation and 
evolution of stellar systems~\citep[E-MOSAICs, see][]{pfeffer2018} that have achieved sufficient mass 
and spatial resolutions to resolve individual GCs and match their evolving spatial distribution with the 
growth of the host galaxy. In particular,~\cite{kruijssen2019} shown that the spatial distribution 
of it in situ and accreted GCs in Milky-Way-like galaxies differ from each other and deviate 
significantly from smooth radial and azimuthal distributions~(see Figure 1 therein). 

Although a comprehensive model connecting GCs and galaxy formation is still missing, 
major merging (either dissipative or dissipationless), tidal interactions and minor mergers/accretion 
of satellite galaxies have been demonstrated to have important roles in shaping the observed GC systems. 
While the formation of young, metal-rich GCs during mergers is emerging as an important ingredient in the 
make-up of the observed GCs of elliptical galaxies~\citep{bekki2003,brodie2006,wang2013}, 
early works from~\cite{cote1998} and~\cite{pipino2007} found that the global properties of the GC systems
of massive elliptical galaxies (radial distribution, specific frequency and metallicity) can be only explained 
if one assumes that a large fraction of GCs have originated from satellite galaxies and have been stripped 
by the accreting galaxy. More recently, cosmological simulations have shown
that in the Milky Way the fraction of GCs accumulated through accretion of satellite galaxies can be as large 
as 50\% of the total observed GCs~\citep{kruijssen2020}. The importance of satellite accretion in the build-up
of the massive GCs populations has been emphasized by the observation of localized, two-dimensional (2D) 
inhomogeneities and streams in the GC systems of elliptical galaxies. Significant features in the projected 
spatial distribution of GCs have been observed in NGC4261~\citep{bonfini2012,dabrusco2013}, 
NGC4649~\citep{dabrusco2014a}, NGC4278~\citep{dabrusco2014b} 
and NGC4365~\citep{blom2012}.~\cite{dabrusco2015} observed multiple spatial structures in the distributions 
of the GC systems of the ten brightest Virgo cluster galaxies. 

In this paper, we analyze the 2D spatial distribution of the GCs in ten among the brightest galaxies 
in the Fornax cluster, using a method based on the K-Nearest-Neighbor (KNN) technique and 
Monte Carlo simulations that has been previously applied to several other 
galaxies~\citep{dabrusco2013,dabrusco2014a,dabrusco2014b,dabrusco2015}. After the Virgo 
cluster, the Fornax cluster is the second closest rich cluster of galaxies: at its distance of $\approx$20.0 
Mpc~\citep{blakeslee2009}, GCs are marginally resolved and can be detected down to relatively low flux 
levels with reasonable integration 
times in HST imaging. Moreover, our method~\citep{dabrusco2013} for the characterization of the 
spatial distribution of GCs is more accurate for large samples of GCs like those hosted by the bright 
galaxies in the Fornax cluster~\citep{liu2019}.
Similarly to Virgo, the Fornax cluster provides a wide range of 
local densities and allows us to probe different stages of the evolution of massive galaxies in different 
environments. The Fornax cluster has two dynamically distinct components~\citep{drinkwater2001}: the 
main cluster centered on NGC1399 and a subcluster, located SW of NGC1399 and centered on NGC1316, which 
is likely infalling toward the cluster core. Additional studies suggest that 
the core of the main cluster is dynamically evolved, and there are several lines of evidences supporting 
the ongoing infall of single galaxies in its outskirts~\citep{liu2019} and interactions between galaxies in the 
core~\citep{iodice2016}. Moreover,
the displacement of the X-ray emission from the hot intracluster gas relative to the closest Fornax
galaxies suggests that the cluster is not relaxed and may be undergoing a merger~\citep{paolillo2002}.

The paper is organized as follows: in Section~\ref{sec:data} we describe the main properties of
the Fornax cluster galaxies investigated and the archival data used to characterize the spatial
distributions of their GC systems, while Section~\ref{sec:coverage} focuses on the spatial coverage
of the observations. Section~\ref{sec:method} synthetically describes the method used to discover the GC 
spatial structures and determine their properties, and the structures discovered in each of the galaxies studied
in this work are described in Appendix~\ref{sec:appendix1}. The properties of the GC structures are discussed in 
Section~\ref{sec:discussion} and its subsections. Finally, our conclusions are summarized in 
Section~\ref{sec:conclusions}. 

We use cgs units unless otherwise stated. Optical magnitudes used in this manuscript are in the Vega system 
and are not corrected for the Galactic extinction. Standard cosmological parameter values have been used for 
the all calculations~\citep{bennett2014} and we used galaxy distances from~\cite{blakeslee2009}.

\section{Data}
\label{sec:data}

In this paper, we use the catalog of GCs extracted from images obtained with the HST Advanced Camera
for Survey (ACS) for the ACS Fornax Cluster Survey (ACSFCS) project~\citep{jordan2007}, complemented by 
wide-field ACS observations by~\cite{puzia2014} for the region surrounding NGC1399 (more details 
in Section~\ref{subsec:ngc1399} of the Appendix). ACSFCS obtained ACS 
pointings for 43 early-type Fornax cluster galaxies in the F475W (SDSS $g$) and F850LP (SDSS $z$) bandpasses, 
with exposures of 680 and 1130 seconds, respectively.~\cite{jordan2015} extracted a catalog of 
9136 GC candidates using both photometry and half-light radii measured in both filters. The total 
number of {\it bona fide} GCs, 
i.e. with probability $p_{\mathrm{GC}}\!\geq\!0.5$\footnote{$p_{\mathrm{GC}}$ is defined by~\cite{jordan2015}
as the probability of a source not consistent with a point source to be a GC as a function of its magnitude
and half-light radius, given a two components Gaussian 
mixture model in the $z_{850}$-$r_{h}$ plane for a sample of {\it bona fide} GCs observed in Virgo 
cluster~\citep{ferrarese2000} and background galaxies from control
fields~\citep{jordan2009}.}, in the ACSFCS catalog is 6275. We investigate the 
spatial distribution of GCs in ten among the brightest galaxies that are associated with the largest 
numbers of ACSFCS GCs. Nine galaxies in this sample are among the ten brightest Fornax cluster galaxies in 
$B_{\mathrm{T}}$ magnitude (ESO 359-G07 ranks 8-th), with the exception of NGC1336 which ranks 
16th~\citep{jordan2007}. Table~\ref{tab:gcnumbers} 
lists the galaxies in the sample and their main properties. In order to 
investigate the distribution of GCs for red and blue GC subclasses separately, we model the $g\!-\!z$ color 
distribution of each galaxy with two Gaussians. The $g\!-\!z$ color thresholds used to separate the two GCs 
color classes are determined using the Gaussian Mixture Modeling code (GMM)~\citep{muratov2010} as the 
value corresponding to equal probabilities of belonging to the red or blue components 
Gaussians~(Table~\ref{tab:gcnumbers}). Even in the cases of GC color distributions whose 
bimodality is not statistically significant, we assumed as color threshold the $g\!-\!z$ value 
for which the probability of being a red or a blue GC is equal in order to compare the properties 
of the spatial distributions of GCs in different color intervals. 

\begin{deluxetable}{llllllllllll}
	\tablecaption{Properties of the galaxies investigated in this paper bf sorted by decreasing total 
	size of their GCs systems}.
	\label{tab:gcnumbers}
	\tablehead{
	    \colhead{Name\tablenotemark{a}} &
	    \colhead{N$_{GC}$\tablenotemark{b}} &
	    \colhead{N$_{\mathrm{blue}}$\tablenotemark{c}} &
	    \colhead{N$_{\mathrm{red}}$\tablenotemark{d}} &	
	    \colhead{morphology\tablenotemark{e}} &
	    \colhead{\% D$_{25}$ area covered\tablenotemark{f}} &		    
	    \colhead{$B_{T}$\tablenotemark{g}} &	    
	    \colhead{r$_{e}$ [\arcsec]\tablenotemark{h}} &
	    \colhead{D$_{25}$ [r$_{e}$]\tablenotemark{i}} &	   
	    \colhead{PA\tablenotemark{j}} &		  
	    \colhead{($g\!-\!z)_{\mathrm{thresh}}$\tablenotemark{k}}
	    }
	\startdata
    NGC1399	&	1077 (1126) & 426 & 651 & 	E0      &	45\%    &    10.6   &	367.28 & 1.13   &	0	&	1.22                \\
    NGC1316	&	647 (1006)  & 325 & 322 & 	S0 (pec)&  	37\%    &     9.4	&	288.40 & 2.5    &	50	&	1.10$^{\ast\ast}$    \\
    NGC1380	&	424 (558)	& 117 & 307 &	S0/a    &  	78\%    & 	 11.3   &	 86.10 & 3.35   &	7	&	1.12$^{\ast}$        \\
    NGC1404	&	380 (429)	& 169 & 211 &	E2    	&  	100\%   & 	 10.9	&	 86.10 & 2.31   &	0	&	1.17                \\
    NGC1427	&	362 (412)   & 256 & 106 &	E4    	&  	99\%    & 	 11.8   &	 65.31 & 3.33   &	76	&	1.19$^{\ast}$       \\
    NGC1344	&	280 (397)	& 182 & 98  &	E5    	&  	65\%    & 	 11.3   &	 86.14 & 3.31   &	165	&	1.02$^{\ast}$       \\
    NGC1387	&	306 (381)	& 96  & 210 &	SB0    	&  	100\%   & 	 12.3   &	 50.70 & 3.32   &	0	&	1.13                \\
    NGC1374	&	320 (367)	& 154 & 166 &	E0	    &  	100\%   & 	 11.9   &	 52.48 & 2.80   &   0	&	1.09$^{\ast}$        \\
    NGC1351	&	274 (357)	& 132 & 142 &	E5    	&  	100\%   & 	 12.3   &	 65.31 & 2.58   &	140	&	1.08$^{\ast}$        \\
    NGC1336	&	276 (355)   & 91  & 185 &	E4    	&  	100\    & 	 13.3   &	 50.70 & 2.53   &	22	&	1.11$^{\ast}$    \\
	\enddata
    \tablecomments{(a): NGC identifier of the galaxy; 
                   (b): Number of $p_{\mathrm{GC}}\!\geq\!0.5$ ACSFCS GCs used in the experiments described in this paper 
                   (in parenthesis, the total number of ACSFCS candidate GCs); 
                   (c): Number of blue GCs; 
                   (d): Number of red GCs; 
                   (e): Morphology of the galaxy from~\cite{jordan2007,ferguson1989};
                   (f): Fraction of the D$_{25}$ area of each galaxy covered by the ACSFCS observation;
                   (g): $B$ band magnitude from~\cite{jordan2007,ferguson1989};
                   (h): Effective radius of the galaxy in arcseconds from~\cite{lauberts1989}~\footnote{More recently,~\cite{iodice2019} 
                   measured the effective radii of galaxies within the Fornax virial radius using Fornax Deep Survey~\citep[FDS,][]{iodice2016} observations. Their r$_{e}$ 
                   in the $g$ band are all in agreement within 10\% with the~\cite{lauberts1989} values for all galaxies in our sample except NGC1404, for which 
                   effective radius reported by~\cite{iodice2019} is $\sim$201\arcsec.};                    
                   (i): Major diameter of the elliptical isophote of the galaxy corresponding to surface brightness of 
                   25 mag/arcsecond$^2$ from~\cite{corwin1994}, expressed in units of r$_{e}$; 
                   (j): Position angle of the galaxy from~\cite{corwin1994}; 
                   (k): Color ($g\!-\!z$) threshold used to define red and blue GC subclasses. One or two asterisks identify color distributions
                   characterized by bimodality with low or no statistical significance.}
\end{deluxetable} 

\section{Spatial coverage}
\label{sec:coverage}

\begin{figure*}
    \centering
	\includegraphics[width=0.49\linewidth]{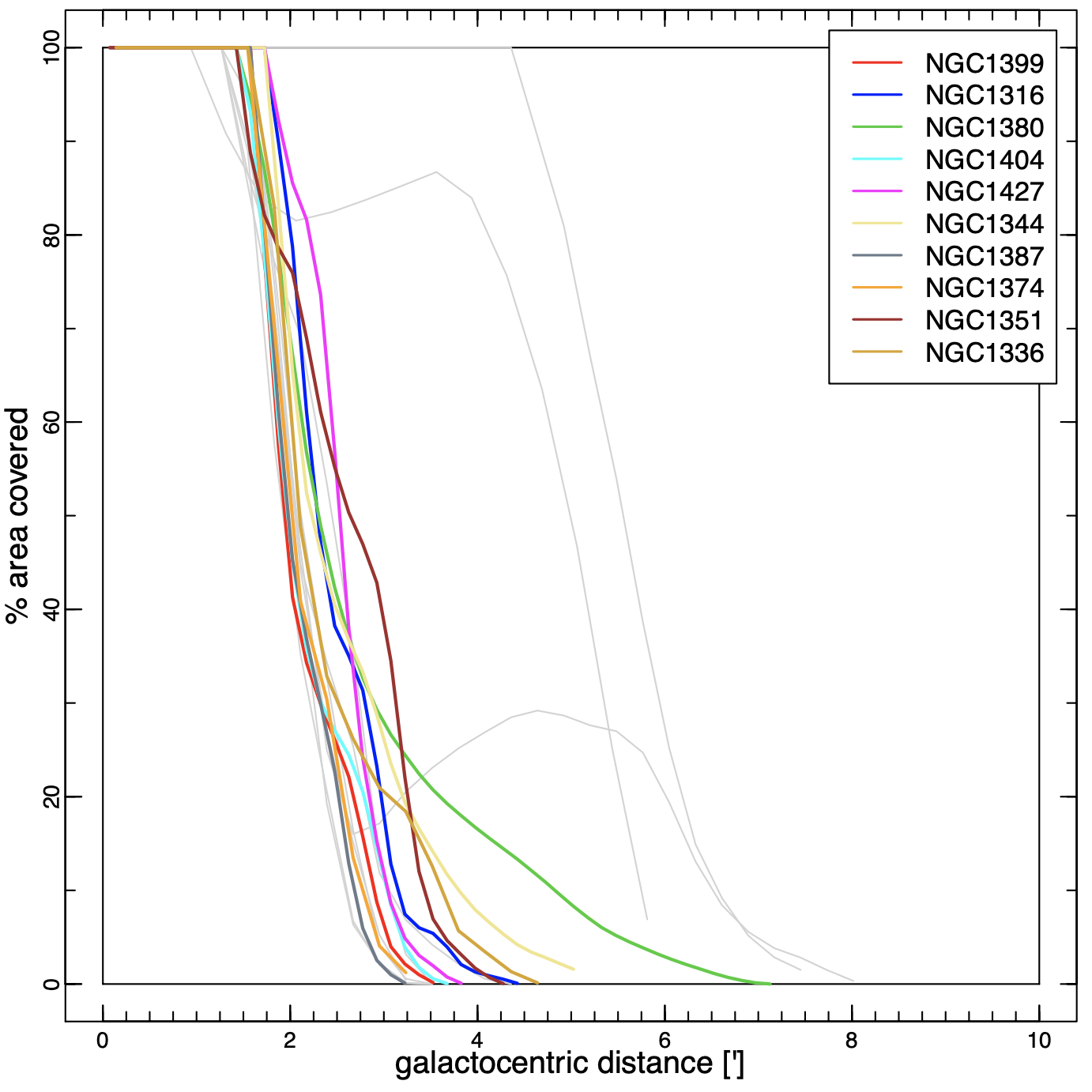}
	\includegraphics[width=0.49\linewidth]{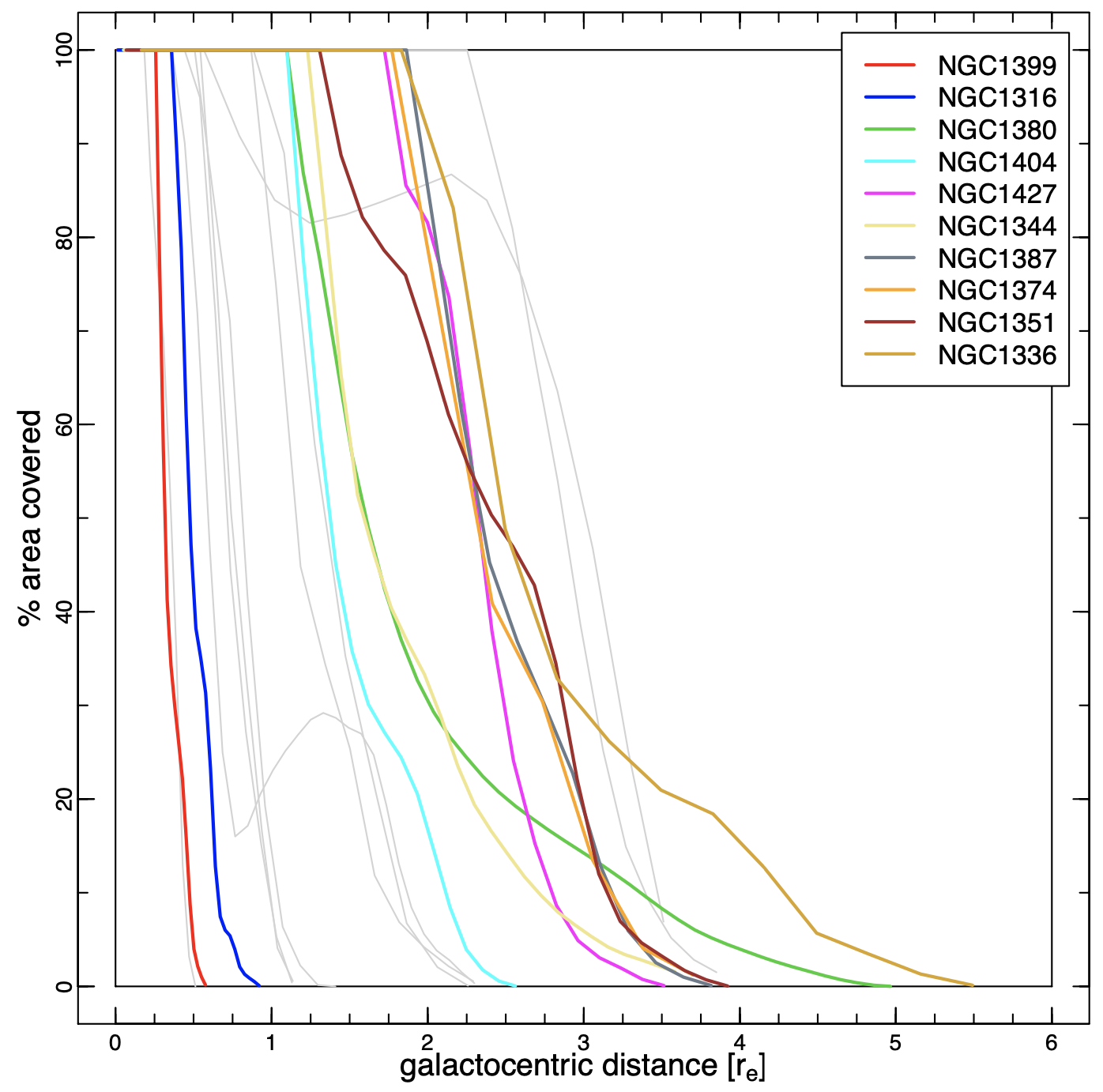}
    \caption{Left: coverage of ACSFCS galaxies studied in this paper as a function of the projected
    galactocentric distance. Right: coverage of ACSFCS galaxies as a function of the galactocentric
    distance in units of effective radius of each galaxy. In both plots, gray lines in the background
    represent the coverage of Virgo cluster galaxies investigated in~\cite{dabrusco2015}. Additional 
    ACS observations of NGC1399 from~\cite{puzia2014} (see Section~\ref{subsec:ngc1399})
    increase the coverage to $\approx 1\ \mathrm{r}_{\mathrm{e}}$ but are not reported in the plot 
    because the GCs selection method is not consistent with the ACSFCS sample.}
 	\label{fig:coverage}
 \end{figure*}

The discovery of structures in the distribution of GCs within galaxies depends critically on the 
spatial coverage of the GC system.~\cite{dabrusco2015} estimated that the available ACS HST observations of 
the ten Virgo cluster galaxies employed in their study covered less than 50\% of the total area of the galaxies 
within the $D_{25}$ elliptical isophote at 
1.5 effective radii for most targets. Figure~\ref{fig:coverage} shows the percentage of the Fornax cluster 
galaxies areas analysed in this paper as a function of the projected galactocentric distance (left) and the 
effective radius (right) of each target galaxy, respectively. While the fraction of galaxy area covered is negligible 
for all targets at $r\!\geq\!3\arcmin$, the fraction of total area covered at $r/r_{e}\!=\!2$ and $r/r_{e}\!=\!2.5$ is 
$\sim 75\%$ and $\sim 55\%$ respectively, and coverage for five galaxies is $\geq\!90\%$ at 2.5 $r_{e}$, thanks 
to the combined effect of the larger distance of the Fornax cluster relative to the Virgo cluster~($\approx$20.0 Mpc vs $\approx$
16.5 Mpc according to~\citealt{blakeslee2009}) and the relatively smaller effective radii of seven of the ten 
galaxies~(Table~\ref{tab:gcnumbers}). 

\section{Method}
\label{sec:method}

The spatial distribution of GCs around the ten galaxies studied in this paper has been 
characterized using the method described in~\cite{dabrusco2013,dabrusco2015}. This method employs 
residual maps obtained by comparing the density map derived from the observed spatial 
distribution of GCs with the density maps of simulated random spatial distributions of 
GCs that follow the spatial distribution of the observed stellar surface brightness 
of the host galaxy. The procedure for determining the residual 
maps of the GC distributions can be summarized as follows: 

\begin{itemize}
    \item Density maps of the observed spatial distribution of GCs are obtained with the $K$-Nearest Neighbor
    (KNN) method~\citep{dressler1980} on a fixed rectangular grid covering the area of the galaxy 
    where GCs are observed. The density at the center of each cell is defined as:
    \begin{equation}
        D_K\!=\!\frac{K}{A_D(d_{K})}
        \label{eq:density}
    \end{equation}
    where $K$ indicates the $K$-th closest GC (or {\it neighbor}) and $A_D(d_{K})\!=\!\pi\cdot\!d_{k}^{2}$ 
    is the area of the circle with radius equal to the distance of the $K$-th neighbor. 
    \item Multiple simulated realizations of the observed GC system are generated using a Monte Carlo 
    approach. The radial and azimuthal positions of the simulated GCs are drawn from the elliptical 
    radial distribution of the observed GCs with flattening and orientation set to the observed 
    values of the diffuse stellar light of the host galaxy. The total and radial bin-by-bin numbers 
    of simulated GCs match the observed values. 
    \item Density maps of the observed GC 
    distribution are generated for each distinct realization of the simulated GCs spatial distribution 
    on the same spatial grid and for each different value of the $K$ parameter;
    \item The residual map is obtained by subtracting, on a cell-by-cell basis, from the observed density map 
    the average density of all the simulated maps, so that the residual value
    R$_{i}$ of the $i$-th cell of the grid is defined as: 
    \begin{equation}
        R_{i}\!=\!\frac{(O_{i}-<\!S\!>_{i})}{<\!S\!>_{i}}
        \label{eq:residual}
    \end{equation}
    where $O_{i}$ and $<\!S\!>_{i}$ are the observed density and the average of the density 
    from all the simulations in cell $i$. Details on the determination of the cell size
    for the results discussed in this paper can be found in Section~\ref{sec:spatialdistribution}.
\end{itemize}

The simulated GCs are drawn from the radial distribution of observed GCs in elliptical bins 
obtained with a fixed bin size of 15\arcsec. The properties of the residual maps change 
negligibly with bin sizes in the $[10\arcsec,20\arcsec]$ interval. Bins larger than 20\arcsec\  
introduce step-like artefacts in two dimensional distribution of simulated GCs, while below 10\arcsec\ 
the number of empty bins increases very rapidly for galactocentric distance larger than 1\arcmin.5.

$K$ is a free parameter of the KNN method: it measures the expected scale of the investigated 
spatial structures and the density contrast of these structures over the average local density. 
Different $K$ values highlight spatial features at different scales: small $K$ values allow 
the exploration of small features, while larger values of $K$ bring out more extended structures. The loss 
of spatial resolution for large $K$ values is balanced by the smaller relative fractional error 
which is proportional to the inverse of the square root of $K$:

\begin{equation} 
    \frac{\sigma_{D(K)}}{D(K)}=\frac{1}{\sqrt{K}} 
    \label{eq:fracerror}
\end{equation}

Moreover, larger values of $K$ are more suitable to detect structures located within the D$_{25}$ of 
the galaxies, where the overall density of GCs is larger, because only high-contrast structures 
can be reliably detected over this high-density background. Smaller $K$ values, on the other hand, 
are more apt at detecting structures in regions where the total number of GCs is smaller, as in 
the outskirts of the host galaxies.

The distributions of simulated density values for each cell of the final residual maps are well
approximated by Gaussians, thus simplifying the estimation of the statistical significance of 
each cell. The spatial structures in the residual maps are determined by selecting
sets of $N\!\geq\!30$ adjacent cells with non-negative residual values. The total significance 
of each observed spatial structure (which differs from the significance of the single cells included in the
structure) is determined by estimating the frequency of structures with comparable features 
(number of cells, average residual value and shape, see Section~\ref{subsec:spatialstructures_morphology}) 
across residual maps obtained from single, distinct simulated distribution of GCs. The search of simulated 
structures with properties similar to those of the observed ones is 
performed over the whole area of each host galaxy where GCs are observed (i.e. is not limited 
to the specific range of galactocentric distances where the structures are detected), therefore 
yielding a lower limit to the statistical significance of the structures. 
The search of structures in each simulated residual distribution is performed similarly to the 
observed GC residual distribution: according to equation~\ref{eq:residual}, the value of the residual 
in the $i$-th cell of the $S^{j}$ simulated distribution of GCs is defined as: 

\begin{equation}
    R_{i}(S^{j})\!=\!\frac{(S^{j}_{i}-<\!S\!>_{i})}{<\!S\!>_{i}}
    \label{eq:residualsimulated}
\end{equation}

The same algorithm used for the detection of structures in the observed residual map is then applied to 
each of the single simulated residual maps. The statistical significance of each observed residual
structure is then expressed as the fraction of simulated density maps where mock residual structures 
with comparable features, have been detected.

While the efficiency of the detection of GCs as a function of the radial position relative to the 
center of the host galaxy varies because of the specific shape of the density profile of the host and 
the surface brightness
profile that changes the background over which GCs are detected, the method described above is self-consistent 
because it is based on the GC observed radial density profile used to seed the simulations. For this reason, 
GCs structures detected at different galactocentric distances with the same significance are associated with 
the same relative density contrast with respect to the underlying GC population. Given the differences 
between the density profiles and GC detection efficiencies in different hosts, the comparison of significance 
of structures observed in different host galaxies is not meaningful, especially in the core of the 
galaxies where the shape of the stellar light brightness profile can affect significantly the density of
observed GCs. 

\section{The spatial distribution of Globular Clusters in Fornax cluster galaxies}
\label{sec:spatialdistribution}

In this paper, we investigate the spatial distributions of GCs observed in ten 
galaxies among the brightest in the Fornax cluster. The descriptions of the GC structures 
observed in each galaxy and their relation with the known properties of the GC system of 
the host can be found in Appendix~\ref{sec:appendix1}. To maximize the statistical significance of 
our results, the entire GC system (blue + red) was used to detect the GC spatial structures.
The statistical significance of each structure was then evaluated also for the red and blue GC distributions,
(see Table~\ref{tab:gcstructures}) assuming the shape and size
of the structures as determined in the residual maps for the whole GCs samples. The properties of 
the structures for all GCs are 
discussed in the context of the features of the color-based residual maps in Appendix~\ref{sec:appendix1}. 
The residual maps were estimated on a regular grid of cells with approximately equal angular extents 
along the Right Ascension and Declination axes. 
The cell sizes range from $\sim\!3$\arcsec to $\sim\!5$\arcsec along both axes for different host 
galaxies, corresponding to physical linear sizes from 0.3 to 0.5 kpc that lead to a factor $\sim2$ maximum 
difference between physical areas of structures containing the same number of cells.
This choice guarantees that 
residual maps of different GC distributions account for similar number of cells in the regions where GCs 
are observed, so that a meaningful comparison 
is possible across residual maps of different galaxies investigated here and between 
Fornax and Virgo galaxies (see Section 3 of~\citealt{dabrusco2015}). The adoption of a 
weakly galaxy-dependent cell size ensures that the residual maps for each galaxy 
are insensitive to the varying fraction of the host galaxy area imaged relative to its total extension 
(see column $f$ in Table~\ref{tab:gcnumbers}). Following~\cite{dabrusco2015}, we compared the residual 
maps obtained by picking ten different, linearly spaced cell sizes (along both axes) within a $\pm30\%$ interval 
centered on the adopted cell size to rule out significant systematic effects on the spatial 
features observed.

The residual maps were generated for 10$^5$ distinct simulated distributions for each galaxy and GCs 
color class (all, red and blue) with $K=\{2, 5, 10, 15, 20\}$. Since the ACSFCS data employed in this 
paper cover both the central, dense regions of the galaxies as well as very low density outskirts, 
the results presented in this paper are those obtained with an intermediate value $K\!=\!10$. Moreover,
$K\!=\!10$ produces structures whose typical scale, measured as the average of the structures'
largest transversal size, matches the average spatial scale of the GCs spatial structures detected in 
the spatial distribution of the ACS VCS by~\cite{dabrusco2015}. With this $K$ value, our approach can 
detect spatial 
structures of scales ranging between $\approx$1 kpc to $\approx$30 kpc across different density 
and background regimes. Large values of $K$ tend to produce residual maps where distinct spatial structures 
are meshed and their spatial properties averaged out, 
while for lower values the residuals are dominated by small scale density peaks that make it very 
difficult to determine the shape, extension and orientation of the structures. 

Each detected spatial structure can be characterized by 1) the statistical significance~(see 
Section~\ref{sec:method} for details), 2) the size measured as the number of connected cells contained 
in a structure, 3) the total number of observed GCs N$^{\mathrm{(obs)}}_{\mathrm{GCs}}$ 
within the structure, 4) the excess number of GCs N$^{\mathrm{(exc)}}_{\mathrm{GCs}}$ and 5) the physical 
area. Based on their size, the spatial structures have been classified in 
intermediate ($30\!\leq\!N_\mathrm{pxl}\!\le\!60$) and large ($N_\mathrm{pxl}\!\geq\!60$). In the remainder 
of the paper, large and intermediate structures detected in each galaxy will be labeled with a capital or 
small letter respectively, uniquely associated with the host followed by a unique numerical index (e.g. A2 for 
the second among the large structures in NGC1399 and c1 for the first among the intermediate structures in 
NGC1380). The excess number of GCs N$^{\mathrm{(exc)}}_{\mathrm{GCs}}$ is calculated by subtracting to the 
observed number of GCs N$^{\mathrm{(obs)}}_{\mathrm{GCs}}$ in each spatial structure the number of expected 
GCs N$^{\mathrm{(exp)}}_{\mathrm{GCs}}$ 
in the same structure, where N$^{\mathrm{(exp)}}_{\mathrm{GCs}}$ is the average number of GCs located in the 
same set of cells included in the observed structure across the simulated GCs distributions used to estimate
the residual maps~(Section~\ref{sec:method}). Since the detection of spatial structures is performed 
on the residual maps based on the observed and simulated spatial densities of the GC distribution, 
N$^{\mathrm{(exc)}}_{\mathrm{GCs}}$ can be negative in some lower-significance structure because of the uneven 
distribution of GCs within the spatial boundaries of the structures. The smooth boundaries of the spatial 
structures 
shown in the plots in the following Sections and the Appendix~\ref{sec:appendix1}, have been determined by 
applying a smoothing algorithm to the rectangular boundaries derived from the set of cells included in each 
structure. The values of the main properties for all intermediate and large structures detected are listed in 
Table~\ref{tab:gcstructures}. 

\begin{deluxetable}{cccccccccccc}
	\tablecaption{Properties of the intermediate and large GCs spatial structures.}
	\tabletypesize{\tiny}
	\label{tab:gcstructures}
	\tablehead{
	    \colhead{Galaxy\tablenotemark{a}} &
	    \colhead{Structure\tablenotemark{b}} &
	    \colhead{Class\tablenotemark{c}} &
	    \colhead{Significance\tablenotemark{d}} &	
	    \colhead{Size\tablenotemark{e}} &
	    \colhead{N$^{\mathrm{(obs)}}_{\mathrm{GCs}}$\tablenotemark{f}} &	    
	    \colhead{N$^{\mathrm{(exc)}}_{\mathrm{GCs}}$\tablenotemark{g}} &	   
	    \colhead{Physical size\tablenotemark{h}} &
	    \colhead{$\overline{g\!-\!z}$\tablenotemark{i}} &
	    \colhead{$\overline{g\!-\!z}_{\mathrm{bluest}}$\tablenotemark{j}} &
	    \colhead{Class\tablenotemark{k}} &	
	    \colhead{M$_{\mathrm{dyn}}$\tablenotemark{l}} 
	    }
	\startdata
    NGC1399	&	A1    & All, red$^{\ast}$, {\bf blue}$^{\ast}$                      &  3.4$\cdot10^{-5}$    & 219 (3504)  & 110   & 37.4  & 25   & 1.25   & 0.93   & RS &   9.94 \\
    	    &	A2    & {\bf All}, {\bf blue}$^{\ast}$                              &  $<\!10^{-5}$         & 272 (4352)  & 151   & 47.9  & 43.5  & 1.25  & 0.93   & RS &   10.04 \\
    	    &	A3    & {\bf All}, red, {\bf blue}$^{\ast}$                         &  4.1$\cdot10^{-5}$    & 85 (1360)   & 21    & 8.1   & 13.6  & 1.25  & 1.09   & AD &   9.30  \\
    	    &	a1    & All, blue                                                   &  1.3$\cdot10^{-4}$    & 55 (880)    & 14    & 3.5   & 8.8   & 1.23  & 0.82   & - &   -  \\
    	    &	a2    & {\bf All}, {\bf red}                                        &  1.5$\cdot10^{-5}$    & 50 (800)    & 17    & 7     & 8     & 1.28  & 1.05   & - &   -   \\
    	    &	a3    & All, {\bf red}                                              &  6.3$\cdot10^{-4}$    & 58 (928)    & 16    & 6     & 9.3   & 1.16  & 0.8    & - &   -  \\   
    	    &	a4    & All, {\bf red}                                              &  7.2$\cdot10^{-4}$    & 42 (672)    & 15    & 2.7   & 6.7   & 1.23  & 0.83   & - &   -  \\   
    NGC1316	&	B1    & {\bf All}$^{\ast}$, {\bf blue}$^{\ast}$                     &  $<\!10^{-5}$         & 434 (6944)  & 118   & 42.4  & 69.4  & 1.08  & 0.82   & Hy &   $^{\ast}$  \\
    	    &	B2    & {\bf All}$^{\ast}$, {\bf red}$^{\ast}$, blue                &  1.7$\cdot10^{-5}$    & 133 (2128)  & 22    & 7.2   & 21.3  & 1.09  & 0.84   & AD &   9.25  \\
        	&	B3    & {\bf All}$^{\ast}$, {\bf red}$^{\ast}$, {\bf blue}          &  1.9$\cdot10^{-5}$    & 304 (4864)  & 49    & 21    & 48.6  & 1.14  & 0.94   & RS &   9.70   \\
    	    &	b1    & {\bf All}, blue                                             &  7.3$\cdot10^{-5}$    & 55 (880)    & 11    & 5.6   & 8.8   & 1.09  & 0.94   & - &   -  \\
    	    &	b2    & All, {\bf red}, blue                                        &  5.4$\cdot10^{-4}$    & 37 (592)    & 17    & 8.6   & 5.9   & 1.19  & 0.97   & - &   -   \\
    	    &	b3    & All, blue                                                   &  7.2$\cdot10^{-4}$    & 54 (864)    & 15    & 3.8   & 8.6   & 1.17  & 0.68   & - &   -  \\
    	    &	b4    & All, {\bf red}                                              &  8.6$\cdot10^{-4}$    & 30 (480)    & 7     & 2.7   & 4.8   & 0.98  & 0.72   & - &   -  \\
    NGC1380	&	C1    & {\bf All}, blue                                             &  4.6$\cdot10^{-5}$    & 162 (2592)  & 5     & 2.4   & 25.9  & 0.98  & 0.78   & AD &   8.79   \\
     	    &	C2    & {\bf All}, red, {\bf blue}                                  & $<\!10^{-5}$          & 211 (3376)  & 20    & 8.1   & 33.8  & 1.14  & 0.98   & RS &   9.30   \\
    	    &	C3    & All$^{\ast}$, {\bf red}$^{\ast}$, blue                      & 9.8$\cdot10^{-5}$     & 202 (3232)  & 41    & 12.6  & 32.3  & 1.23  & 0.92   & Hy &   $^{\ast}$ \\  
    	    &	C4    & {\bf All}, {\bf red}, blue                                  & 2.3$\cdot10^{-5}$     & 177 (2832)  & 21    & 6.8   & 38.3  & 1.19  & 0.9    & AD &   9.22   \\
    	    &	c1    & All$^{\ast}$, red, {\bf blue}$^{\ast}$                      & 2.7$\cdot10^{-5}$     & 49 (784)    & 27    & 11.4  & 7.8   & 1.20  & 0.99   & - &   -  \\
    	    &	c2    & All, blue                                                   & 8.6$\cdot10^{-4}$     & 34 (544)    & 16    & 6.4   & 5.4   & 1.19  & 0.98   & - &   -  \\
    NGC1404	&	D1    & {\bf All}, {\bf blue}                                       & $<\!10^{-5}$          & 208 (3328)  & 16    & 7.2   & 33.3  & 1.05  & 0.86   & AD &   9.25   \\
    	    &	D2    & All, {\bf red}$^{\ast}$                                     & 4.1$\cdot10^{-5}$     & 97 (1552)   & 9     & 3.9   & 15.5  & 1.20  & 0.9    & AD &   8.99  \\
    	    &	D3    & {\bf All}, {\bf red}$^{\ast}$                               & 6.1$\cdot10^{-5}$     & 71 (1136)   & 9     & 2.7   & 11.4  & 1.24  & 1.1    & AD &   8.84   \\
    	    &	D4    & {\bf All}, red, blue                                        & 3.5$\cdot10^{-5}$     & 206 (3296)  & 18    & 6.8   & 33    & 1.18  & 0.89   & AD &   9.23  \\
    	    &	D5    & {\bf All}, {\bf red}, blue                                  & 4.6$\cdot10^{-5}$     & 63 (1008)   & 28    & 12.6  & 10.1  & 1.24  & 0.97   & AD &   9.48  \\
    	    &	D6    & {\bf All}, {\bf red}, {\bf blue}                            & $<\!10^{-5}$          & 173 (2768)  & 22    & 11.4  & 27.7  & 1.12  & 0.95   & TS &   9.44   \\
    	    &	d1    & All, red                                                    & 2.5$\cdot10^{-4}$     & 33 (528)    & 5     & 1.7   & 5.4   & 1.19  & 0.95   & - &   - \\
    	    &	d2    & All                                                         & 9.3$\cdot10^{-4}$     & 52 (832)    & 11    & 3.8   & 8.3   & 1.35  & 1.11   & - &   -  \\
    NGC1427	&	E1    & {\bf All}, blue                                             & 1.4$\cdot10^{-5}$     & 171 (2736)  & 6     & 2     & 27.4  & 1.12  & 0.97   & AD &   8.70   \\
    	    &	E2    & {\bf All}, {\bf blue}                                       & 3.7$\cdot10^{-5}$     & 112 (1792)  & 29    & 11.8  & 17.9  & 1.08  & 0.97   & AD &   9.46  \\
    	    &	E3    & {\bf All}, {\bf red}$^{\ast}$, {\bf blue}                   & 2.1$\cdot10^{-5}$     & 133 (2128)  & 15    & 6.6   & 21.3  & 1.05  & 0.85   & AD &   9.21   \\
    	    &	E4    & {\bf All}$^{\ast}$, {\bf red}$^{\ast}$, {\bf blue}$^{\ast}$ & $<\!10^{-5}$          & 575 (9200)  & 73    & 29.1  & 92    & 1.09  & 0.88   & Hy &    $^{\ast}$ \\
    	    &	E5    & {\bf All}, {\bf blue}                                       & $<\!10^{-5}$          & 165 (2640)  & 19    & 8.9   & 26.4  & 1.06  & 0.92   & AD &   9.34   \\ 
    	    &	e1    & All, {\bf blue}                                             & 7.0$\cdot10^{-4}$     & 30 (480)    & 20    & 9.6   & 4.8   & 1.14  & 0.94   &  - &   - \\ 
    NGC1344	&	F1    & {\bf All}, {\bf red}, {\bf blue}                            & $<\!10^{-5}$          & 458 (7328)  & 42    & 19.3  & 73.3  & 0.94  & 0.75   & RS &   9.66   \\
    	    &	F2    & {\bf All}, {\bf red}, blue$^{\ast}$                         & 3.1$\cdot10^{-5}$     & 218 (3488)  & 38    & 11    & 34.9  & 1.00  & 0.76   & RS &   9.42   \\
    	    &	F3    & All, blue                                                   & 4.8$\cdot10^{-4}$     & 201 (3216)  & 12    & 4     & 32.2  & 0.94  & 0.78   & RS &   9.00   \\
    	    &	F4    & {\bf All}, (red), blue                                      & 3.1$\cdot10^{-5}$     & 336 (5376)  & 11    & 4.4   & 53.8  & 1.03  & 0.85   & AD &   9.04  \\  
    	    &	f1    & (All), {\bf red}                                            & 4.6$\cdot10^{-3}$     & 37 (592)    & 12    & 5.8   & 5.9   & 1.01  & 0.87   & - &   -  \\  
    NGC1387	&	G1    & {\bf All}, red, ({\bf blue})                                & 3.4$\cdot10^{-5}$     & 458 (7328)  & 20    & 3.7   & 73.3  & 1.22  & 0.88   & TS &   9.27   \\
    	    &	G2    & (All), ({\bf red}), (blue)                                  & 4.8$\cdot10^{-4}$     & 150 (2400)  & 2     & -0.9  & 24    & 1.47  & -      & AD &   $\downarrow$  \\
    	    &	G3    & {\bf All}, red, blue                                        & 6.4$\cdot10^{-4}$     & 61 (976)    & 18    & 6.9   & 9.8   & 1.31  & 0.98   & AD &   9.22 \\
    	    &	G4    & {\bf All}$^{\ast}$, ({\bf blue})                            & 5.9$\cdot10^{-5}$     & 192 (3072)  & 15    & 3.8   & 30.7  & 1.25  & 0.75   & Hy &   $^{\ast}$   \\
    	    &	G5    & {\bf All}, {\bf red}, blue                                  & 4.2$\cdot10^{-5}$     & 91 (1456)   & 22    & 7.2   & 14.6  & 1.24  & 0.93   & AD &   9.2  \\
    NGC1374	&	H1    & {\bf All}, {\bf blue}                                       & 8.4$\cdot10^{-5}$     & 95 (1520)   & 6     & 1.6   & 15.2  & 1.12  & 0.89   & TS &   8.61  \\
    	    &	H2    & {\bf All}, {\bf red}, {\bf blue}                            & $<\!10^{-5}$          & 536 (8576)  & 42    & 19.9  & 85.8  & 1.12  & 0.93   & RS &   9.67  \\
    	    &	H3    & All, red, blue$^{\ast}$                                     & 5.8$\cdot10^{-5}$     & 87 (1392)   & 32    & 14.5  & 13.9  & 1.18  & 0.96   & AD &   9.54  \\
    	    &	H4    & {\bf All}$^{\ast}$, red                                     & 2.3$\cdot10^{-5}$     & 196 (3136)  & 21    & 8.8   & 31.4  & 1.15  & 0.92   & TS &   9.33  \\
    	    &	h1    & All$^{\ast}$, red, (blue)                                   & 7.8$\cdot10^{-4}$     & 37 (592)    & 15    & 5.6   & 5.9   & 1.14  & 0.92   &  - &   - \\
    NGC1351	&	I1    & {\bf All}$^{\ast}$, (red), {\bf blue}$^{\ast}$              & 1.8$\cdot10^{-5}$     & 368 (5888)  & 20    & 5.9   & 58.9  & 1.03  & 0.84   & TS  &   9.16  \\
   	        &   I2    & {\bf All}, red$^{\ast}$, blue                               & 4.5$\cdot10^{-5}$     & 266 (4256)  & 48    & 18.6  & 42.6  & 1.15  & 0.91   & RS &   9.65  \\
    	    &	I3    & (All)                                                       & 5.3$\cdot10^{-5}$     & 113 (1808)  & 3     & 0.9   & 18.1  & 0.99  & -      & AD &   8.39  \\
    	    &   i1    & (All), (red), (blue)                                        & 4.7$\cdot10^{-3}$     & 37 (592)    & 0     & -0.7  & 5.9   &  -    & -      & - &   - \\
    	    &	i2    & {\bf All}, {\bf red}                                        & 7.5$\cdot10^{-5}$     & 46 (736)    & 9     & 2.9   & 7.4   & 1.10  & 0.83   & - &   - \\
   	        &   i3    & (All), (blue)                                               & 3.7$\cdot10^{-3}$     & 47 (752)    & 1     & 0     & 7.5   & 1.17  & -      & - &   -   \\
    NGC1336	&	J1    & {\bf All}, (red), {\bf blue}                                & $<\!10^{-5}$          & 309 (4944)  & 25    & 8.9   & 49.4  & 1.02  & 0.86   & AD &   9.34  \\
    	    &	J2    & All, (blue)                                                 & 5.4$\cdot10^{-5}$     & 167 (2672)  & 3     & -0.9  & 26.7  & 0.92  & -      & AD &   $\downarrow$  \\
        	&	J3    & {\bf All}$^{\ast}$, {\bf red}$^{\ast}$, {\bf blue}$^{\ast}$ & 7.5$\cdot10^{-5}$     & 256 (4096)  & 70    & 27.3  & 41    & 1.11  & 0.9    & Hy &   $^{\ast}$  \\
        	&	J4    & All, (blue)                                                 & 4.7$\cdot10^{-3}$     & 109 (1744)  & 5     & 1.4   & 17.4  & 1.09  & 1.01   & TS &   8.55  \\ 
        	&	j1    & All, {\bf red}                                              & 6.5$\cdot10^{-3}$     & 33 (528)    & 1     & 0.3   & 5.3   & 0.91  & -      &  - &   - \\
 	\enddata
    \tablecomments{(a): Name of the host galaxy; 
                   (b): Label of the GC structures; 
                   (c): Significance of each structure in the residual maps generated for all, red and blue GCs: 
                   boldface is used for GC color classes where the structure has high statistical significance, 
                   plain text is used for classes where the structure is  
                   still reliably detected with lower significance and plain text within parenthesis indicate
                   low significance. GC color classes where the structure is not detected are not reported. 
                   Asterisks indicate structures whose shape or size across suggest the presence of multiple substructures;
                   (d): Statistical significance of the structure~(see Section~\ref{sec:method} for details);
                   (e): Size of the structures measured in pixels and in square arcseconds (in parentheses); 
                   (f): Total number of GCs within the structures; 
                   (g): Number of excess GCs within the structures~(see Section~\ref{subsec:spatialstructures_progenitors} for details); 
                   (h): Approximate physical area of the structure (kpc$^{2}$);
                   (i): Average $g\!-\!z$ color of all GCs in the structure;
                   (j): Average $g\!-\!z$ color of N$^{\mathrm{(exc)}}_{\mathrm{GCs}}$ bluest GCs in the structure;
                   (k): Classification of the structure (Sec.~\ref{subsec:spatialstructures_morphology}): AD for Amorphous Dweller, 
                   TS for Tangential Streamer, RT for Radial Streamer and Hy for Hybrid;
                   (l): Logarithm of the dynamical mass of the progenitor of the large structures calculated according to the eq.~\ref{eq:dynamicalmass} for the 
                   full sample (see details in Sec~\ref{subsec:spatialstructures_progenitors}). Asterisks and arrows indicate hybrid/composite structures 
                   (Sec.~\ref{subsec:spatialstructures_morphology}) and structures with negative N$^{\mathrm{(exc)}}_{\mathrm{GCs}}$ respectively
                   for which M$_{\mathrm{dyn}}$ is not calculated.}
\end{deluxetable} 

\section{Discussion}
\label{sec:discussion}

The GCs spatial structures discovered in the investigated Fornax galaxies as discussed in 
Section~\ref{sec:spatialdistribution}, are described in the context of the characteristics 
of the GC systems of the hosts in Appendix~\ref{sec:appendix1}. In what follows, we will 
investigate possible dependencies of the
intrinsic properties (size, shape, significance) of these GCs structures, on the geometry of 
the host galaxies, the color distribution of the overall GC populations, and the galaxy and GCs density 
of their environments. 

\subsection{The properties of the GC spatial structures}
\label{sec:spatialstructures}

\begin{figure}[h]
    \centering
	\includegraphics[width=0.49\linewidth]{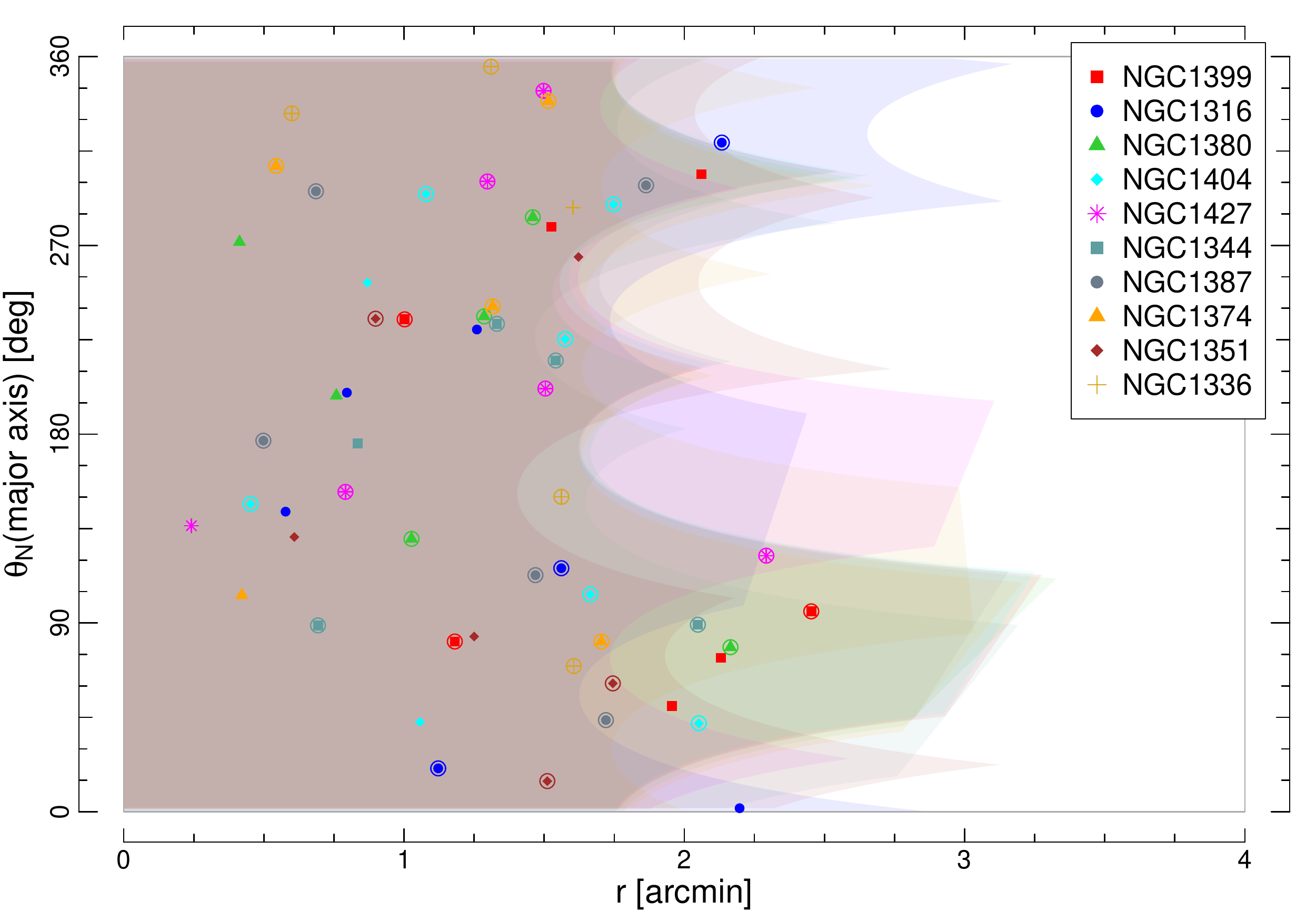}
	\includegraphics[width=0.49\linewidth]{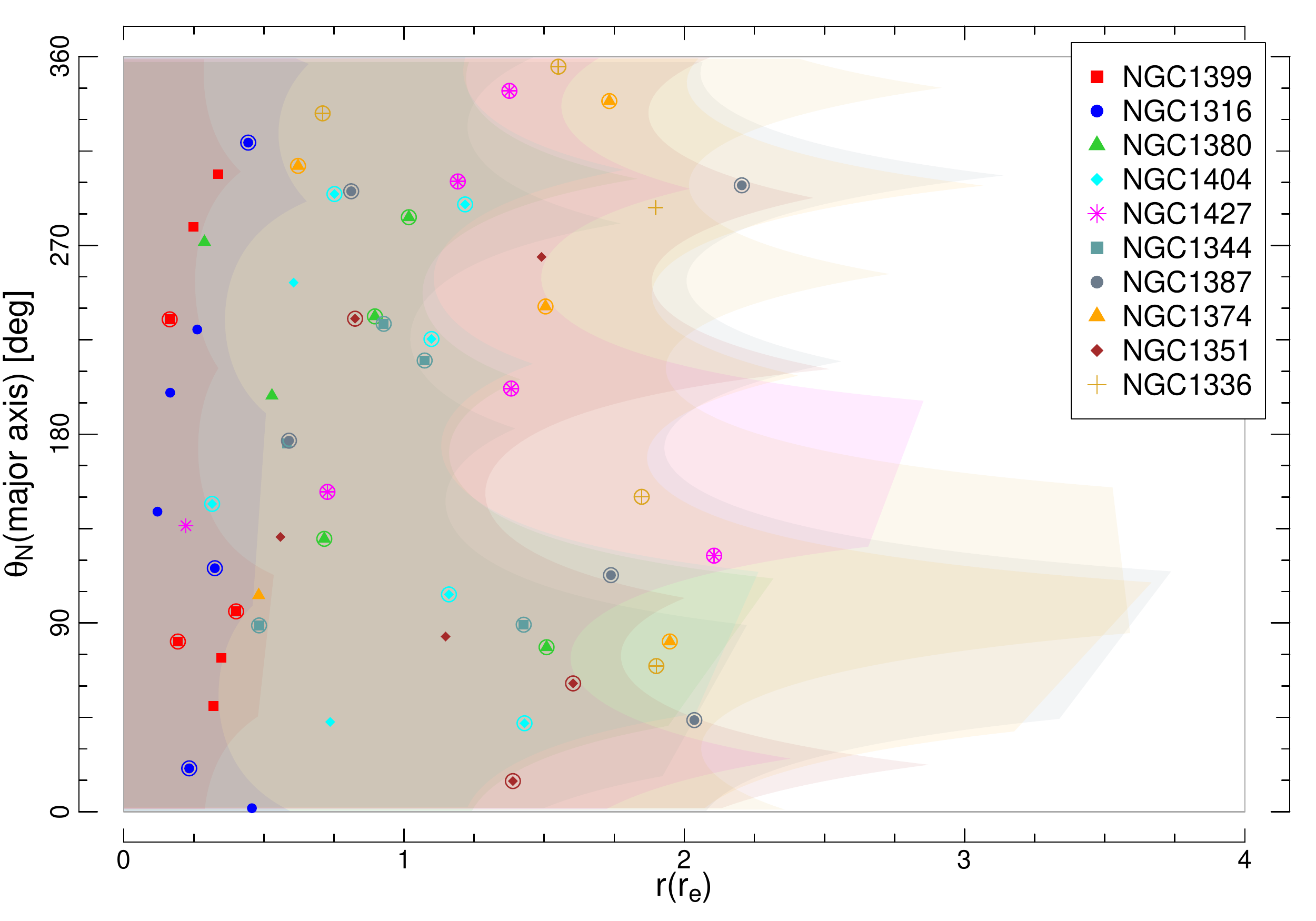}
    \caption{Distribution of the over-density structures shown in Table~\ref{tab:gcstructures} 
    in the galactocentric radius versus azimuthal angle plane (measured clockwise from the S direction of 
    the major axis of the galaxy). Circled symbols identify large GC structures. In the left and right panels 
    the galactocentric radius is expressed 
    as a projected angular distance and measured in units of the 
    effective radii r$_{e}$ of each galaxy respectively. The shaded background areas in both panels 
    represent the regions of the plane that can be accessed with the sample of GCs used in this paper 
    for each galaxy. These regions follow the same color-coding used for the points. In both plots, 
    each different symbol is located in
    the significance-weighted central positions of the intermediate and large GCs spatial structures 
    discussed in Section~\ref{sec:spatialdistribution}.}
    \label{fig:rthetamaps}
\end{figure}

We studied the spatial distribution of the structures detected in the GCs residual maps of the ten 
Fornax cluster galaxies described in detail in Section~\ref{sec:spatialdistribution} with respect 
to the geometry of their host galaxies. Figure~\ref{fig:rthetamaps} shows the positions of the 
significance-weighted centers of all GCs 
structures in the radial vs azimuthal coordinates plane relative to the center of each galaxy.
The radial distance is the galactocentric distance measured as the projected angular separation 
(left panel) or in units of effective radii r$_{e}$ (right panel); the azimuthal position is 
defined as the angular separation from the local S direction of the major axis. Different symbols
and colors indicate different host galaxies and large structures ($\geq$ 60 cells) are circled. 
The areas for each galaxy covered by the ACSFCS observations, are displayed in the background. 
Figure~\ref{fig:rthetamaps} shows that GCs structures 
are inhomogeneously distributed in this plane: few structures are detected 
at radial distances smaller than 0\arcmin.5, likely because of the combined effects of 
the size threshold applied to indvidual intermediate structures ($\geq$ 30 cells) and 
the very inefficient detection of GCs in the core of the galaxies due to the very 
high background. The underpopulated areas in the right panel of Figure~\ref{fig:rthetamaps}
are mostly occupied instead by large, spatially extended structures at $r(r_{e})\approx 1$. 

\begin{figure}[h]
    \centering
	\includegraphics[width=\linewidth]{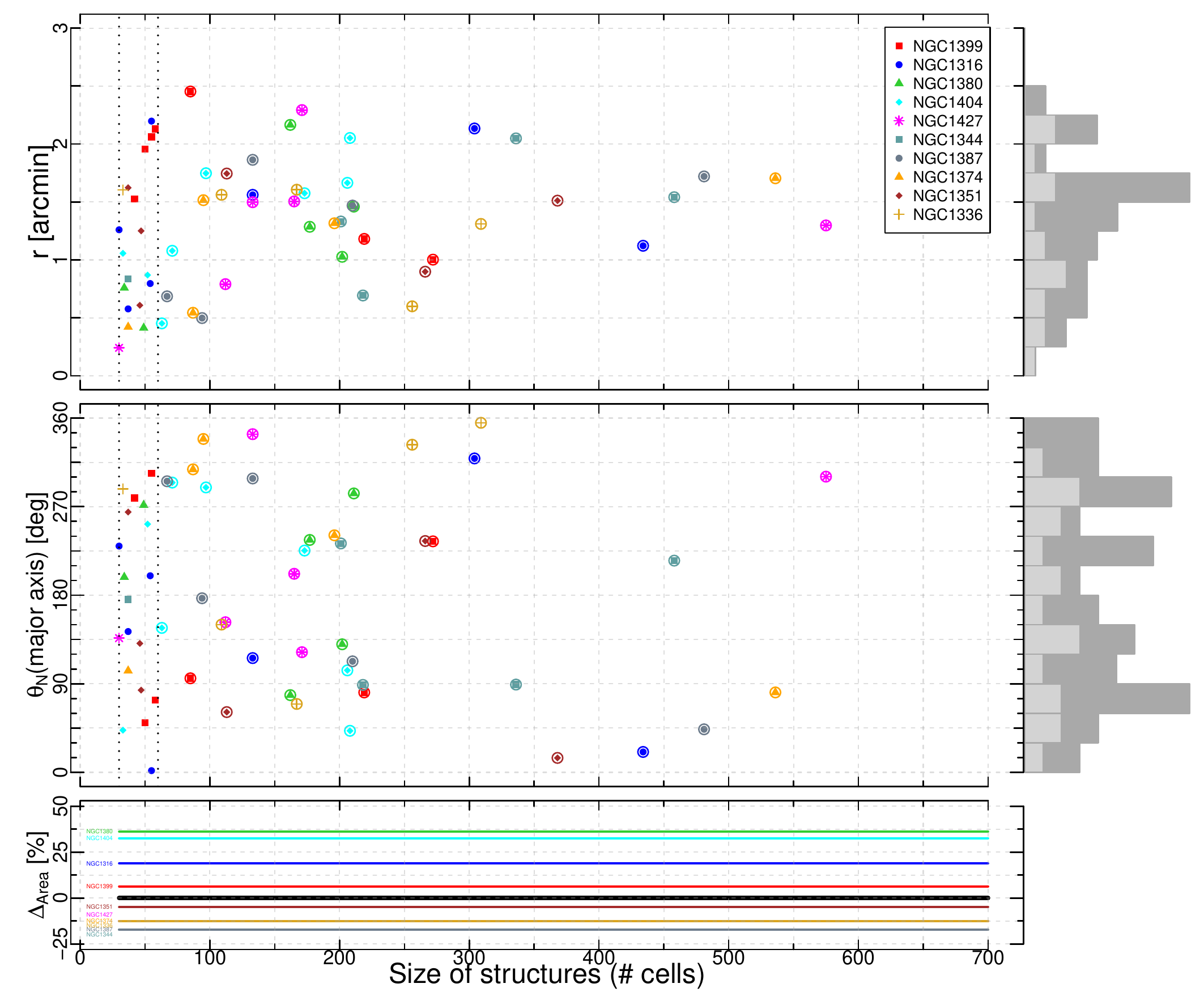}
    \caption{Size of the GC spatial structures detected in the Fornax cluster galaxies vs galactocentric
    distances (upper panel) from the center of the host galaxies, and azimuthal position mid panel) 
    measured clockwise from the S direction of the major axis of the galaxy, and (lower panel)
    percentage difference between physical size of structures detected in different galaxies, by 
    taking into account the host's distance and exact cell angular size as discussed in Section~\ref{sec:spatialdistribution},
    and the physical size estimated for average values of the parameters (black line). These differences
    are independent of the size of the structures and are displayed in this panel only for visibility.
    In the upper and mid panels, the structures are color-coded
    based on their host galaxy and their size is measured as number of cells, while their positions are 
    calculated 
    as the significance-weighted geometric centers. Large structures are enclosed by circles.
    The black dotted lines separate intermediate from 
    large structures as defined in Section~\ref{sec:spatialdistribution}. The histograms on the right 
    display the distribution of intermediate (light gray) and large (dark gray) structures.}
    \label{fig:sizestructures_spatialcoordinates}
\end{figure}

Figure~\ref{fig:sizestructures_spatialcoordinates} shows separately the galactocentric distances and 
azimuthal positions of all GCs structures investigated in this paper as a function of the size of 
the structures measured in number of cells. As mentioned in Section~\ref{sec:spatialdistribution}, 
the linear size of the cells in the residual maps ranges from $\sim\!3$\arcsec to $\sim\!5$\arcsec 
along both axes depending on the actual distance of the host galaxy. These angular sizes correspond 
to physical sizes from 0.3 to 0.5 kpc along the axes and to physical transversal sizes from $\sim$0.4 
to $\sim$0.5 kpc.
The intermediate structures (upper panel) appear to be
homogeneously distributed along the radial range spanned by the observations, while 
large structures peak between 1\arcmin~and 2\arcmin~from the centers of their host galaxies, with the 
nine largest structures ($\geq 300$ cells) all located between 1\arcmin~and 2\arcmin.1, although this could 
be an effect of the available observed area, since structures identified as large in this work cannot be
located too close to the edges of the field where they would be truncated. At distances 
larger than 2\arcmin.5, the spatial coverage of the ACSFCS observations rapidly declines (cp. left 
panel of Figure~\ref{fig:coverage}), likely leading to a significant incompleteness in both radial 
and angular coverage. The angular distribution of intermediate GC spatial structures is roughly 
homogeneous (lower panel of Figure~\ref{fig:sizestructures_spatialcoordinates}), while the large 
structures peak around the minor axis direction of their host galaxies, with $\sim$ 65\% 
of them located within $\theta\!=\![90^{\circ},270^{\circ}]\pm45^{\circ}$ (lower panel).
Unlike~\cite{dabrusco2015}, who found that a majority of the total area of the GC structures
was located along the major axes of the host galaxies in the Virgo cluster, we observe that 
$\sim\!70\%$ of the total area of the large structures in the Fornax cluster galaxies 
are located within $\pm 20\degree$ from the minor axis of their hosts and $\sim\!60\%$ 
of the area of all structures is observed in the same azimuthal range. 

\begin{figure}[h]
    \centering
	\includegraphics[width=0.96\linewidth]{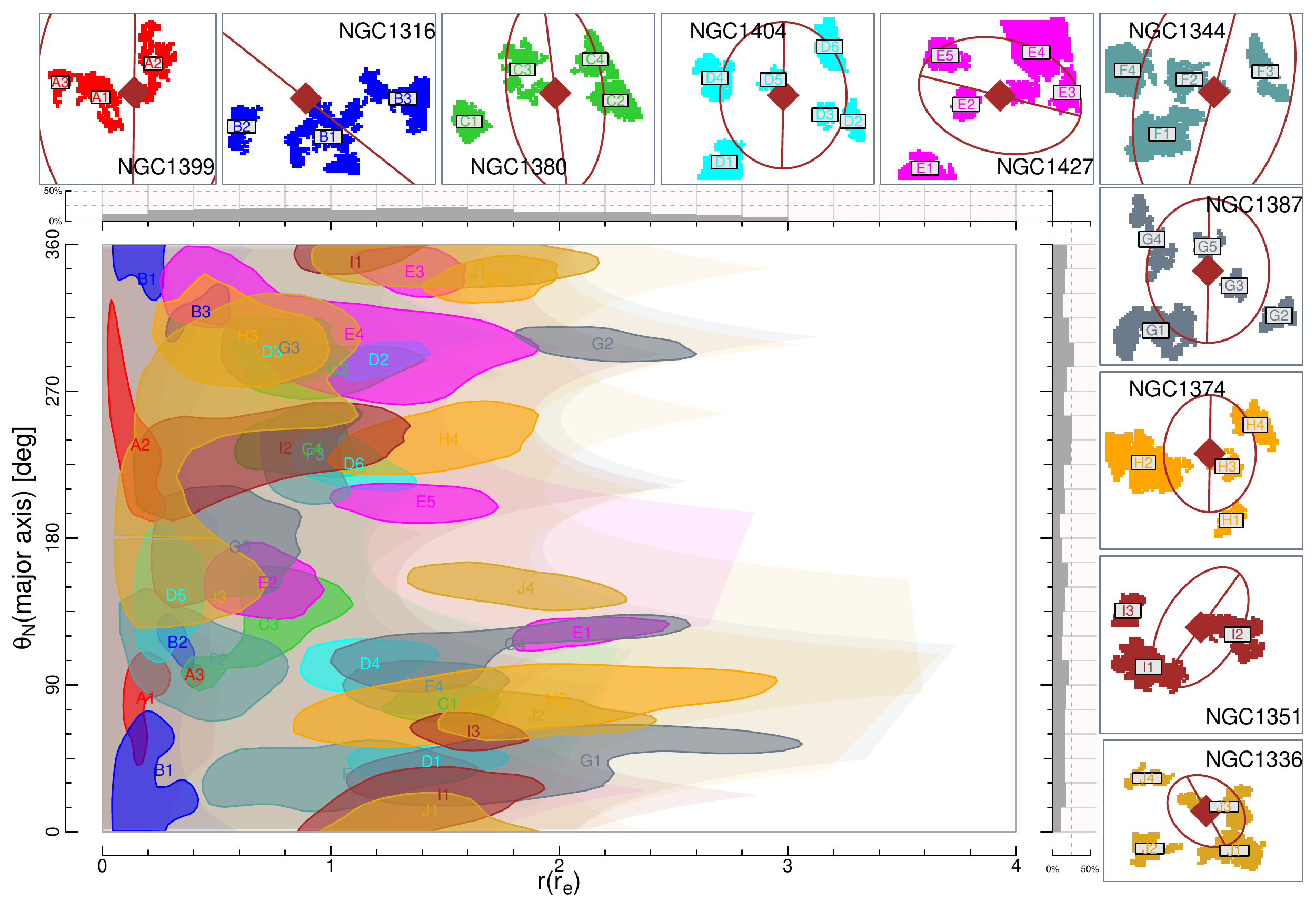}
    \caption{Boundaries of the large GC over-density structures in the galactocentric radius 
    (expressed in units of the effective radii r$_{e}$) versus the azimuthal angle plane (measured
    clockwise from the S direction of the major axis). The marginal histograms display the total percentage
    of area accessible with the ACSFCS data (shaded background region) that is covered by large 
    GCs structure along the galactocentric and azimuthal distances intervals. The galaxy plots 
    show the positions of the structures relative to each host, color-coded by galaxy.}
    \label{fig:rthetamaps_largestructures}
\end{figure}

Figure~\ref{fig:rthetamaps_largestructures} shows the whole areal extension of large  
residual structures in the same galactocentric vs azimuthal distance plane of 
Figure~\ref{fig:rthetamaps}. The vertical marginal histogram in the plot shows that 
large GCs structures occupy a larger fraction of the total available area sampled by 
ACSFCS observations along the minor axes of the hosts (close or larger than 25$\%$), 
while the fraction occupied along the major axis is $\sim\!10\%$. 

\begin{figure}[h]
    \centering
	\includegraphics[width=0.9\linewidth]{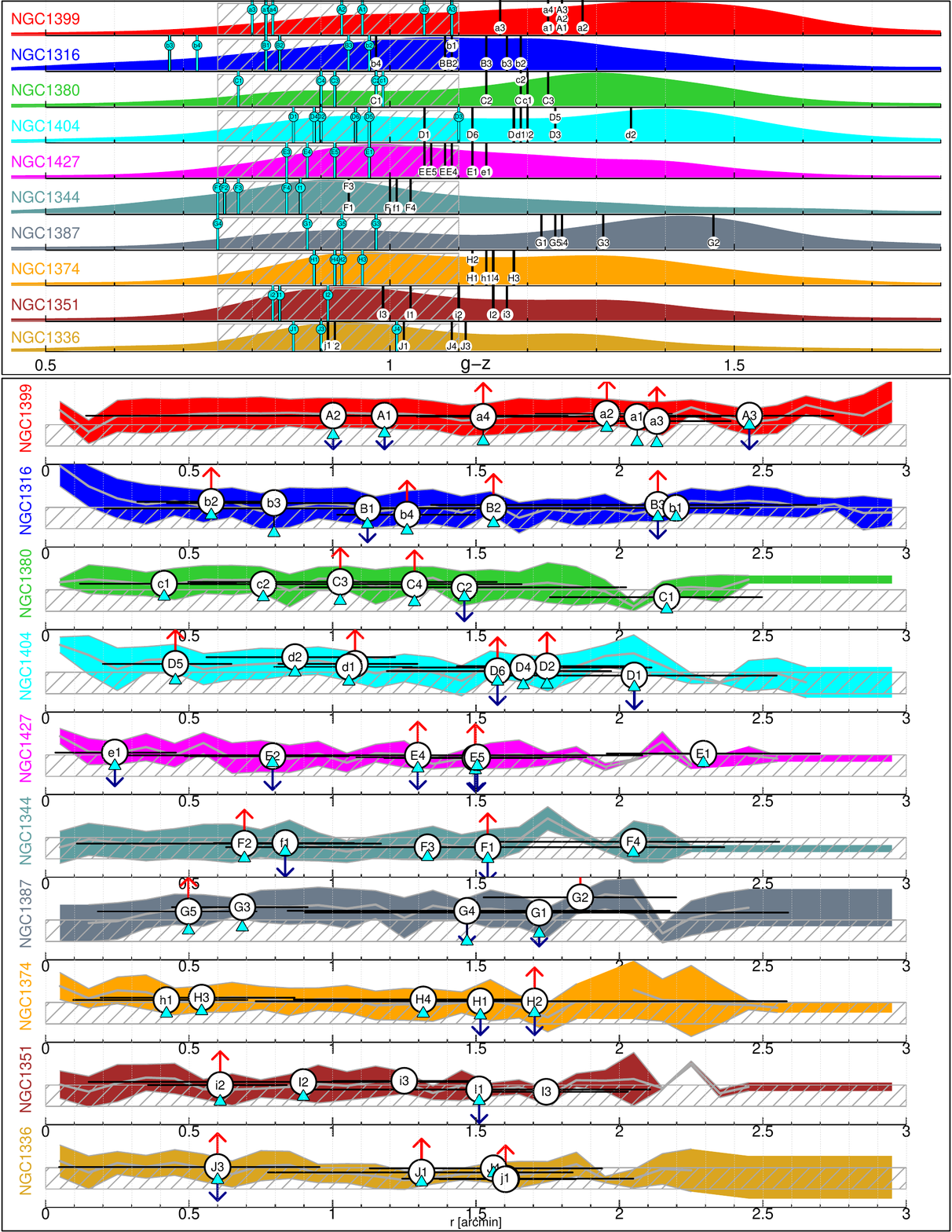}
    \caption{Upper panel: $g\!-\!z$ color distributions of ACSFCS GCs for the ten galaxies investigated 
    in this paper; the vertical black lines show the average colors of large and intermediate GCs 
    residual structures based on all GCs within the boundaries of the structures, while the 
    vertical blue lines display the average color of the bluest N$^{\mathrm{(exc)}}_{\mathrm{GCs}}$ GCs 
    within the boundaries of each structure}. Lower panel: 
    radial $g\!-\!z$ color profiles of the host galaxies GC systems (GCs located within the boundaries of the 
    large and intermediate structures have been removed) with the mean colors of
    all the GCs (white circles) and the bluest N$^{\mathrm{(exc)}}_{\mathrm{GCs}}$ GCs (blue triangle)
    included in each structure. The colored areas represent the uncertainty of the radial profiles, while
    the black horizontal segments display the full radial extension of GCs structures. The blue and red
    arrows pointing down and up, respectively, indicate whether each structure has large statistical 
    significance in the blue and red GCs color subclasses~(cp. with column C in Table~\ref{tab:gcstructures}. 
    In both panels, the shaded areas highlight the color interval occupied by the GCs systems of 
    Virgo cluster dEs from~\cite{peng2006} for comparison.
    \label{fig:structures_color_comparison}
\end{figure}

Figure~\ref{fig:structures_color_comparison} displays the average $g\!-\!z$ colors of all GCs located
within the boundaries of all intermediate
and large GC structures, superimposed to the general color distribution of the GC population of the host 
galaxies (upper panel), and to the radial color profiles of the GC systems obtained after removing 
the GCs within the boundaries of the GC structures (lower panel). The colors of GCs structures in 
NGC1399, NGC1316, NGC1344, NGC1427 and NGC1374 do not vary significantly and are very close to the average 
colors of the overall GC populations of their hosts, suggesting comparable metallicities and similar 
formation mechanisms. The colors of GC structures in NGC1380, NGC1404, NGC1387, NGC1351 and NGC1336 span 
a larger interval of colors hinting at possible different formation mechanisms at play. In particular, 
the structures with the largest color offset relative to the mean value of the general GC color 
distribution of the host are G2 in NGC1387 (redder than the host GC systems at the same radial 
distances), D1 and 
d2 in NGC1404 (bluer and redder respectively than the host GC populations at the same radial distances) 
and J2 and j1 in NGC1336 (both redder than the GCs at observed at the same galactocentric
distances). In the case of G2, the spatial structure is located in the SW corner of the region of host
galaxy observed by the ACSFCS along a direction opposite to the direction connecting NGC1387 to NGC1399 
and not associated with an observed overdensity of galaxies. Similarly, D1 is located in the SE corner
of NGC1404 field and opposite to the direction connecting its host to NGC1399, in a region of high 
density of ICGCs~(see Figure~\ref{fig:core_gc_patchwork}). 

In the remainder of this paper, we assume that the GCs associated with residual structures are the 
relics of the GC systems of satellite galaxies accreted by the host during its 
assembly~(see details in Sec~\ref{subsec:spatialstructures_progenitors}). Under this hypothesis, 
the GCs observed within the boundaries of the structures belong either to the GC system 
of the structure's progenitor or to the host galaxy. Given that the colors of the GC 
systems of dwarfs in clusters of galaxies are typically bluer 
than the GCs of massive ETGs (the host) in the same environment~\citep{peng2006}, the average of the colors of 
all GCs located within the boundaries of each structure is an upper limit on 
the real average color of GCs of the progenitor of the structure. For this reason, in order to provide a lower limit to the colors of 
residual structures GCs, we also calculated the average colors of the N$^{\mathrm{(exc)}}_{\mathrm{GCs}}$ 
bluest GCs in each structure~(reported in 
column j of Table~\ref{tab:gcstructures}). These values are shown in Figure~\ref{fig:structures_color_comparison} 
in the context of the general color distribution of all GCs in each host and their color radial 
profiles as blue lines (upper panel) and blue triangles (lower panel). The ``blue'' limits 
on the colors of the GC structures (shaded regions of both panels in Figure~\ref{fig:structures_color_comparison}) 
mostly lie in the interval of colors of the GCs systems of galaxies with $M_{\mathrm{B}}\!\geq\!-18$ 
from Figure~3 of~\cite{peng2006}.

\subsection{Morphological classification of GCs spatial structures}
\label{subsec:spatialstructures_morphology}

\begin{figure}[h]
    \centering
    \includegraphics[width=0.48\linewidth]{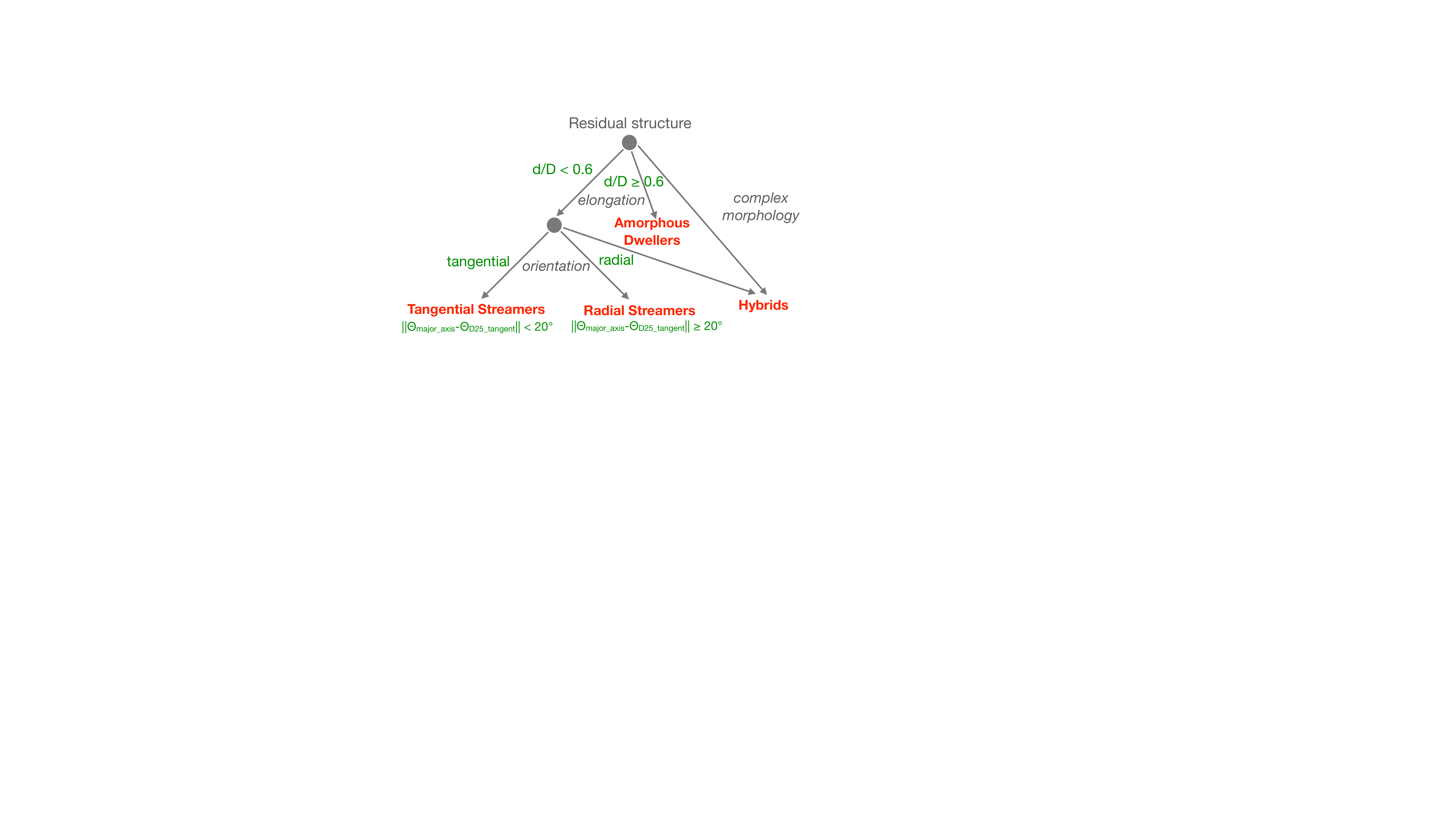}
    \includegraphics[width=0.48\linewidth]{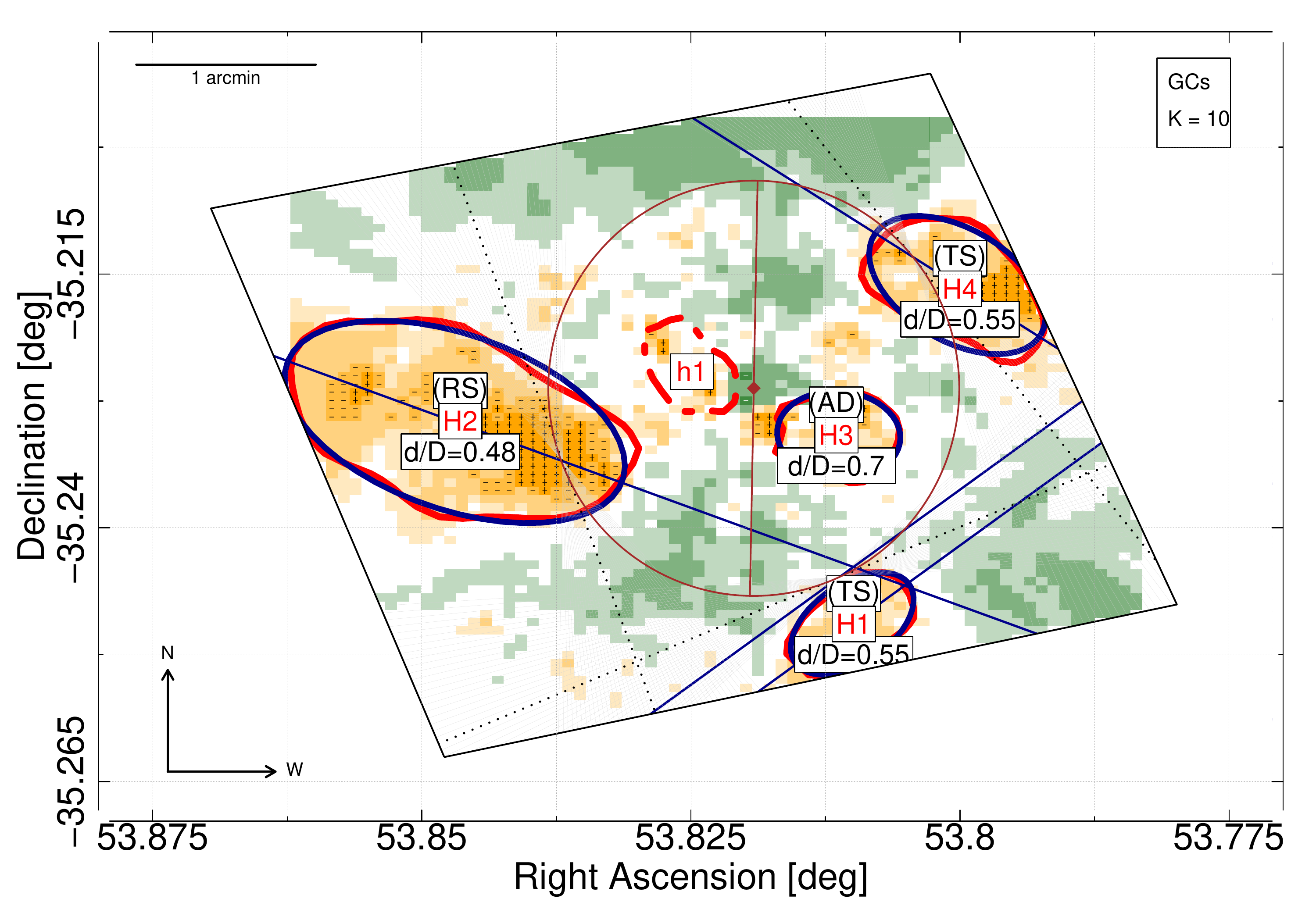}
    \caption{Left panel: Decision tree used to determine the classification of the large
    residual structures investigated in this paper. Right panel: $K\!=\!10$ residual map for the distribution
    of all GCs observed in NGC1374. The best-fit ellipses of the large residual structures and their major axes 
    are shown with solid blue lines. The shaded gray regions represent the $\pm 20\degree$
    cones centered on the direction of the tangent to the elliptical D$_{25}$ elliptical isophote in the 
    intersection between the major axis of the fitted ellipses for the GC residual structures and the D$_{25}$ 
    of the host galaxy. The cones are used to classify elongated structures in TSs or RSs according to their
    orientation. In the insets, the label, the axial ratio of the best-fit ellipse and the morphological
    classification of each structures are reported.}
    \label{fig:morphological_classification}
\end{figure}

Based on the morphology and the orientation of the spatial structures detected in the distribution
of all GCs observed in the galaxies investigated in this paper (as discussed in Section~\ref{sec:method}, 
we did not perform the detection of structures in the residual maps obtained from red and blue GCs), we 
define four classes as shown in left 
panel of Figure~\ref{fig:morphological_classification}: ``Amorphous Dwellers'' (ADs), ``Radial Streamers'' (RSs), 
``Tangential Streamers'' (TSs) and ``Hybrids''. 
The structures are first split according to their shape: the full set of cells belonging to each structure 
is fit with an elliptical model whose major and minor axes are free to vary, and the structures with minor 
axis to 
major axis ratio d/D$\geq\!0.6$ (where d and D are the minor and major axes of the best-fit elliptical model) 
are identified as ADs. While most detected GCs structures have
shapes that are not well modeled by an ellipse, this step permits to separate flattened from generally unflattened 
shapes. The elongated structures are further split 
according to their orientation: if the direction of their fitted major axis is contained within a $\pm 20\degree$ 
cone centered on the direction of the tangent to the D$_{25}$ elliptical isophote of the host galaxy in the 
intersection of the structures with (or the closest point to) the D$_{25}$, they are classified as TSs. Given
an arbitrary reference direction, this condition translates to 
$||\Theta_{\mathrm{major\_raxis}}-\Theta_{\mathrm{D25\_tangent}}||\leq\!20$ where $\Theta_{\mathrm{major\_axis}}$ 
is the angular
distance of the major axis of the fitted ellipse of the GC residual structure from the reference direction, and 
$\Theta_{\mathrm{D25\_tangent}}$ is the angular distance of the tangent to the D$_{25}$ in the intersection between
the ellipse major diameter (or its closest point) and D$_{25}$.
All other elongated structures are labeled as RSs. The fourth class, ``Hybrids'', includes spatial structures 
whose large size and/or complex morphology do not permit a straightforward classification in one of the 
classes defined above and suggest that they are composite. The right panel of 
Figure~\ref{fig:morphological_classification} shows an example of morphological classification for the four 
large residual structures observed in NGC1374, where at least one structure for each class (excluded Hybrids) 
has been observed.

\begin{figure}[h]
    \centering
	\includegraphics[width=0.49\linewidth]{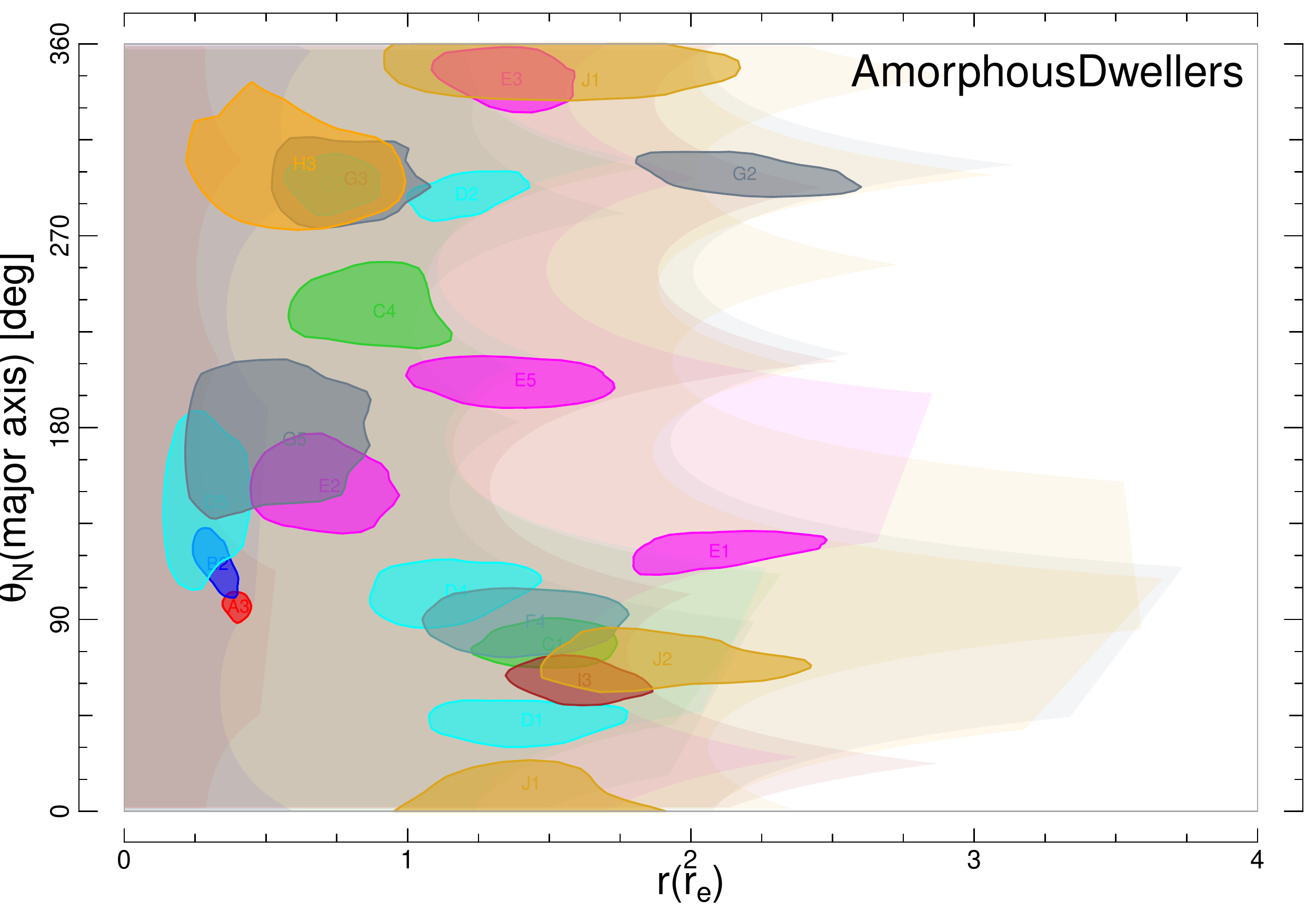}
	\includegraphics[width=0.49\linewidth]{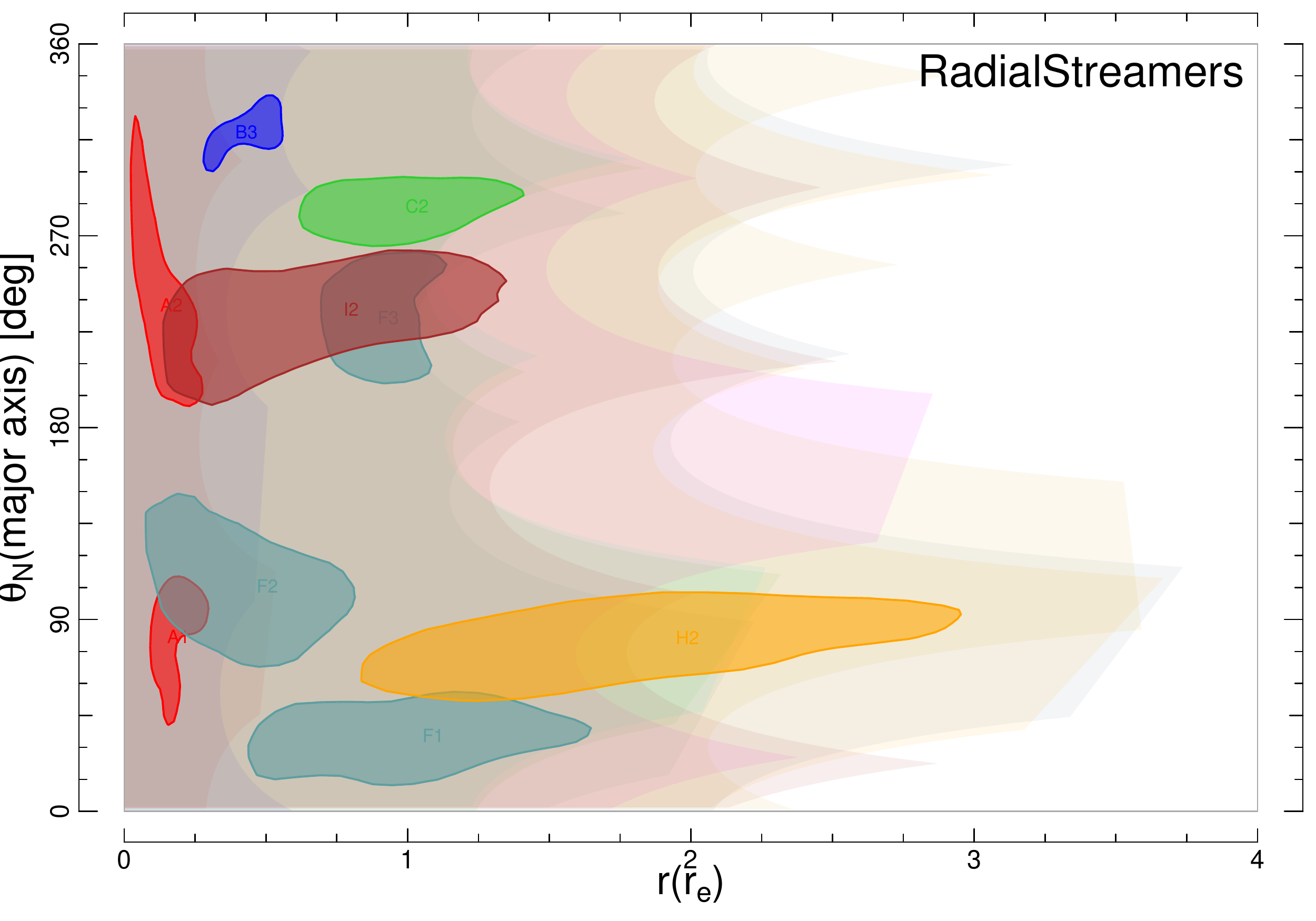}\\
	\includegraphics[width=0.49\linewidth]{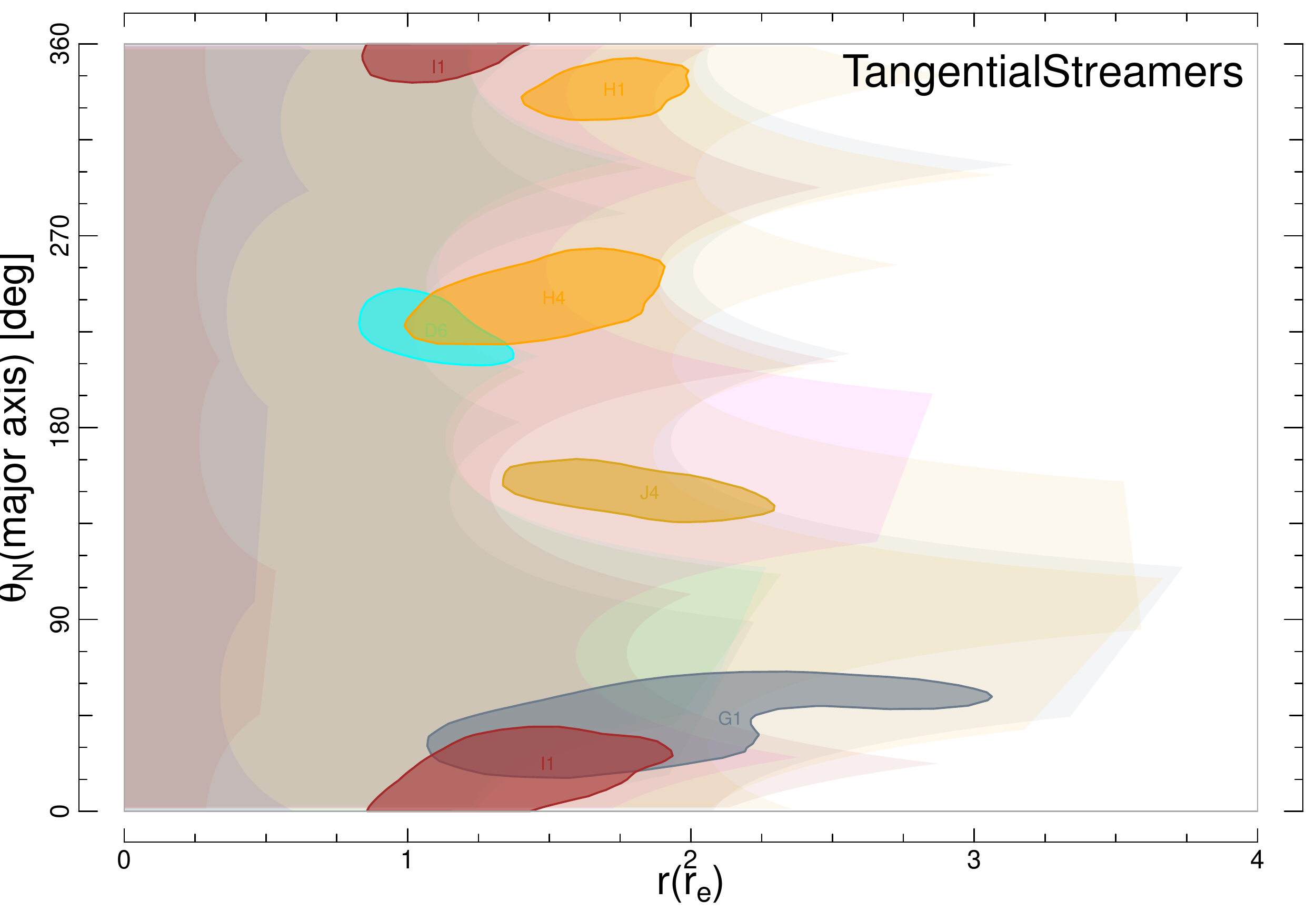}
	\includegraphics[width=0.49\linewidth]{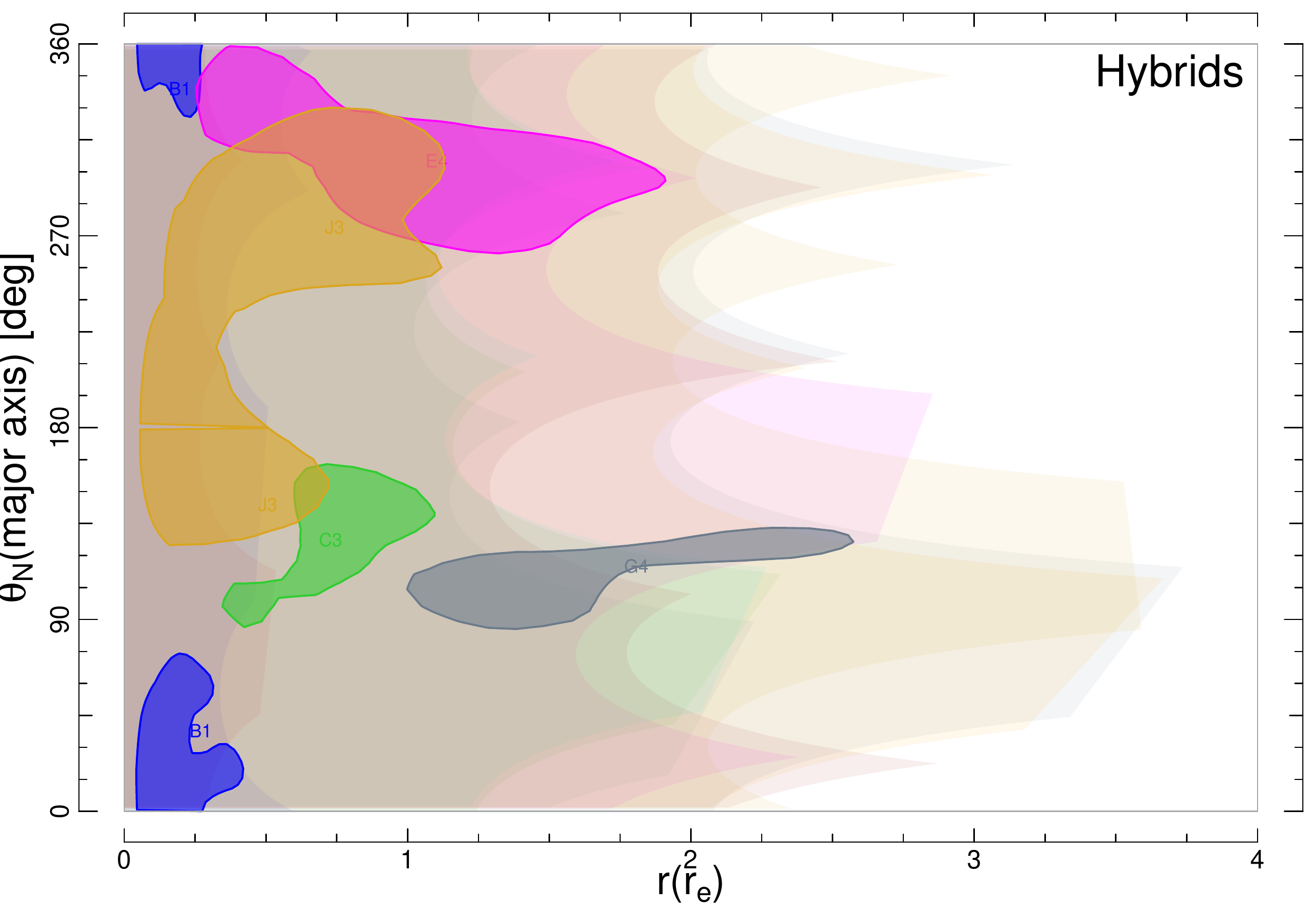}
    \caption{Boundaries of the large GC spatial
    structures in the galactocentric distance (expressed in units of the effective radii r$_{e}$) versus 
    the azimuthal angle plane (measured clockwise from the S direction of the major axis) for different 
    morphological classes. 
    As in Figure~\ref{fig:rthetamaps}, the structures are color-coded by galaxy membership and 
    the shaded background areas represent the regions of the galaxies accessible with the ACSFCS data.}
    \label{fig:rthetamaps_largestructures_classification}
\end{figure}

Figure~\ref{fig:rthetamaps_largestructures_classification} 
shows the large spatial structures in the galactocentric vs azimuthal distances plane split by their classes. 
These plots highlight the following results:

\begin{itemize}
    \item ADs (upper left panel) are almost equally distributed along the major and minor axes of their host 
    galaxies: $\sim\!55\%$ of the total area of ADs is located within $\pm30\degree$ of 
    the $\Theta\!=\!\{90\degree,270\degree\}$ directions along the azimuthal axis. They are observed
    along the whole interval of radial distances investigated.
    \item RSs (upper right panel) span a large range along the galactocentric distance axis and are more 
    likely not located along the major axes of their host galaxies, with only 
    $\sim12\%$ of their total areas within $\pm30\degree$ from the 
    $\theta\!=\{0\degree, 180\degree\}$ directions.
    \item TSs (lower left panel) are, by definition, only located at radial distances $\geq\!1$ r$_{e}$.
    Their azimuthal distribution shows that they display a slight preference for the direction of the 
    major axis ($\sim31\%$ of their total areas within $\pm30\degree$ from the 
    $\theta\!=\{0\degree, 180\degree\}$ direction), and are not observed along the minor axes of the host
    galaxies.
    \item The Hybrids structures (lower right panel) usually occupy large intervals along the azimuthal axis, 
    with the exception of G4 and C3, and they can almost straddle the directions of both axes (B1 and E4). 
    Only G4 is entirely located at galactocentric distance larger than 1 r$_{e}$, while E4 covers the largest
    radial interval, between 0.25 and 1.8 r$_{e}$.
\end{itemize}

The class of a structure depends on both its intrinsic properties, i.e. the position of the 
structure relative to the host galaxy and its shape, and on the orientation of both the host galaxy and 
the structures relative to the line of sight. Under different projections, the same three-dimensional 
spatial structure could be classified as ADs, TSs or RSs. Assuming that each observed GC structure is the relic 
of the GC system of a single satellite accreted by the main galaxy, different classes identify 
different initial orbit along which the satellite was moving relative to the 
host, and the line of sight, and its orbital phase. 

\subsection{Position and properties of the GCs spatial structures in the cluster of galaxies}
\label{subsec:spatialstructures_position}

\begin{figure}
    \centering
	\includegraphics[width=0.75\linewidth]{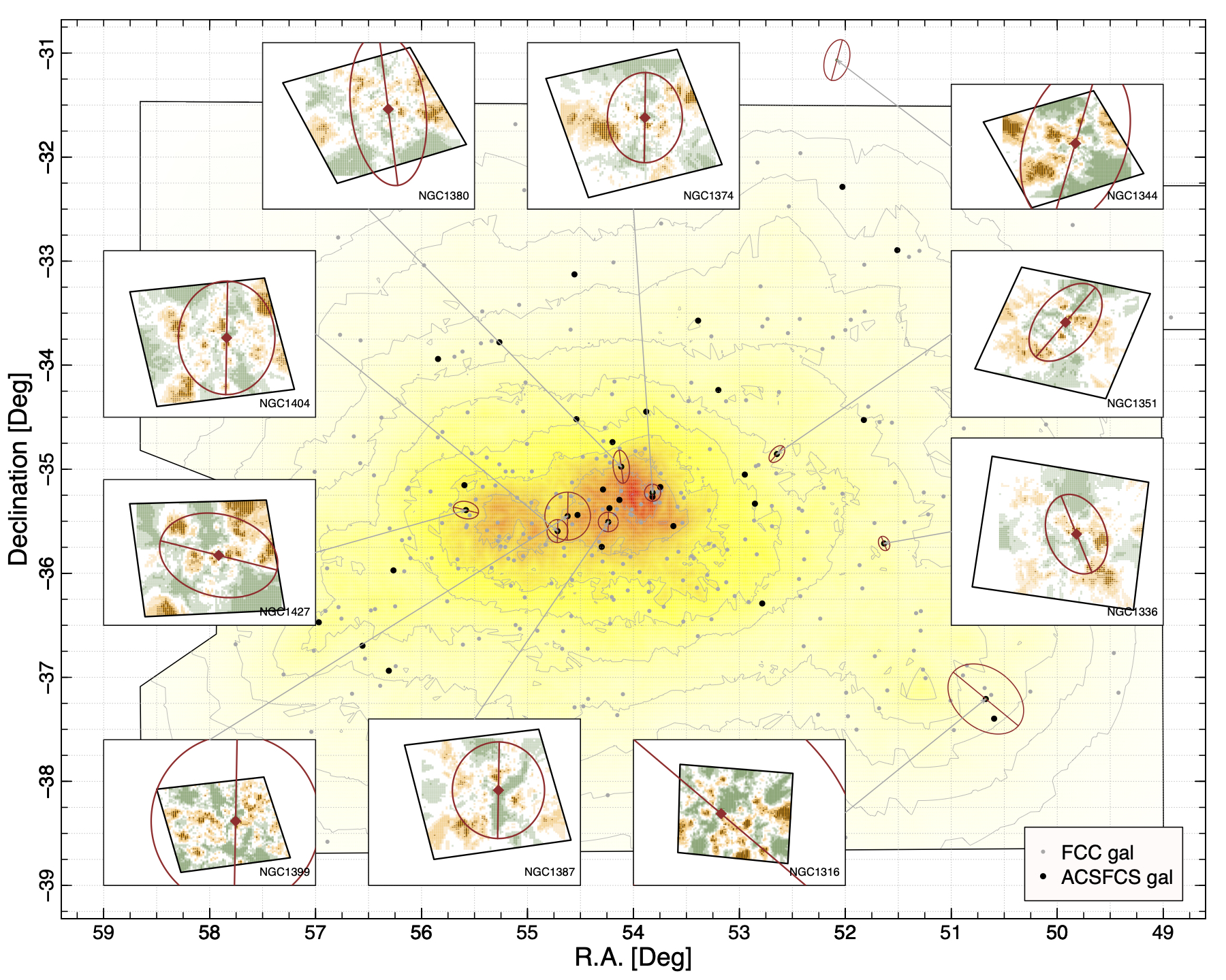}\\
	\includegraphics[width=0.75\linewidth]{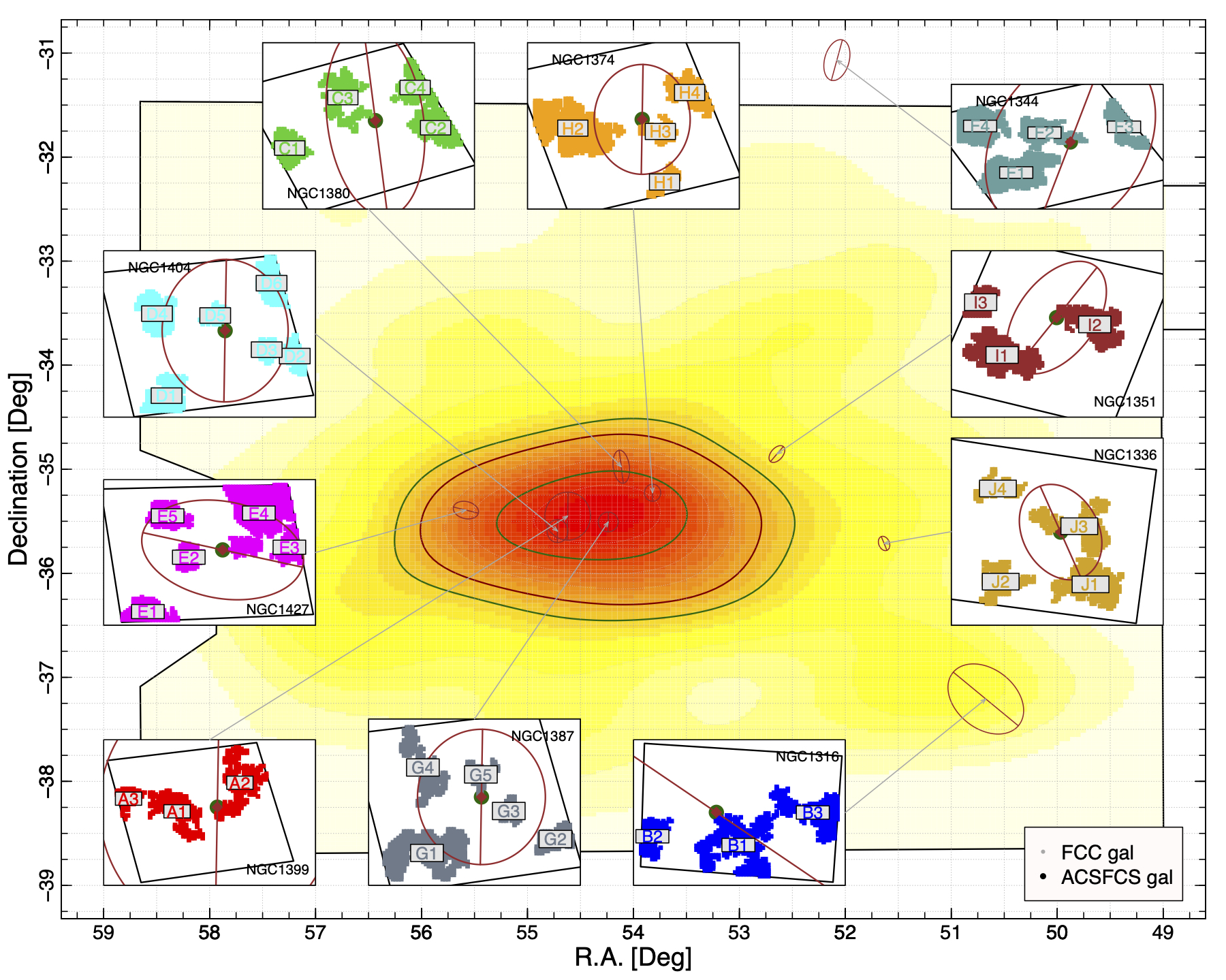}	
    \caption{Upper panel: KNN density maps ($K=\!=\!21$) of the spatial distribution of FCC 
    galaxies~\citep{ferguson1989} (gray points) with insets displaying the $K\!=\!10$ KNN residual 
    maps of the distribution of ACSFCS GCs for the galaxies investigated in this paper. 
    Lower panel: KDE density map of the spatial distribution of
    FCC galaxies with insets showing the large GC spatial structures for the ten galaxies studied 
    in this paper. In both plots, the gray lines represent 10 isodensity contours of the galaxy 
    density distribution logarithmically spaced between 
    2\% and 99\% of the peak density. In the lower plot, the thick green lines indicate the 
    isodensity contours corresponding to the 40\% and 
    80\% of the KDE peak density used to separate low, intermediate and high galactic density regions
    in the cluster; the thick red line represents the 50\% isodensity contours. Black points 
    show the positions of ACSFCS galaxies. The D$_{25}$ 
    diameters of the elliptical isophotes of the ACSFCS hosts investigated are scaled up for visibility.}
    \label{fig:cluster_galaxy_insets}
\end{figure}

Figure~\ref{fig:cluster_galaxy_insets} shows the location of the ten galaxies discussed in this 
paper within the Fornax cluster, compared with the density maps derived with two different methods 
from the positions of galaxies that are likely cluster members included in the Fornax Cluster 
Catalog (FCC)~\citep{ferguson1989}, which is complete at $\sim$90\% level at the B$_{T}\!=\!19$ 
magnitude for Es and dEs within the core of the Fornax cluster. 
The upper plot in Figure~\ref{fig:cluster_galaxy_insets} shows the full residual maps
of the ACSFCS GCs distribution overlayed to the FCC galaxy density calculated with the KNN 
method with $K\!=\!21$ to highlight relatively small scale spatial structures in the 
projected galaxy density.
The lower panel displays only the large GCs structures overplotted to the density map of FCC galaxies obtained
with the Kernel Density Estimation\footnote{The Kernel Density Estimation is a non-parametric method to 
estimate the probability density function (pdf) of a random variable. The pdf subtending the observed data is
reconstructed by assuming a functional form for the kernel
and by fitting the free parameter {\it bandwidth} to the observed distribution of observation through 
minimization of the mean integrated squared error.} (KDE) method: 
the two solid green lines represent the isodensity contours corresponding to the 40\% and 80\% of 
the density value at the peak that we used to separate low, intermediate and high galaxy density 
regions in the cluster. Using these thresholds, NGC1374, NGC1387, NGC1399 and NGC1404 are located
in the high density regions; NGC1427 and NGC1380 lie in the intermediate density region and 
the remaining galaxies (NGC1316, NGC1351 and NGC1336) are in the low density area. NGC1344 is 
located outside of the footprint of the FCC catalog and is not considered in this analysis.

\begin{figure}
    \centering
    \includegraphics[width=0.94\linewidth]{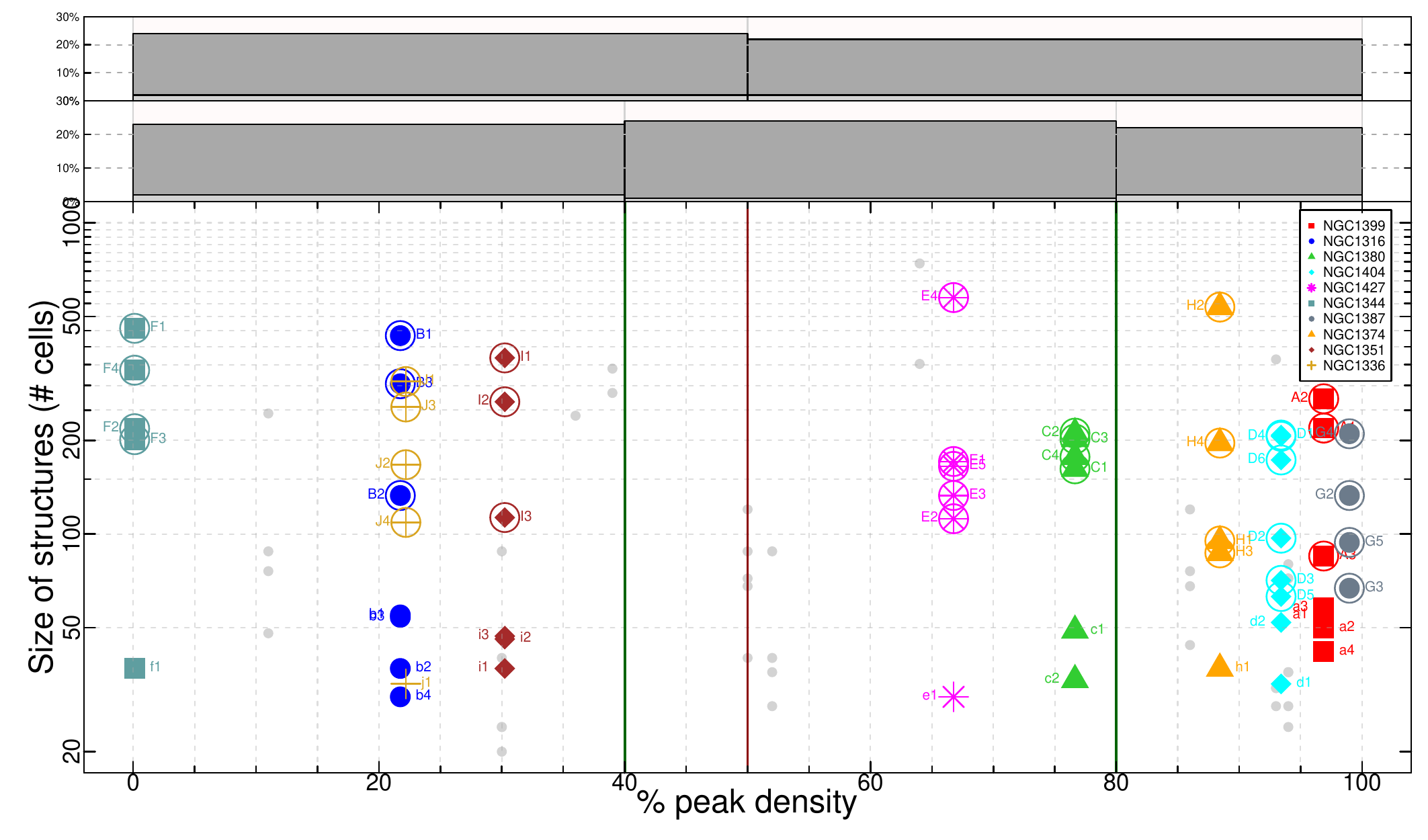}
    \caption{Size of intermediate and large spatial structures in the ACSFCS GCs distribution as a 
    function of the FCC KDE galaxy density normalized to the peak, color-coded according to the host galaxy.
    Gray points in the background display the spatial structures detected in Virgo 
    cluster~\citep{dabrusco2016}. Large structures are enclosed by circles. 
    The marginal histograms show the fraction of total area covered by the 
    ACSFCS observations occupied by intermediate (light gray) and large (dark gray) GC spatial structures
    in low, intermediate and high galaxy density regions (lower histogram) and in low and high galaxy 
    density regions (upper histogram) of the Fornax cluster defined by the green and red lines 
    in Figure~\ref{fig:cluster_galaxy_insets}, respectively.}
    \label{fig:size_vs_normdensity}
\end{figure}

Figure~\ref{fig:size_vs_normdensity} shows the size of all GCs spatial structures
as a function of the galaxy density in the location of the host galaxy (normalized to 
the peak density) derived from the FCC catalog with the KDE method~(Figure~\ref{fig:cluster_galaxy_insets}).
The data are inconclusive regarding the existence
of a correlation between size of structures and local galaxy density (Pearson's sample 
correlation parameter $r=-0.1$ with $p$-value 0.442); unlike in the case of Virgo 
cluster~\citep{dabrusco2016}, large structures are found with similar frequency in both 
high and low density regions of the cluster of galaxies, although intermediate structures 
are more frequently detected in galaxies in the lower density area of the 
cluster (8 on 17 total structures outside of the 50\% isodensity contour vs 10 
on 35 inside the 50\% isodensity contour). The marginal histograms in Figure~\ref{fig:size_vs_normdensity}
show the percentage of the areas occupied by GCs structures relative to the total 
imaged areas of the galaxies when the cluster is divided again in high and low galaxy density regions 
(upper histogram) and high, intermediate and low galaxy density regions (lower histogram) using 
the isodensity contours associated with the 50\% and 40\%, 80\% of the peak galaxy 
density respectively (red and green lines in Figure~\ref{fig:cluster_galaxy_insets}). 
The data available do not permit to draw conclusion even when correlations between the 
local galaxy density and the average or maximum size of all GC structures detected in the same
structures are investigated.

Assuming that the spatial features of the overdensities in the distribution of GCs trace the 
accretion history of the hosts and depend on the mass ratios and the time passed since the 
accretion events, we should expect large GC structures to be observed more frequently than 
intermediate structures in lower density regions of the cluster, 
where the accretions of large satellites and minor mergers have occurred more recently and/or 
more often than in high density areas. The absence of conclusive evidence regarding the lack 
of correlation between the galaxy density and the properties of the GCs structures in the Fornax 
cluster does not allow us to draw final conclusion about this aspect. Should the lack of 
a correlation be confirmed and deemed statistically significant with additional data, it
may either indicate that the scenario above is not always applicable or that it strongly 
depends on the global history of the cluster where the observed galaxies reside. The Fornax cluster 
is traditionally thought to be more relaxed than the Virgo cluster, a notoriously dynamically 
young and unrelaxed cluster, from the analysis of the velocity 
distribution of member galaxies~\citep{drinkwater2001}, although there are emerging evidences 
that point towards a more lively recent history, including a potential ongoing 
merger~\citep[based on the asymmetry of the intracluster diffuse X-ray emission][]{paolillo2002}, 
recent infall of NGC1399~\citep[from the spatial variation of the Intra-Cluster Matter temperature in the 
cluster core][]{murakami2011} and the recent accretion of a galaxy group~\citep[from the
properties of the diffuse stellar light and GCs in the N-NW area within the Fornax virial 
radius][]{iodice2019}.

Moreover, the properties of the observed GCs structures likely depend also on additional parameters 
(i.e., the geometry of the accretion events, the gas content of the satellites and the ratio
of early-to-late type galaxies that all can determine the degree of asymmetry in the distribution 
of GCs formed in tidal tails and streamers) that, in the case of the Fornax cluster, might have had a 
significant effect in shaping the GC populations.

\begin{figure}
    \centering
	\includegraphics[width=0.94\linewidth]{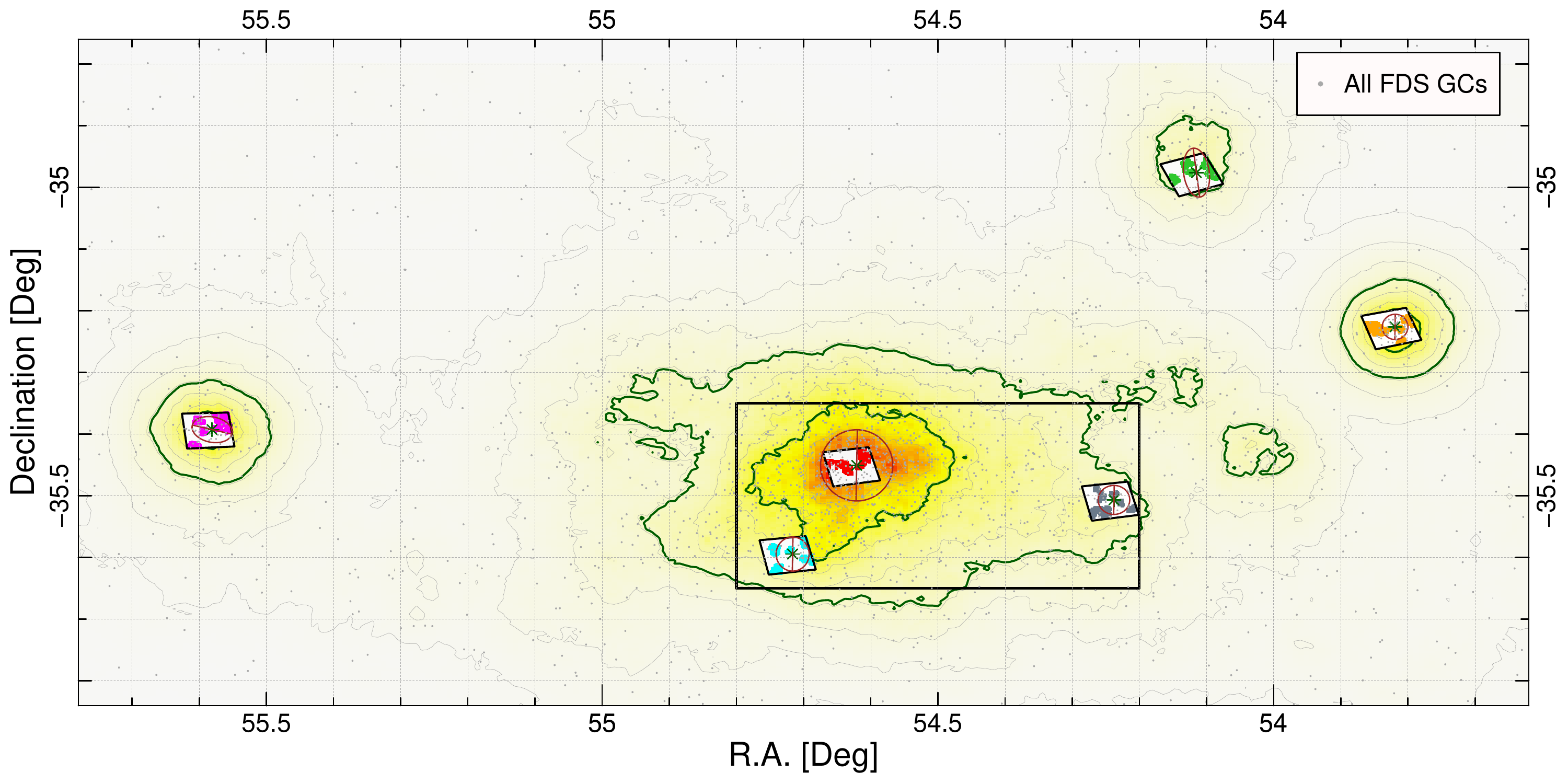}\\    
	\includegraphics[width=0.94\linewidth]{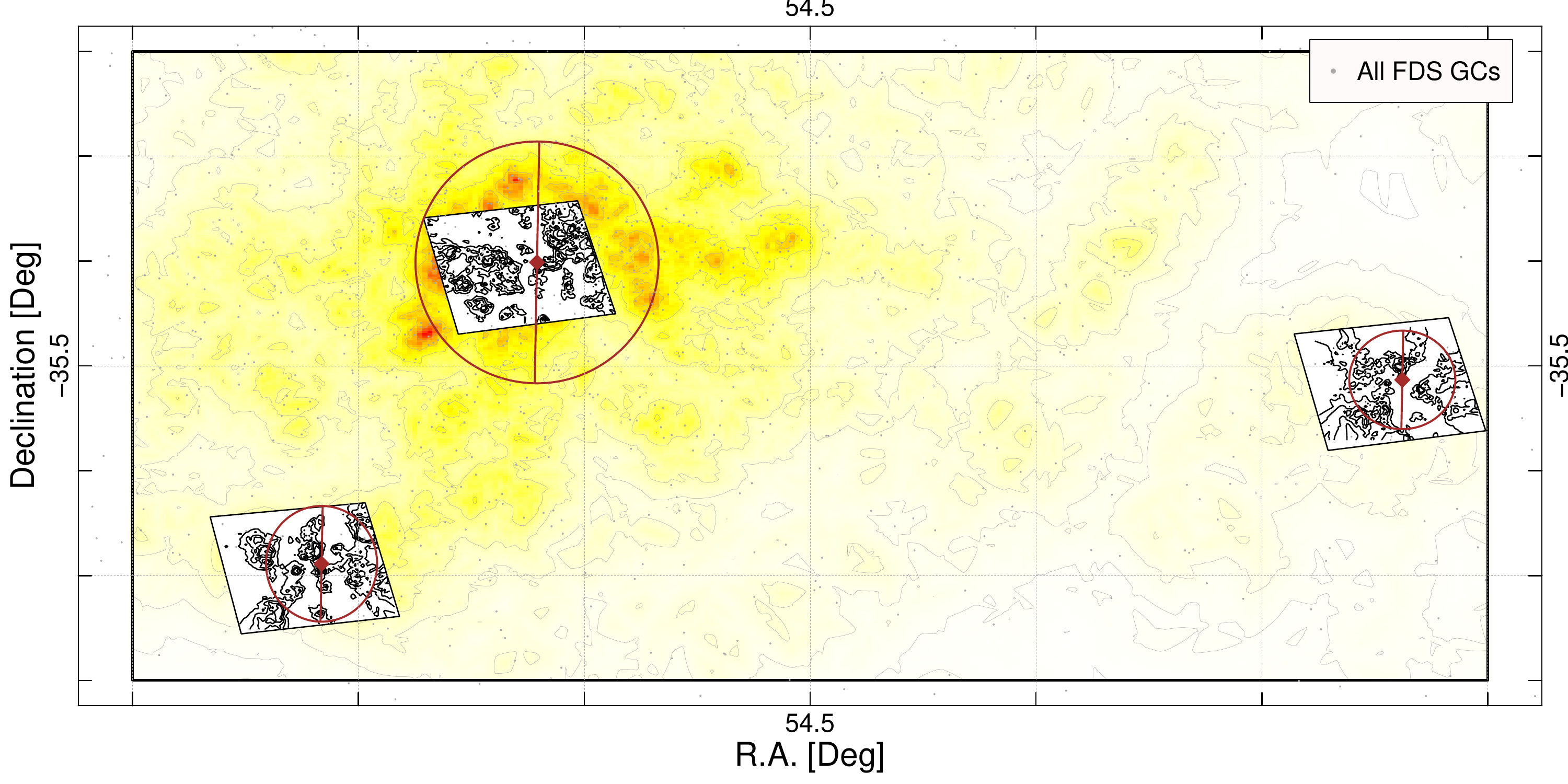}
    \caption{Upper panel: Density map of the spatial distribution of GC candidates from the 
    Fornax Deep Survey (FDS) collaboration~\citep{cantiello2020} (gray points) in the core of 
    the Fornax cluster, with the large $K\!=\!=10$ residual GCs structures of the galaxies
    investigated in this paper and derived from the spatial distributions of ACSFCS GCs 
    overplotted. Ten isodensity contours 
    of the FDS GCs density distribution, logarithmically spaced between 2\% and 99\% 
    of the peak density, are shown (gray lines). The three green lines represent the 5\% and
    20\% of the peak GC density and separate low, intermediate and high density regions of the 
    cluster.
    Lower panel: zoom of the FDS GCs density distribution showing the high density region 
    centered between NGC199 and NGC1387, with density contours of the full $K\!=\!=10$ 
    residual maps derived from the distribution of ACSFCS GCs in NGC1399, NGC1404 and NGC1380. 
    Ten logarithmically-scaled isodensity contours of the GC distribution inside this region 
    are shown (gray lines).}
    \label{fig:core_gc_patchwork}
\end{figure}

Recent investigations of the Intra-Cluster Globular Clusters (ICGCs) in 
Fornax~\citep{cantiello2020,dabrusco2016}, based on observations from the Fornax Deep Survey (FDS) 
survey~\citep{iodice2016}, have confirmed the existence of an abundant population of ICGCs
in the core of this cluster.~\cite{cantiello2020} report the observation 
of an elongated overdensity extending $\sim$ 10 Mpc, centered around NGC1399 and stretching 
in the W-E direction with a ($\sim\!10\degree$) tilt. Figure~\ref{fig:core_gc_patchwork} 
shows the density map of the~\cite{cantiello2020} catalog of candidate FDS GCs, 
estimated with the KNN method with $K\!=\!10$ (this value is chosen to highlight spatial structures
of scale similar to that of the GCs residual structures detected in the spatial distribution of 
GCs in the ten galaxies investigated in this paper), around NGC1399, where the insets display the 
density contours from the full residual maps obtained from all GCs detected 
in NGC1399, NGC1404 and NGC1387 with $K\!=\!10$ (lower panel) and the large structures only 
(upper panel) detected in the $K\!=\!10$ residual maps of ACSFCS
GCs for all galaxies located in the core of the cluster of galaxies. 
The qualitative agreement between the positions of the overdensities within the ACSFCS 
footprints and the FDS GCs distribution is particularly evident along the higher-density 
``bridges'' connecting the central galaxies to NGC1404 and NGC1387 and the complex structures 
in the outskirts of NGC1399. These similarities suggest a continuity in the spatial properties
of the different populations of GCs in the core of the Fornax cluster and hints at the possibility
that the large-scale ICGCs spatial features discovered by~\cite{cantiello2020} extend within
the core of the cluster and are coherent with the smaller scale spatial structures observed
in the ACSFCS GCs distribution within few effective radii. More detailed modeling of the 
GC systems of the host galaxies and the Fornax ICGCs population would be needed to explore this 
possibility, but there is growing evidence of the existence of a connection between the 
anisotropies of the GCs spatial distributions at very small galactocentric distances and 
on scales typical of the core of clusters for other galaxies, for example in the cases of 
NGC4365~\citep{dabrusco2015,blom2014} and NGC4406~\citep{dabrusco2015,lambert2020}.

\begin{figure}[h]
    \centering
	\includegraphics[width=0.49\linewidth]{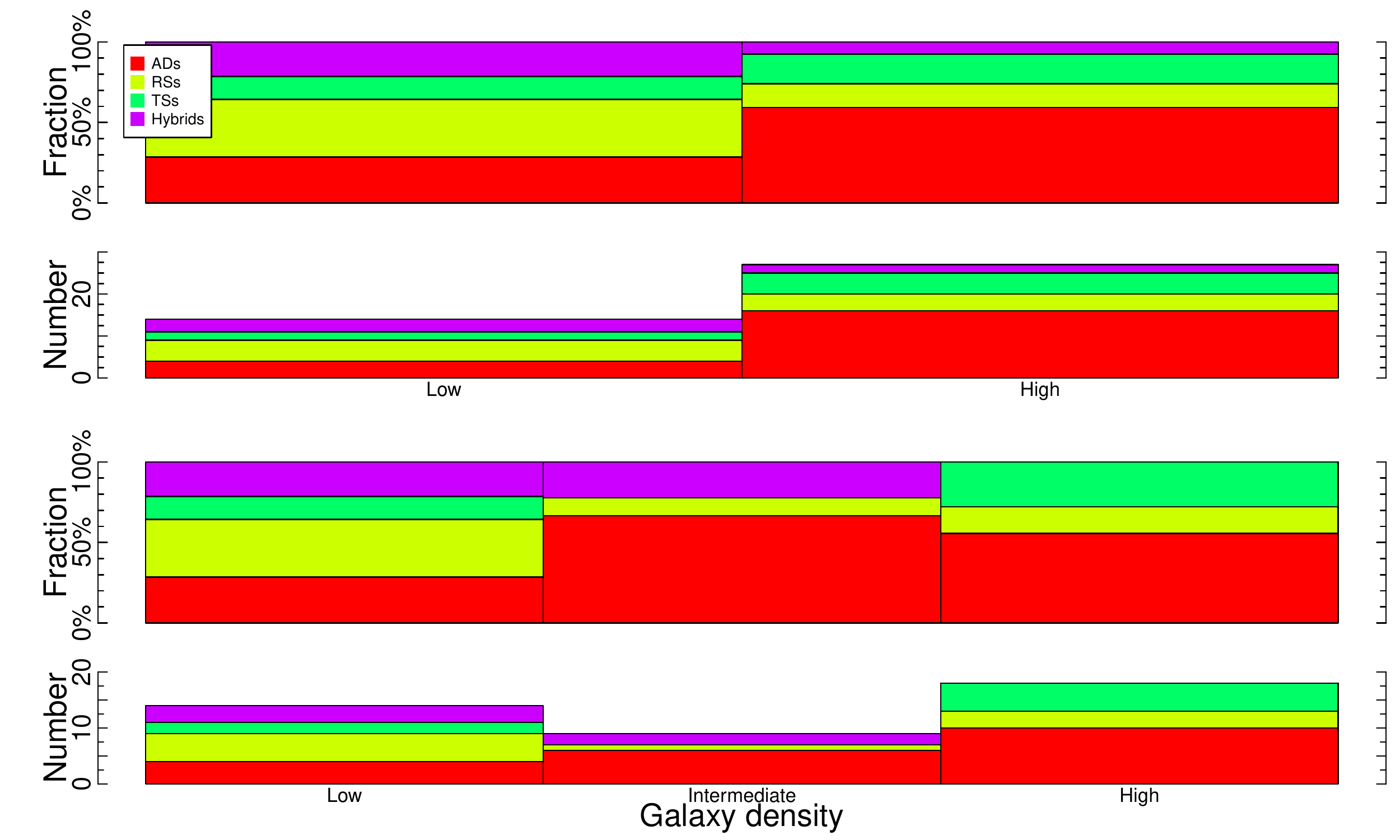}
	\includegraphics[width=0.49\linewidth]{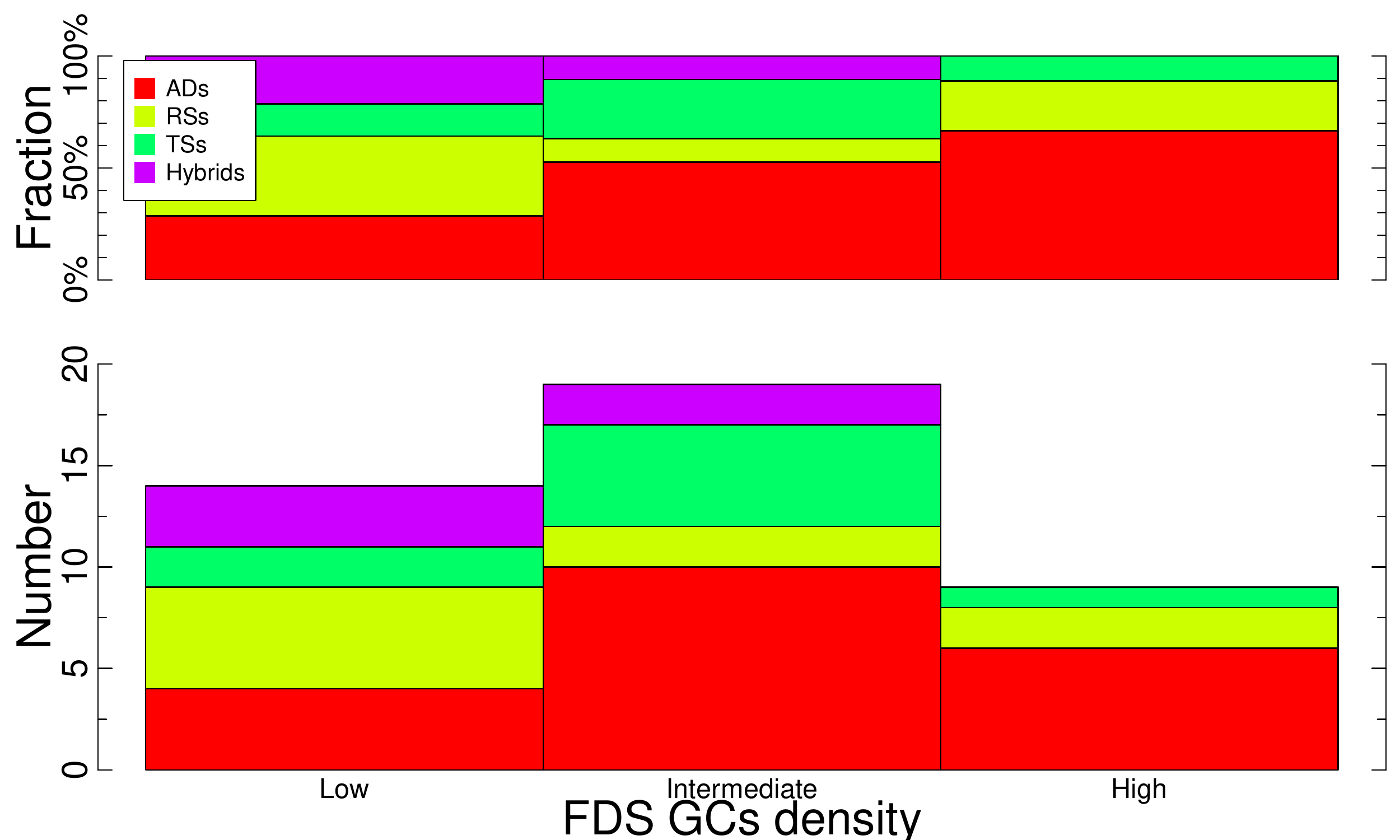}
    \caption{Numbers (lower panels) and fractions (upper panels) of large GC structures split
    by morphological classes as defined in Section~\ref{subsec:spatialstructures_morphology},
    in different galaxy (left) and GC (right) density regions of the Fornax cluster. Percentages 
    and absolute numbers of structures for galaxy density regions (left panel) are displayed for
    both the low/intermediate/high regions separation (lower plots) and the low/high regions 
    (higher plots), respectively marked by 
    the thick green lines and the thick red line in Figure~\ref{fig:cluster_galaxy_insets}.}
    \label{fig:barplots}
\end{figure}

The frequency of large spatial structures of different morphological classes (described 
in Section~\ref{subsec:spatialstructures_morphology}) as a function of the type of density environment 
where their hosts reside, is displayed in Figure~\ref{fig:barplots}. In this plot, the 
absolute number and fraction of GC structures of different classes in low, intermediate
and high density regions based on the spatial distributions of both FCC galaxies (left) and FDS 
ICGCs (right), are shown. The number and fraction of ADs increases from the low to the high
galaxy density regions (lower left plot in Figure~\ref{fig:barplots}), while RSs and TSs are more 
frequent in the high and low regions; Hybrids are only observed in galaxies located
outside of the high galaxy density area instead. Similar trends can be observed when galaxies
are split in low and high density areas (separated by the isodensity contour corresponding to 
50\% of the peak density), as shown by the upper left plot.

Also in the case of density regions based on the FDS GC distribution (right plot in
Figure~\ref{fig:barplots}), we notice that the fraction of ADs increases from low to 
high density regions and Hybrids structures are only detected outside of the highest
density area. The fraction and number of TSs, instead, is largest in galaxies in the 
low GCs density area and lowest in the high-density galaxies. 

Assuming that all GC structures are formed through the interaction
of the host galaxy with satellite galaxies, the observed differences 
in the frequency of different types of GC structures as a function of the galaxy 
density in the cluster of galaxies indicate different properties 
of the seed population of satellites. In the higher galaxy density regions, 
the larger fraction of ADs may be caused by earlier mergers and 
accretion events that resulted in more tightly bound and regularly shaped 
GCs overdensities~\citep[as described by][using E-MOSAICS simulations of MW-sized galaxies]{pfeffer2020} 
that also tend to be located at smaller galactocentric distances than structures 
of the same type observed in hosts in medium and low density regions. Another effect 
potentially shaping the morphology of GC structures observed in the high density region is 
the destruction and/or spatial degradation of coherent structures due to 
gravitational interaction with neighbors and the deeper cluster potential. 
The observed class and position of the GCs structures can also be the footprint of anisotropies 
in the distribution of galaxies relative to the geometry of the cluster, namely
the alignment of the orbits of the satellite with the cluster major axis reported for 
nearby clusters~\citep{knebe2004} and non-random orientation of satellite galaxies 
relative to their hosts~\citep{agustsson2006,wang2021}.

\subsection{Progenitors of the GCs structures}
\label{subsec:spatialstructures_progenitors}

The $\Lambda$CDM model of hierarchical galaxy formation~\citep{white1978,dimatteo2005} 
predicts that galaxies continuously evolve through merging and accretion of satellites. 
Observational evidence of this process in the local Universe is abundant and 
convincing. The observations of the Sagittarius Stream~\citep{ibata1994}, MW companions
undergoing tidal disruption~\citep{belokurov2006} and of streams and 
dwarf galaxies in the halo of M31~\citep{mcconnachie2006} support this model. More recently, 
data from the Gaia mission allowed the discovery of ancient merger events 
that contributed to the build up of the stellar mass currently observed in the Milky 
Way~\citep{helmi2018,belokurov2018}. At larger distances, ongoing accretion of satellite galaxies
has been inferred through the kinematical signature left on the GC systems of the 
hosts~\citep{strader2011,romanowsky2012,blom2012}. 

Following~\cite{dabrusco2015}, we assume that all structures detected in the spatial distribution of 
the GCs in the ten Fornax ETGs studied in this paper are the relics of the GC systems of accreted 
satellite galaxies, detected over the smooth and relaxed GC distribution of the host. The only 
exception is represented by GC structures classified as Hybrids: 
given their large sizes and peculiar morphologies, Hybrids are more likely to be the 
either composite structures or resulting from different physical mechanisms, like major 
dissipationless mergers or wet dissipation mergers that might have triggered the formation of 
young, metal-rich GCs along the major axis of the newly formed galaxy~\citep[as observed in Virgo by][]{wang2013}.
In particular, in disk-disk major mergers, the increased pressure in metal-rich molecular clouds 
triggers the collapse and subsequent formation of GCs concentrated along the major axis of the 
remnant~\citep{bekki2002,brodie2006}.

Moreover, other effects that may contribute to their features are the existence of a sizeable 
disk component in the spatial distribution of metal-rich GCs~\citep[as observed in Virgo cluster galaxies by][]{wang2013}
and the chance superposition of multiple distinct structures. For this reason, 
Hybrids GC structures have been excluded from the analysis performed in this Section. 

\begin{figure}[h]
    \centering
	\includegraphics[width=0.9\linewidth]{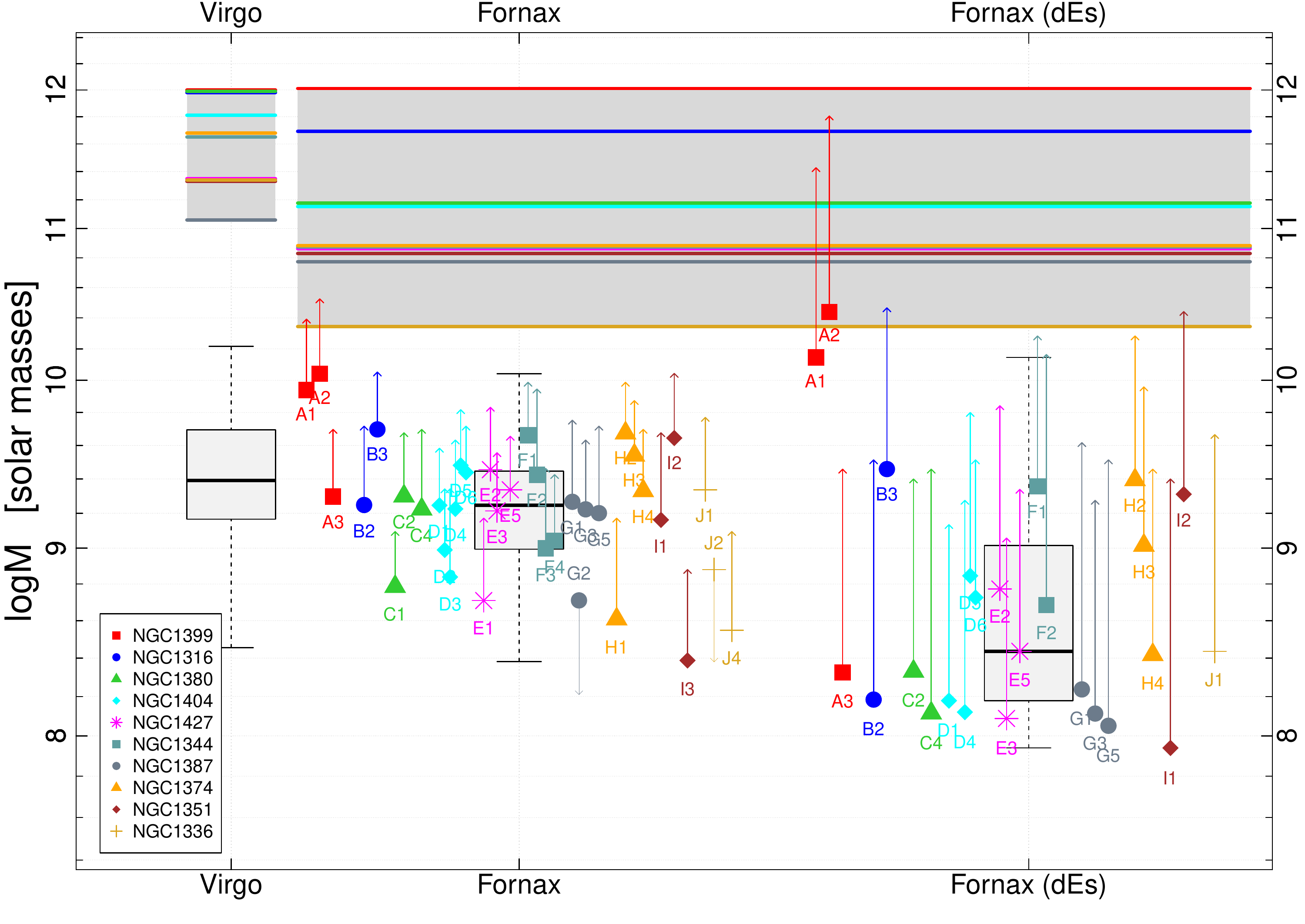}
    \caption{Distribution of dynamical masses of progenitors of the large GCs spatial structures
    observed in the Fornax cluster galaxies. For each spatial structure,
    the expected and maximal mass of the progenitor satellite galaxies are determined by applying the 
    relation between dynamical mass and total number of GCs of a galaxy determined by~\cite{harris2013}
    (see Section~\ref{subsec:spatialstructures_progenitors}) to the excess number of GCs
    N$^{\mathrm{(exc)}}_{\mathrm{GCs}}$ (base of the arrows) and the total observed number of 
    GCs in the structure N$^{\mathrm{(obs)}}_{\mathrm{GCs}}$ (tip of the arrows). The masses of the progenitors 
    have been calculated using both the~\cite{harris2013} correlations for the general sample of galaxies 
    and for dEs only. Hybrids structures are excluded, while structures with negative N$^{\mathrm{(exc)}}_{\mathrm{GCs}}$ 
    are displayed as arrows pointing downwards with the base showing the maximal mass of the progenitor
    derived from their N$_{\mathrm{obs,GC}}$. The dynamical 
    masses of the host galaxies~\citep{harris2013} are shown as horizontal lines in the grayed 
    out region. The boxplot based on the distribution of satellite progenitors of GCs 
    spatial structures in Virgo cluster galaxies~\citep{dabrusco2015} is also shown for reference.}
    \label{fig:massesprogenitors}
\end{figure}

\cite{harris2013} determined observational correlations between the dynamical mass of the host galaxy and 
the total number of GCs for a full sample of galaxies of different morphological types and a subsample of 
dwarf ellipticals (dEs)~(Table 2 and Figure 9 therein): 

\begin{eqnarray}
    \mathrm{Full\ sample}\rightarrow\mathrm{N}_{\mathrm{GCs}}\!=\!\mathrm{M}_{\mathrm{dyn}}^{1.04\pm0.03} \nonumber \\
    \mathrm{dEs}\rightarrow\mathrm{N}_{\mathrm{GCs}}\!=\!\mathrm{M}_{\mathrm{dyn}}^{0.37\pm0.09}    
     \label{eq:dynamicalmass}
\end{eqnarray}

The relation for the full sample is based on galaxies with $\mathrm{M}_{\mathrm{dyn}}\!\geq\!10^{10}\mathrm{M}_{\Sun}$
and average $\log{(\mathrm{M}/\mathrm{M_{\Sun}})}\!=\!11.2$, while the dEs correlation is observed
for a sample of dwarfs ellipticals with average $\log{(\mathrm{M}}/\mathrm{M_{\Sun})}\!=\!9.2$ and covering a 
much larger interval at low masses. By inverting these relations, we calculated the dynamical masses of the 
progenitor galaxies of the GC structures from the number of GCs in the structure for both the general and 
dEs samples~(Figure~\ref{fig:massesprogenitors}). Dynamical masses 
$\log{(\mathrm{M}}/\mathrm{M_{\Sun})}\!<\!7.7$ resulting from Eq.~\ref{eq:dynamicalmass} for dEs because were not
considered in the following analysis because they 
lie outside the range of masses occupied by the galaxies used by~\cite{harris2013} to determine the correlations. 

For each structure, we also determined the ``maximal'' mass of the progenitor using the same procedure 
and assuming that all the observed GCs within a structure belong to the GCS of the potential progenitor 
(i.e. $\mathrm{N}^{\mathrm{(exc)}}_{\mathrm{GC}}\!\equiv\!\mathrm{N}^{\mathrm{(obs)}}_{\mathrm{GC}}$). 
Before calculating the dynamical 
mass of the progenitors, both the excess and total numbers of GCs in each structure have been corrected for 
completeness using the Fornax cluster GCLF~\citep{villegas2010}. 

Figure~\ref{fig:massesprogenitors} shows the distribution of the masses of the progenitors derived from 
the excess and total GCs numbers in the large spatial structures of the Fornax cluster galaxies, compared 
to the distribution of M$_{\mathrm{dyn}}$ of the 
progenitors of the spatial GCs structures detected in a sample of bright Virgo cluster 
galaxies by~\cite{dabrusco2015}. The masses of the progenitors are shown both for the general and dEs
relations~(Equations~\ref{eq:dynamicalmass}). Maximal masses are indicated in Figure~\ref{fig:massesprogenitors} 
by the tips of the arrows pointing upwards. The range of dynamical masses of GC structures in the Fornax 
galaxies for the general prescription is compatible with the distribution of masses for GC over-densities 
in the Virgo cluster of galaxies, although the total interval covered is slightly wider. In general, 
the progenitors of GCs structures in the most massive galaxies have larger $\mathrm{M}_{\mathrm{dyn}}$ 
than in less massive galaxies. As expected, the average mass of dEs progenitors is lower and the covered
range larger than in the case of the general correlation, and we do not see an increase of masses of the 
progenitors with the mass of the host. 

\begin{figure}[h]
    \centering
	\includegraphics[width=0.95\linewidth]{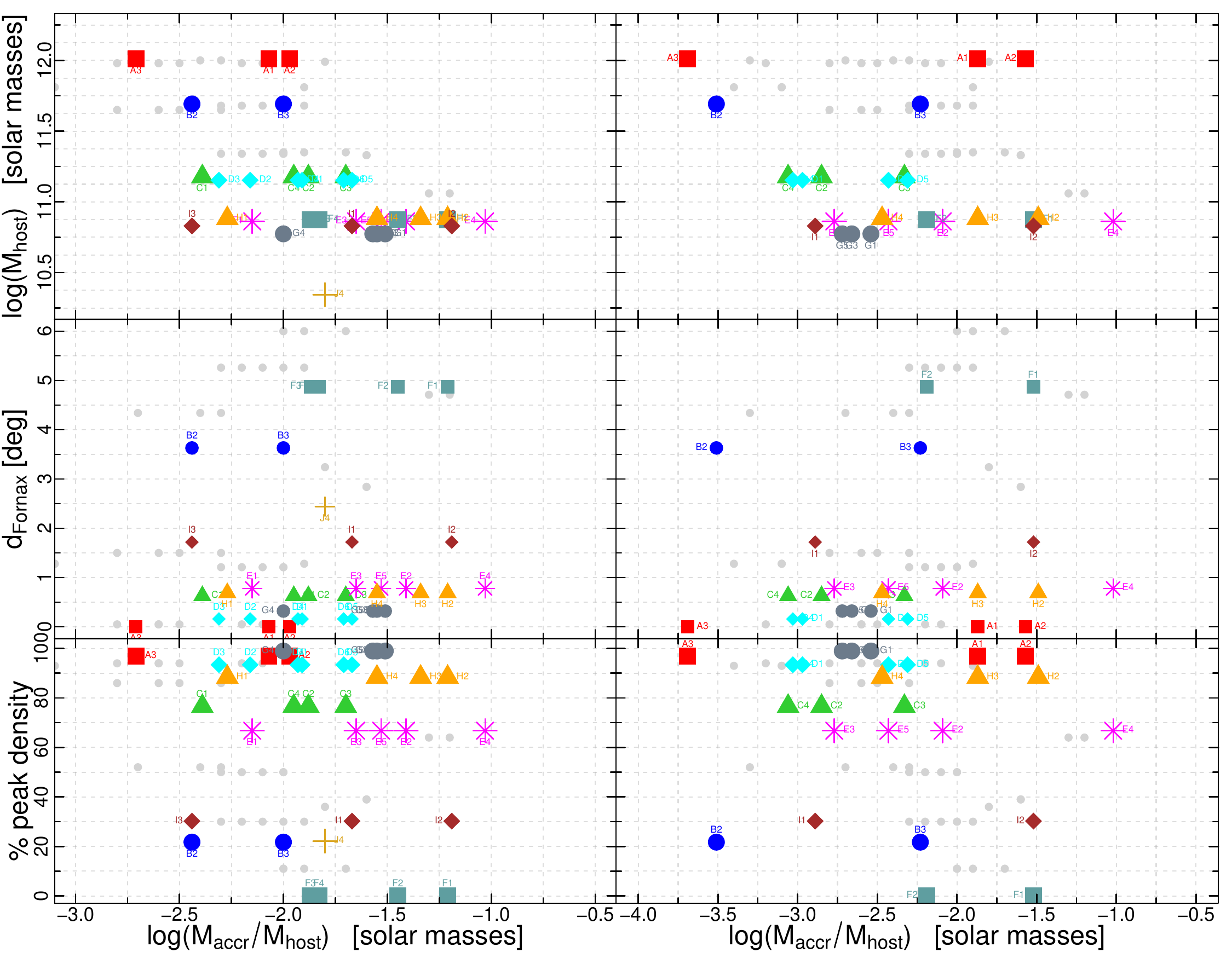}
    \caption{Distribution of the mass ratios of the progenitors of the 
    large GCs spatial structures to the total dynamical mass of their host galaxies vs
    the host M$_{\mathrm{dyn}}$ (top panels), distance from the center of the Fornax cluster, assumed to 
    coincide with the position of NGC1399 (mid panels), and percentage of the peak density derived from the 
    spatial distribution of FCC galaxies~\citep{ferguson1989} (low panels). Left and right columns
    show the same plot for masses of progenitors calculated with the general (left) and dEs relation (right) 
    in Eq.~\ref{eq:dynamicalmass}. The gray circles in the background
    represent the values from GC structures detected in the Virgo cluster galaxies~\citep{dabrusco2015}.}
    \label{fig:massratios}
\end{figure}

The mass ratios for the progenitors of the large GC spatial structures in the Fornax cluster
galaxies, calculated with the general and the dEs correlations are shown in Figure~\ref{fig:massratios} 
(left and right columns, respectively). Larger mass ratios are observed for less 
massive galaxies for both prescriptions (upper panels), similar to the trend observed for progenitors 
of GC structures in Virgo cluster galaxies~\citep{dabrusco2015}. This confirms that the mass of different
galaxies grows by accreting satellites of different masses, or that the observed GCs structures 
in different hosts are relics of different stages of a similar galaxy mass growth process. 
The lack of a clear pattern observed for 
mass ratios as a function of both the distance of the host from the center of the cluster and the 
local two-dimensional galaxy density (mid and lower panels) suggests that the environment does 
not significantly affect the mass ratios of the satellite accretion inferred from the current 
spatial distribution of GCs. 

\subsection{The physical nature of GC structures and projection effects}
\label{subsec:physicalnature}

The results presented in this paper are based on the assumption that all the 
the statistically significant GC structures may be associated to either the accretion of satellite galaxies or 
mergers undergone by the host galaxy. Since the full knowledge of the three-dimensional 
distribution of GCs around the host galaxy is unobtainable, indirect approaches to confirm the nature of 
the structures are necessary. 

One way to confirm the physical nature of the GC residuals structures is to search 
for coherent substructures in the phase space of GCs spatially associated to the 
2D residual GCs structures sufficiently distinct from the surrounding host galaxy GC system 
distribution in the phase-space. However, the available data do not allow this investigation:
GC structures reside all at smaller galactocentric radii than the thirteen {\it cold streams} discovered 
by~\cite{napolitano2021} in a sample of spectroscopically observed GCs in the core of the Fornax 
cluster. We also compared the positions of GCs with line-of-sight velocity
measurements from both catalogs of spectroscopically observed GCs presented by~\cite{napolitano2021} 
and~\cite{chaturvedi2021} (including all the data available in the literature), and we found at best five 
GCs within a single GC residual structure and most structures with no corresponding GC (both catalogs 
only cover the core of the Fornax cluster). The low statistics does not allow to draw conclusions on the 
nature of the structures. 

Even if a larger number of GCs with measured velocities were available, the approach described above would 
be potentially hindered by the difficulty in obtaining spectra of GCs at small galacto-centric distances 
(due by the very bright galaxy background) and in disentangling GCs belonging to the system of the progenitor 
from the underlying host galaxy populations. A subtler issue that can also affect this diagnostic is the 
bias caused the fact that bright GCs, for which high quality spectra are more easily achieved, to be redder 
and, thereby, more likely to belong to the host galaxy than to the less massive progenitor GC, whose GC 
systems tend to have bluer average color~\citep{peng2006}. Finally, we can expect to be able to distinguish 
substructures in the phase space only for GCs whose progenitors' orbits had a large component along the 
line-of-sight direction.

Some of the structures presented in Section~\ref{sec:spatialdistribution} might not be physical and be the 
result of chance superposition of GCs due to projection effects. Such bias cannot be excluded {\it a priori}, 
but several considerations that significantly mitigate the risk of misidentifying chance superpositions 
as GC structures can be made. 

\begin{itemize}
    \item It is reasonable to assume that the importance of projection effects decreases with 
    decreasing galactocentric distance as the density of GCs shrinks.~\cite{woodley2011} investigated 
    the presence of physical groups of GCs and planetary nebulae (PNs) in NGC5128, a giant elliptical 
    galaxy, and assessed through simulations that the probability of observing a fake two-dimensional 
    subgroup of GCs as the result of project effects is $<1\%$, while the probability of observing real 
    subgroups overlap is $\sim\!4\%$ over the whole galaxy. All simulated fake subgroups were observed 
    at $<2 r_{\mathrm{e}}$.
    \item Most GCs structures discussed in this paper have complex morphologies
    that cannot be explained with simple projection effects, assuming a smooth 3D spatial distribution of GCs, 
    even when considering distinct components associated to the bulge, halo and disk of the host galaxies. 
    Some of the structures have large sizes and extend to radii larger than $r_{\mathrm{e}}$.
    \item Projection effects cannot explain the differences observed in the properties 
    of the same GC structures in different color classes, if one assumes that the 
    radial profiles of red and blue GCs only differ in their slope and extent.
\end{itemize}

\section{Conclusions}
\label{sec:conclusions}

We studied the 2D spatial distributions of the GC systems of 10 among the most massive Fornax cluster galaxies 
using the GC catalogs extracted from the ACSFCS data~\citep{jordan2007}. We characterized the GC structures detected 
by estimating their statistical significance, size, shape, morphological classification and position relative to the 
host galaxies for both the total GC samples and the red and blue color sub-classes, separately. Our results can be 
summarized as follows.

\begin{itemize}
    \item We detected 60 GCs spatial structures in the ten galaxies investigated, confirming that 
    large-scale overdensities in the GC systems of massive ETGs are common~\citep{dabrusco2015}. Among these
    structures, 17 are classified as intermediate structures and 43 as large structures based on their size. 

    \item The spatial distribution of the observed GC structures, in general, is radially and azimuthally 
    homogeneous except for the innermost ($r\leq 0\arcmin.5$) regions of the host galaxies. The largest and 
    statistically most significant structures are typically located at galactocentric distances between 
    1\arcmin and 2\arcmin, while intermediate structures are homogeneously distributed at all probed 
    radial positions. Similarly, $65\%$ of the significance-weighted centers of the large structures and 
    $70\%$ of the total area of large clusters are found within $45\degree$ and $20\degree$ 
    from the direction of the minor axis respectively, while the centers of the intermediate structures
    are uniformly distributed along the azimuthal direction. Large structures also tend to occupy the largest
    fraction of total observed area along the minor axes directions. 

    \item We proposed a classification of the GC structures based on their morphology, position and orientation 
    relative to the geometry of the host galaxy, and investigated their distribution relative to the host galaxy. 
    We found that amorphous, roughly circular and small 
    structures (ADs) are homogeneously distributed along both the radial and azimuthal axes, while elongated structures
    tend to be less likely located along the major axes of the hosts than the minor axes when their orientation is radial 
    (RSs), while they spread evenly both radial and angular axes when their orientation is tangential (TSs). A small
    number of very large, morphologically complex and possibly composite structures (Hybrids) occupy large azimuthal 
    and radial intervals, straddling the directions of both the minor and major axes of the host galaxies. 
    
    \item We have estimated the average color of the GCs in the residual structures by either considering all 
    GCs located within the geometrical boundaries of each structure or by only averaging over the bluest 
    N$^{\mathrm{(exc)}}_{\mathrm{GCs}}$ GCs. The first approach produced mean colors that are likely skewed 
    towards red because of the contribution of the typically red GCs belonging to the massive host galaxy, while
    the second recipe provides a lower estimate of the real color of the GCs belonging to the progenitors of the 
    structure. Based on the distribution of the average colors of all the GCs included in the GC spatial 
    structures, two 
    families of galaxies can be distinguished: NGC1399, NGC1316, NGC1427, NGC1344 and NGC1374 where the colors of the 
    GC structures are very similar and tightly distributed around the average color of the general GCS of the host, 
    and NGC1380, NGC1404, NGC1387, NGC1351 and NGC1336 whose GC structures colors show a larger variance and are 
    more widely scattered along the GC color distribution of the host. The analysis of the statistical significance 
    of the large GC structures in red and blue GCs subclasses shows that in the first class of galaxies the majority 
    of large structure are more significant in the red GCs than in the blue (8 vs 6, while the remaining structures 
    have comparable significance in both color classes), while the structures in the second group of galaxies tend 
    to be more significant in the blue GCs than red (10 vs 7). On the other hand, the average ``bluest'' colors 
    are all consistent with the interval of colors observed in dwarfs in rich 
    clusters of galaxies~\citep[cp. with][for the GCSs of dwarf galaxies in Virgo cluster]{peng2006}.

    \item Large, statistically significant GC structures are observed in galaxies located in all galaxy 
    density levels within the Fornax clusters, while intermediate structures are more frequent relative
    to the total number of structures detected in hosts in 
    the low galaxy density region of the cluster. This scenario contradicts what was observed in 
    Virgo~\citep{dabrusco2015}, 
    where larger GCs structures are more likely to be detected in relatively low galaxy density regions 
    where accretion of large satellites probably occurred more recently than in the core of the cluster. 
    The data available do not permit to draw statistically robust conclusions regarding the existence
    of a correlation between the size of the structures and the galaxy density in the position of the host.

    \item Similarities in the spatial distributions of ACSFCS GCs and the 
    population of ICGCs are observed in the core of the cluster, where NGC1399, NGC1404 and NGC1387 reside, 
    suggesting a continuity between the spatial properties of the GC populations of these galaxies and the 
    surrounding population of GCs. Split by class, the fraction of ADs is largest for hosts located in the 
    highest galaxy and GCs density regions, while elongated structures (TSs and RSs) are more frequent in 
    the low and high galaxy
    density regions and tend to be found more likely in low GCs density regions. These differences can hint at 
    different geometry of the satellite systems that were the progenitors of the structures currently
    observed in the GCs distribution. 
    
    \item The dynamical masses of the progenitors of the GC structures in the Fornax cluster galaxies, inferred 
    using both the~\cite{harris2013} relations based on the complete sample of galaxies and dEs only, range 
    between $\approx10^8$ and 4$\cdot10^{10}$ M$_{\sun}$. The $M_{\mathrm{dyn}}$ of the progenitors are larger 
    for more massive host galaxies and cover an interval comparable with the range occupied by the dynamical masses 
    of the progenitors of GC structures in the Virgo cluster galaxies~\citep{dabrusco2015}. Conversely, larger mass 
    ratios are observed for the least massive host galaxies, while no clear pattern emerges between the mass ratios 
    and both the projected distance of the host galaxy from the center of the cluster and the local galaxy density. 
\end{itemize}

The results presented in this paper provide additional evidence that that 2D structures are common in the GC 
systems of massive early-type galaxies. The trends of the GCs structure size and orientation relative to the 
geometry of the host galaxy as a function
of the local galaxy density in the Fornax cluster differ from the Virgo cluster galaxies~\citep{dabrusco2015}, 
hinting at a different assembly history for galaxies in the two clusters. Shedding light on the cause of these 
differences, whether they are exclusively informed by the specific assembly history of the each host galaxy or 
they are also influenced by general evolution of the cluster, will require a joint investigation of the spatial 
properties of the GC populations at larger galactocentric radii and of the spatial distribution of the surrounding 
ICGCs. While the interval of effective radii that can be probed by ACSFCS observations extends significantly over
the coverage available for Virgo cluster galaxies, the lack of deep, high spatial resolution data in the 
outskirts ($10\leq r_{e}\leq 5$) for a large sample of ETGs in the nearby Universe is still the main limiting 
factor preventing the observation of recently formed GC overdensities in the halo of their hosts that are 
necessary to draw a complete picture of the properties of the GC structures as a function of their location 
in the host galaxy.

A new generation of cosmological simulations capable of resolving mass and spatial scale typical of 
GCs~\citep{pfeffer2020} provides the exciting opportunity of a directly comparison with the observations 
and to fine-tune our interpretation of the presence of GC structures as a powerful tool to infer the past 
merging/accretion history of the host galaxies and to move along our understanding of how galaxies grow 
and evolve.

\section{Acknowledgments}

\noindent R.D'A. is supported by NASA contract NAS8-03060 (Chandra X-ray Center). M.C. acknowledges support from 
MIUR, PRIN 2017 (grant 20179ZF5KS). M.P. acknowledges the financial support from the ASI-INAF agreement 2017-14-H.O. 
A.Z. acknowledges funding from the European Research Council under the European Union’ s Seventh Framework 
Program (FP/ 2007-2013)/ERC Grant Agreement n. 617001.
This project has also received funding from the European Union's Horizon 2020 research and innovation program
under the Marie Sklodowska-Curie RISE action, grant agreement No 691164 (ASTROSTAT). The SAO REU program is funded 
by the National Science Foundation REU and Department of Defense ASSURE programs under NSF Grant AST-1659473, and 
by the Smithsonian Institution. This research has also made use of results from NASA’s Astrophysics Data System. 
Based on observations with 
the NASA/ESA Hubble Space Telescope, obtained at the Space Telescope Science Institute, which is operated 
by the Association of Universities for Research in Astronomy, Inc., under NASA contract NAS5-26555.

\appendix

\section{GC structures in ten galaxies of the Fornax cluster}
\label{sec:appendix1}

A summary of the properties of the spatial structures detected in the distribution 
of ACSFCS GCs for each galaxy studied can be found below. Details on the procedure
used to detect the GC structures can be found in Section~\ref{sec:spatialdistribution}, 
while Section~\ref{sec:data} provides details on the definition of red and blue GCs color 
classes.

\subsection{NGC1399}
\label{subsec:ngc1399}

\begin{figure*}[ht]
    \centering
	\includegraphics[width=0.49\linewidth]{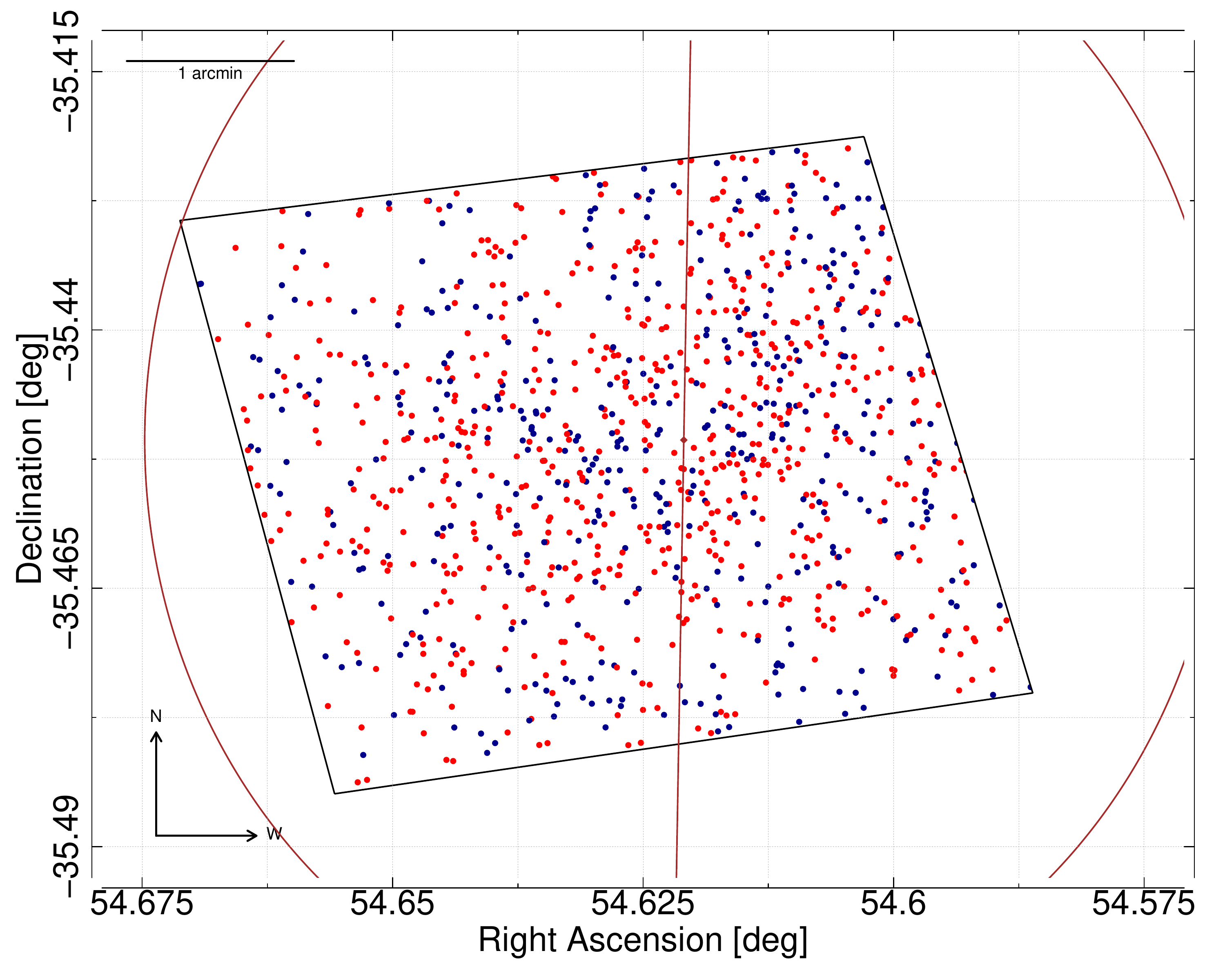}
	\includegraphics[width=0.49\linewidth]{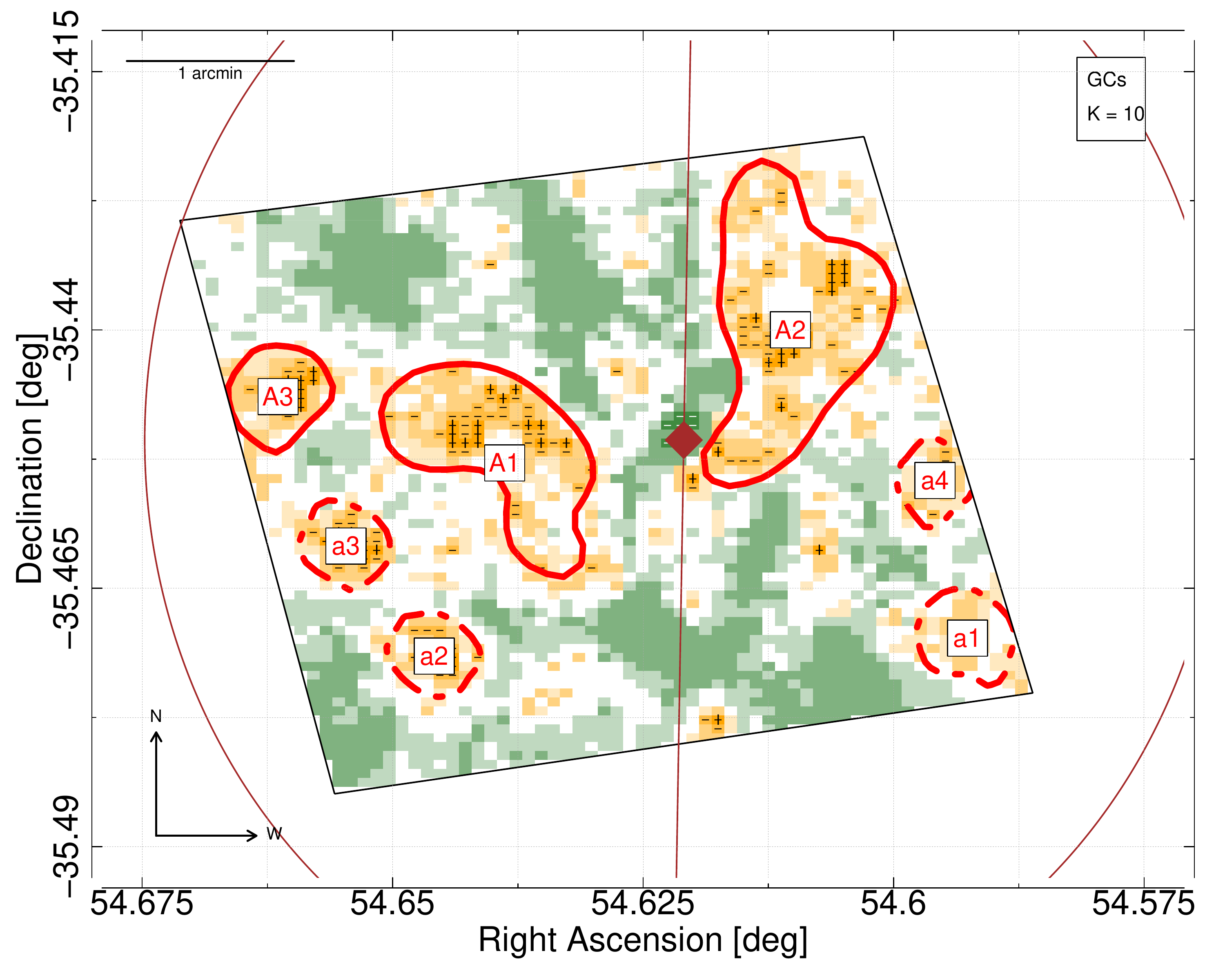}\\
	\includegraphics[width=0.48\linewidth]{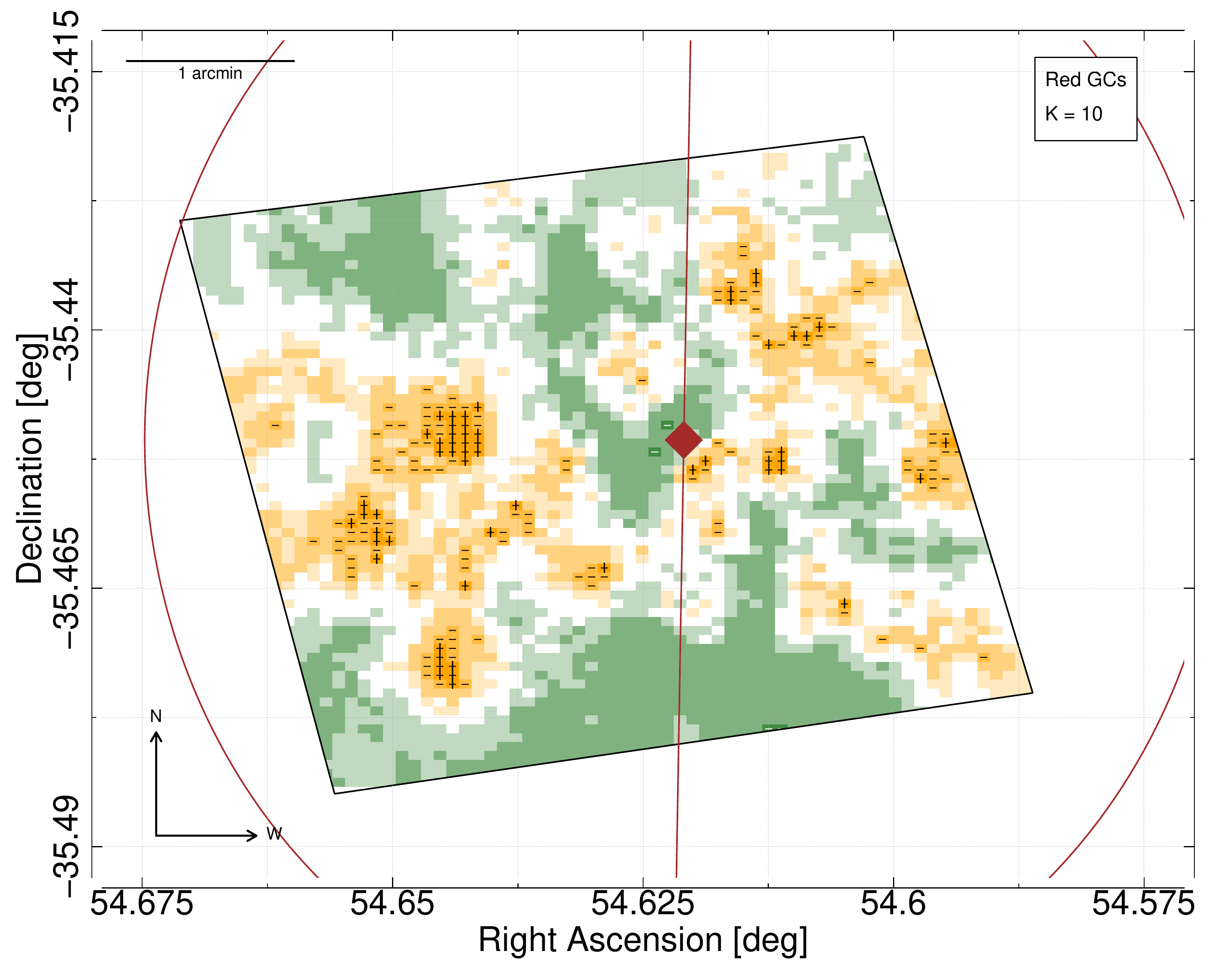}
	\includegraphics[width=0.48\linewidth]{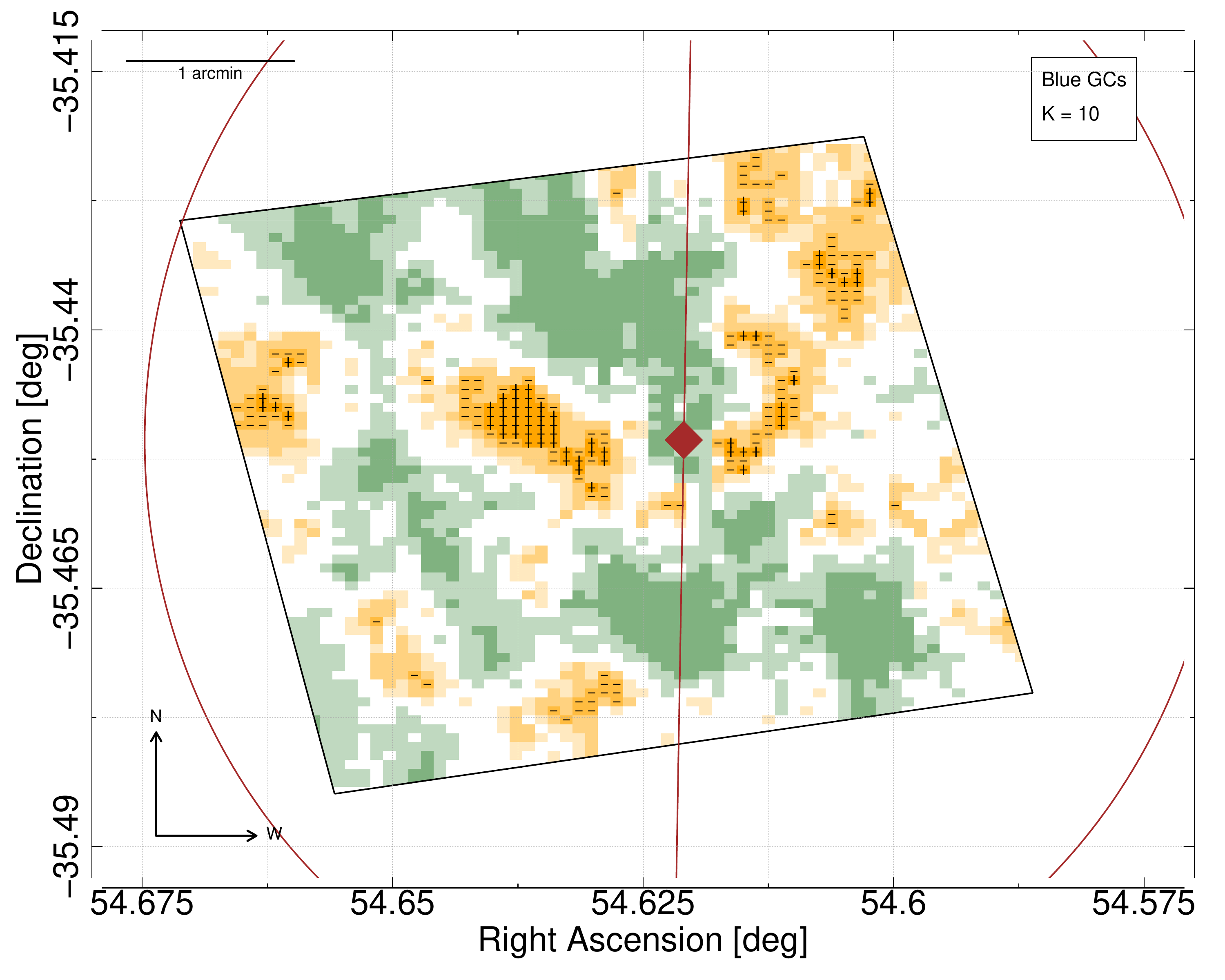}	
    \caption{Spatial distribution of GCs NGC1399. Upper-left: positions of GCs, where blue 
     and red points are associated with blue and red GCs respectively. Upper-right: the 
     residual map generated by the whole sample of GCs, with large (solid lines)
     and intermediate (dotted lines) spatial structures labeled. Orange and green pixels refer to over-dense and 
     under-dense regions of GCs respectively. Darker shades indicate larger statistical 
     significance, horizontal bars and crosses mark cells whose significance is between 2 and 3 
     $\sigma$'s and larger than 3 $\sigma$'s respectively. Lower-left: the residual map for blue 
     GCs; Lower-right: the residual map for red GCs. The ellipse displays the D$_{25}$ elliptical 
     isophote and the red line represents the galaxy major axis.}
 	\label{fig:ngc1399}
 \end{figure*}

NGC1399 is the central galaxy of the Fornax cluster and, in the optical light, is classified as a peculiar 
elliptical galaxy. Given its large size ($r_{e}\!=\!367\arcsec.6$), the single ACSFCS pointing is entirely 
located within the D$_{25}$ diameter of the galaxy. Figure~\ref{fig:ngc1399} shows the distribution of GCs 
in NGC1399 (upper left) and the $K\!=\!10$ residual maps obtained for all, red and blue GCs (upper right, 
bottom left and right panels, respectively). The largest and most significant overdensity in the residual 
maps for all GCs (A2) is located W of the nucleus and extends toward the NW corner of the observed field 
along a mostly radial direction. This structure is also visible in the blue residual map, where it shows 
three apparently disjoint substructures, while in the red GCs map no coherent structure is visible with 
only a few isolated peaks located in the area occupied by A2. 
Four significant spatial structures, potentially connected, are observed E of the center of the galaxy. 
Among these structures, A1 and A3 are the largest: A1 has an elongated shape extending radially 
from the center of the galaxy towards the NE direction, while the other three structures in this region 
(A3, a2 and a3) are approximately circular and are located at larger galactocentric distances than A1 
and closer to the E edge of the observed field. Two smaller, less significant circular structures are 
located along the W boundary of the observed field (a1 and a4). In the E 
half of the observed field, A1 and A3 can be clearly associated to similar overdensity regions in the 
spatial distribution of both red and blue GCs, while a2 and a3 are clearly detected only in the red residual map.
The structures a1 and a4 are most prominent in the red GCs. 
The presence of very significant red GCs spatial structures in the outskirts of the imaged region, 
especially in the E side of the field, reflects the reported abundance of red GCs at relatively large
galactocentric distances possibly due to a recent history of wet mergers experienced by this 
galaxy~\citep{goudfrooij2004}.

\begin{figure*}[ht]
    \centering
	\includegraphics[width=0.75\linewidth]{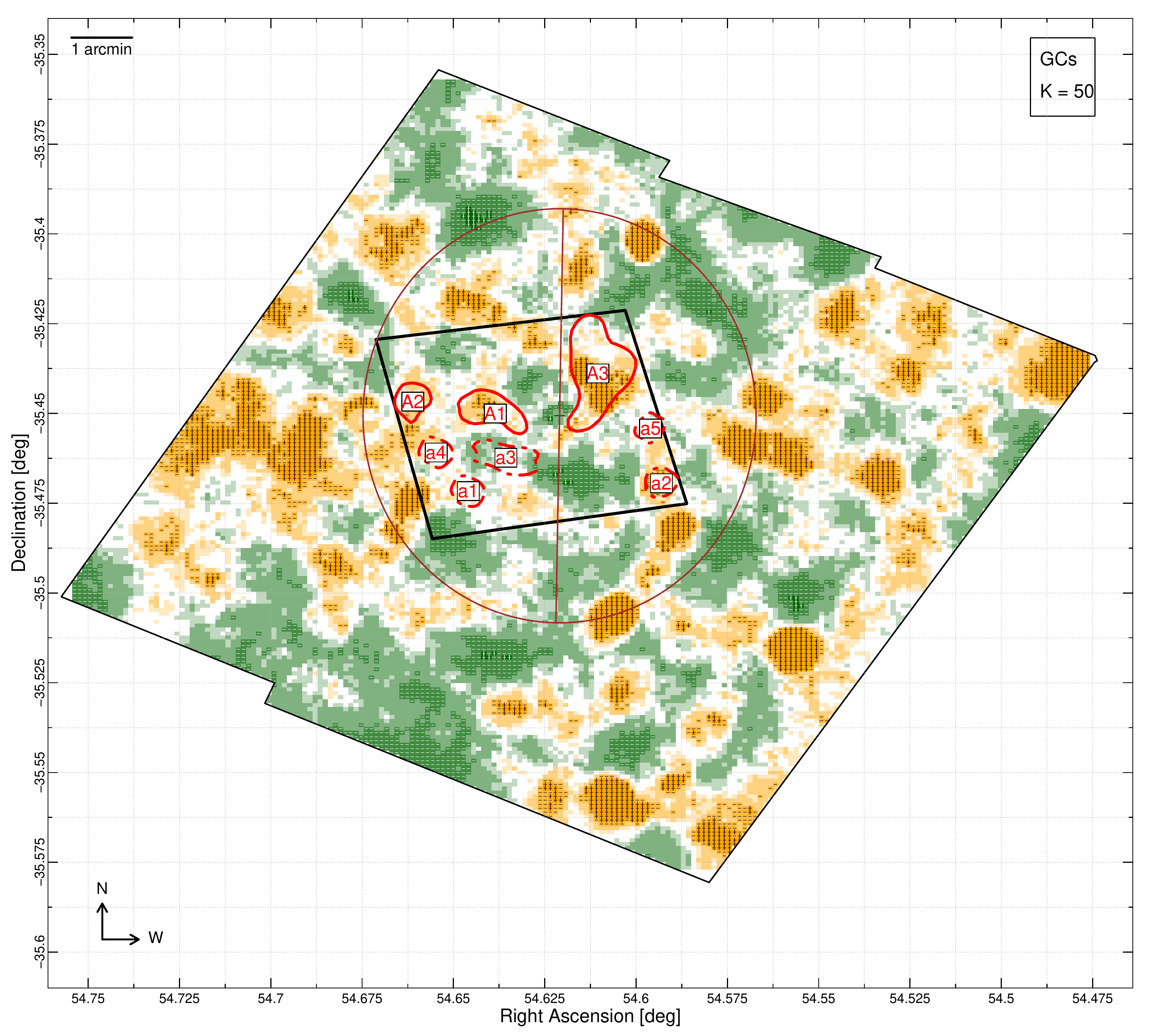}
    \caption{Residual map of the spatial distribution of all GCs from~\cite{puzia2014} ($K\!=\!50)$. The red 
    lines show the large and intermediate GC spatial structures detected in the ACSFCS GCs spatial distribution. 
    The black polygon represents the field-of-view of the ACSCFCS observation in the core of NGC1399. The ellipse 
    displays the D$_{25}$ elliptical isophote and the red line represents the galaxy major axis.}
 	\label{fig:ngc1399_comparison}
 \end{figure*}

\cite{puzia2014} investigated the trends of the GCs structural parameters in the NGC1399 GCS as a function 
of the distance from the center of the galaxy using a catalog of GCs extracted from a mosaic of 9 HST 
ACS observations (PI Puzia) 
in the F606W filter, arranged to cover a large field-of-view reaching a maximum galactocentric distance of 
$\sim8\arcmin.76$, corresponding to $\sim$5.2 $r_{e}$. Figure~\ref{fig:ngc1399_comparison} shows the residual 
map ($K\!=\!50$) derived from the spatial distribution of all~\cite{puzia2014} GCs candidates with overplotted 
the boundaries of the the large and intermediate GC structures detected in the ACSFCS GCs distribution. The
large $K$ value used to derive the residual map of this catalog of GCs qualitatively matches the spatial resolution 
of the ACSFCS maps over a larger field of view and to balance the larger density of GC candidates. 
The ACSFCS overdensities are clearly visible in the residual map of the~\cite{puzia2014} GCs; outside the ACSFCS 
field of view, a prominent area of enhanced density is observed E of the NGC1399 D$_{25}$ elliptical isophote, 
while more structures are located near the S and W corners of the observed field. 

\subsection{NGC1316}
\label{subsec:ngc1316}

\begin{figure}[ht]
    \centering
	\includegraphics[width=0.45\linewidth]{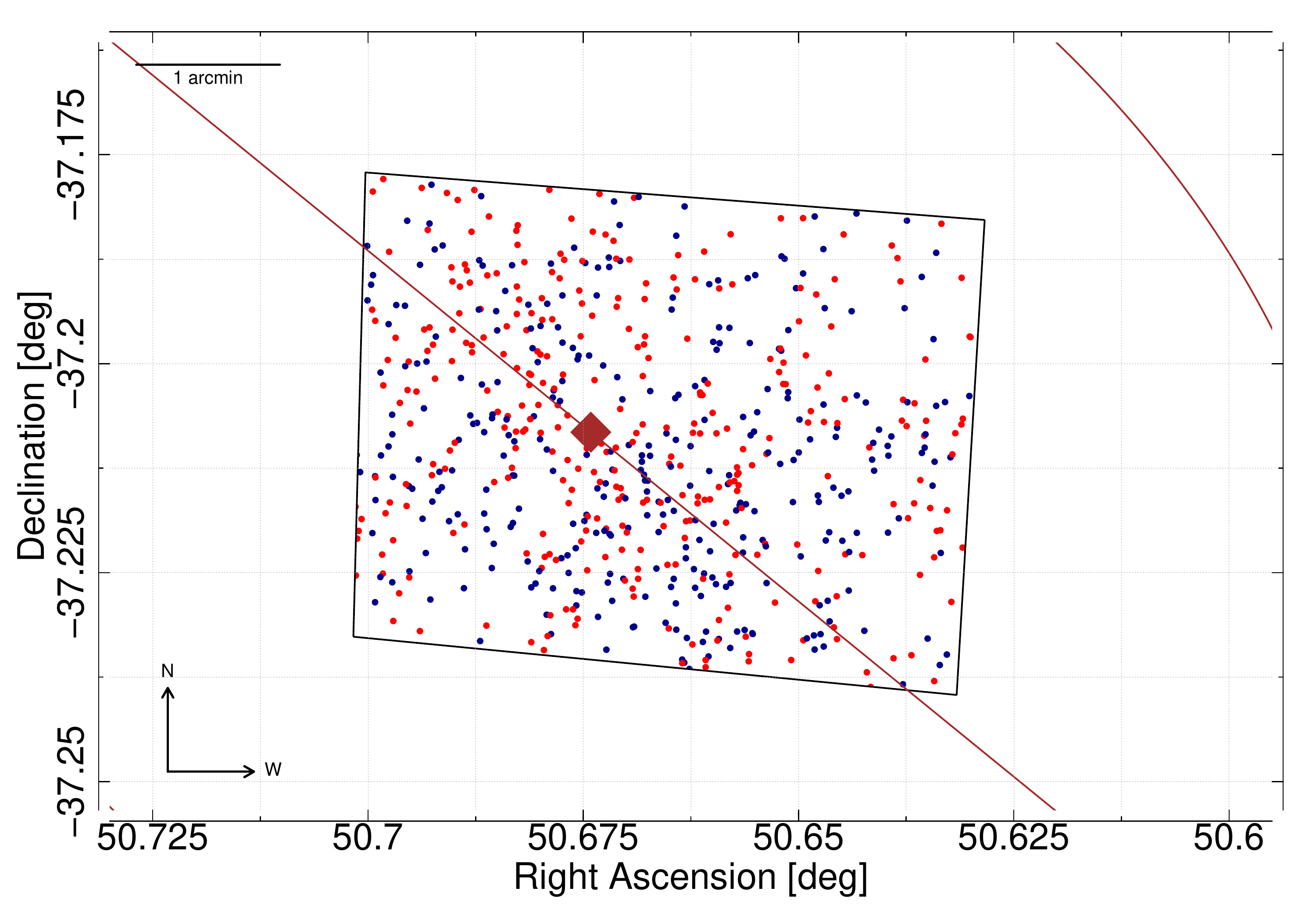}
	\includegraphics[width=0.45\linewidth]{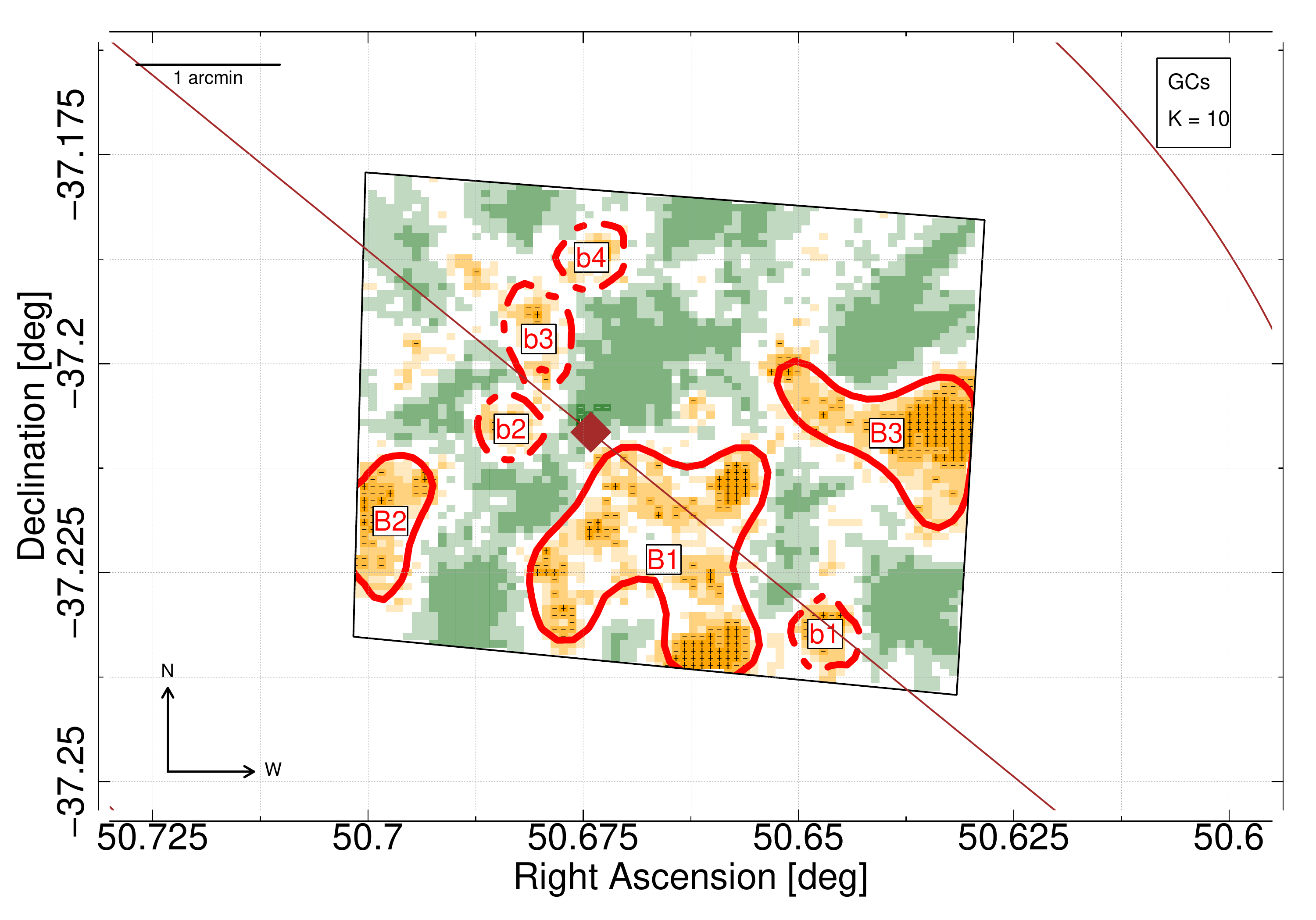}\\
	\includegraphics[width=0.45\linewidth]{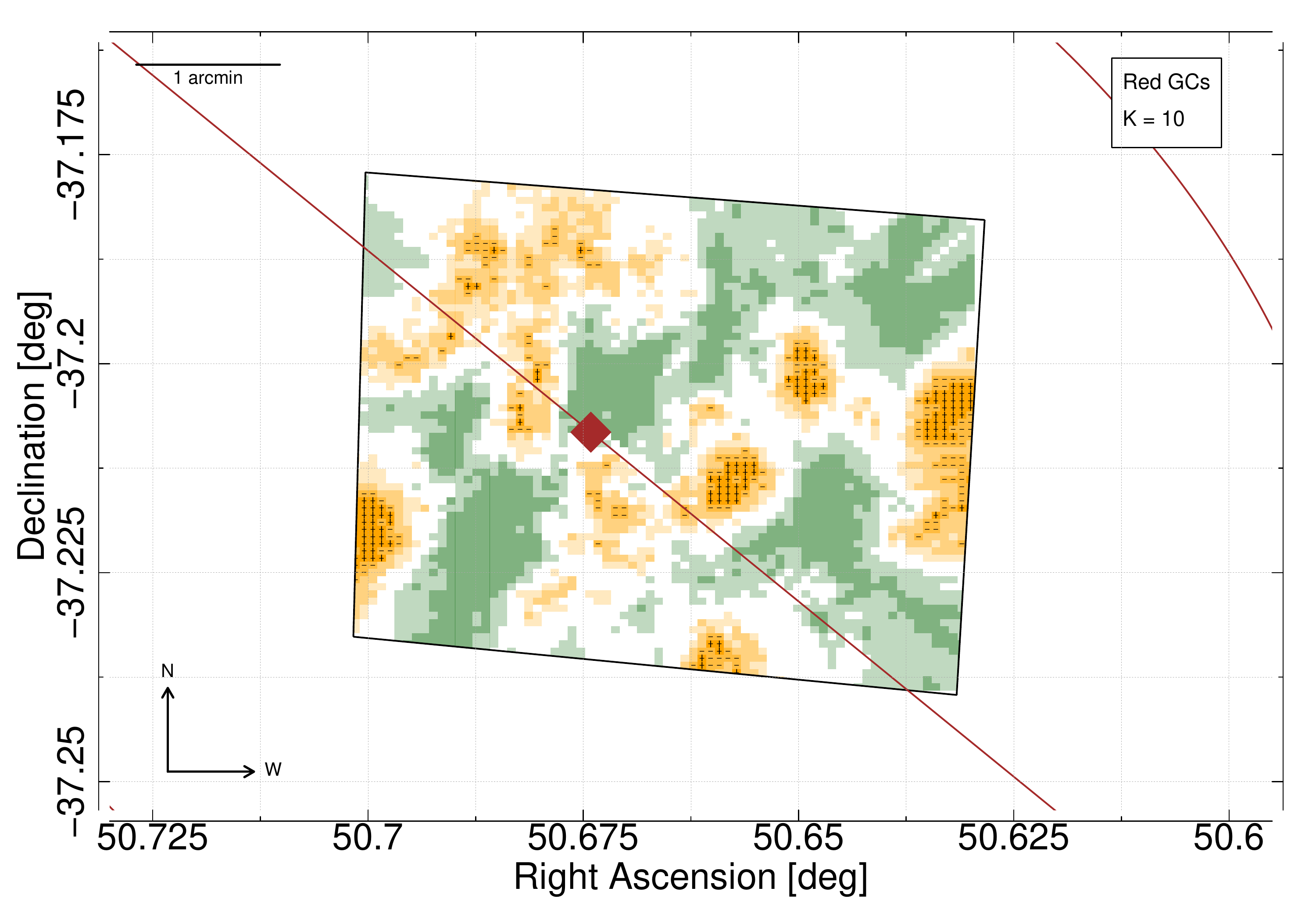}
	\includegraphics[width=0.45\linewidth]{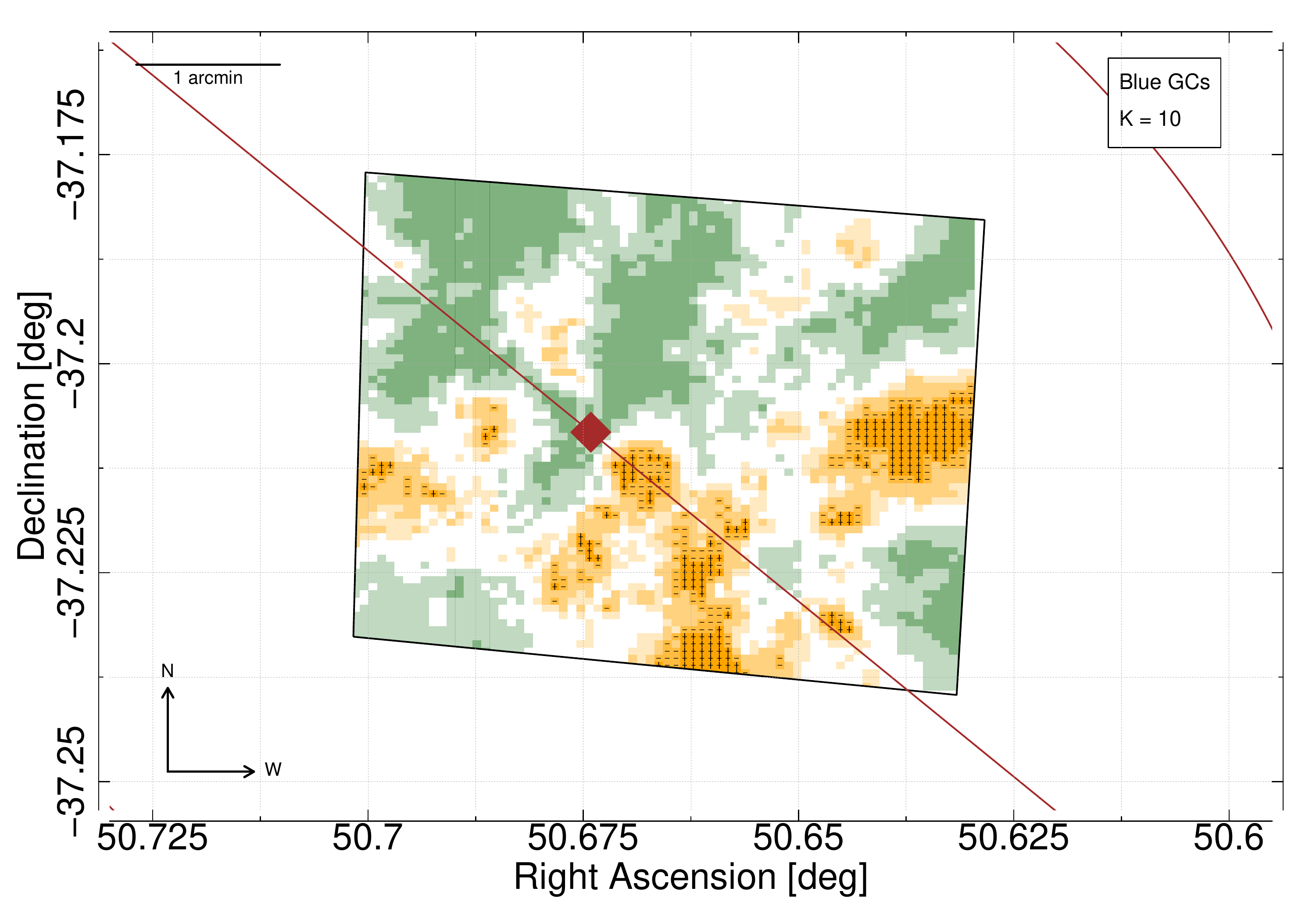}	
    \caption{Spatial distribution of GCs in NGC1316 (see caption of Figure~\ref{fig:ngc1399} for details).}
    \label{fig:ngc1316}
\end{figure}

The investigation of the kinematics of the GC system of the giant elliptical NGC1316 has revealed 
that this galaxy is the likely remnant of a merger occurred $\approx 3$ Gyr ago~\citep{goudfrooij2001,sesto2018}.
~\cite{sesto2016}, in particular, using Gemini ground-based observations, 
have established the existence of four different color components whose 
spatial distributions differ for ellipticities and radial profiles, consistent with the presence of an 
almost spherical low-metallicity halo and a more flattened and chemically enriched bulge.~\cite{goudfrooij2004}
observed different turn-over magnitudes of the luminosity functions of the inner and outer red GCs 
sub-populations, drawing a picture of complex stratification of GC color classes and
confirming the merger scenario according to which the metal-rich GC subsystems formed during mergers can 
dynamically evolve into the normal giant ellipticals red GC systems.

As in the case of NGC1399, the region of NGC1316 where ACSFCS data are available is entirely contained within 
the D$_{25}$ elliptical isophote. The residual map of the general distribution of 
GCs in NGC1316 (upper right panel in Figure~\ref{fig:ngc1316}) features two large, complex regions of significant 
overdensities (B1 and B3) located S and E of the center of the galaxy. B1 partially overlaps the S section of
the major axis of the galaxy and is characterized by a complex morphology and the presence of several 
density peaks that suggest the existence of multiple underlying physical structures. B1 is clearly visible in 
the residual maps of blue GCs while in red GCs only two peaks are detected in the B1 area. B3 extends radially and
is visible in both red and blue GCs, but in red GCs it is split in two separate substructures. N and 
E of the center of the galaxy, four roughly circular structures (B2, b2, b3 and b4) are aligned 
on a radial direction perpendicular to the major axis of the galaxy. B2 is clearly visible in both GCs color 
classes, while the other less significant structures are only clearly detected in the red GCs. The structure 
b1, on the other hand, is located along the S major axis and is only detected in the blue residual map. 
 
\subsection{NGC1380}
\label{subsec:ngc1380}

\begin{figure}[ht]
    \centering
	\includegraphics[width=0.45\linewidth]{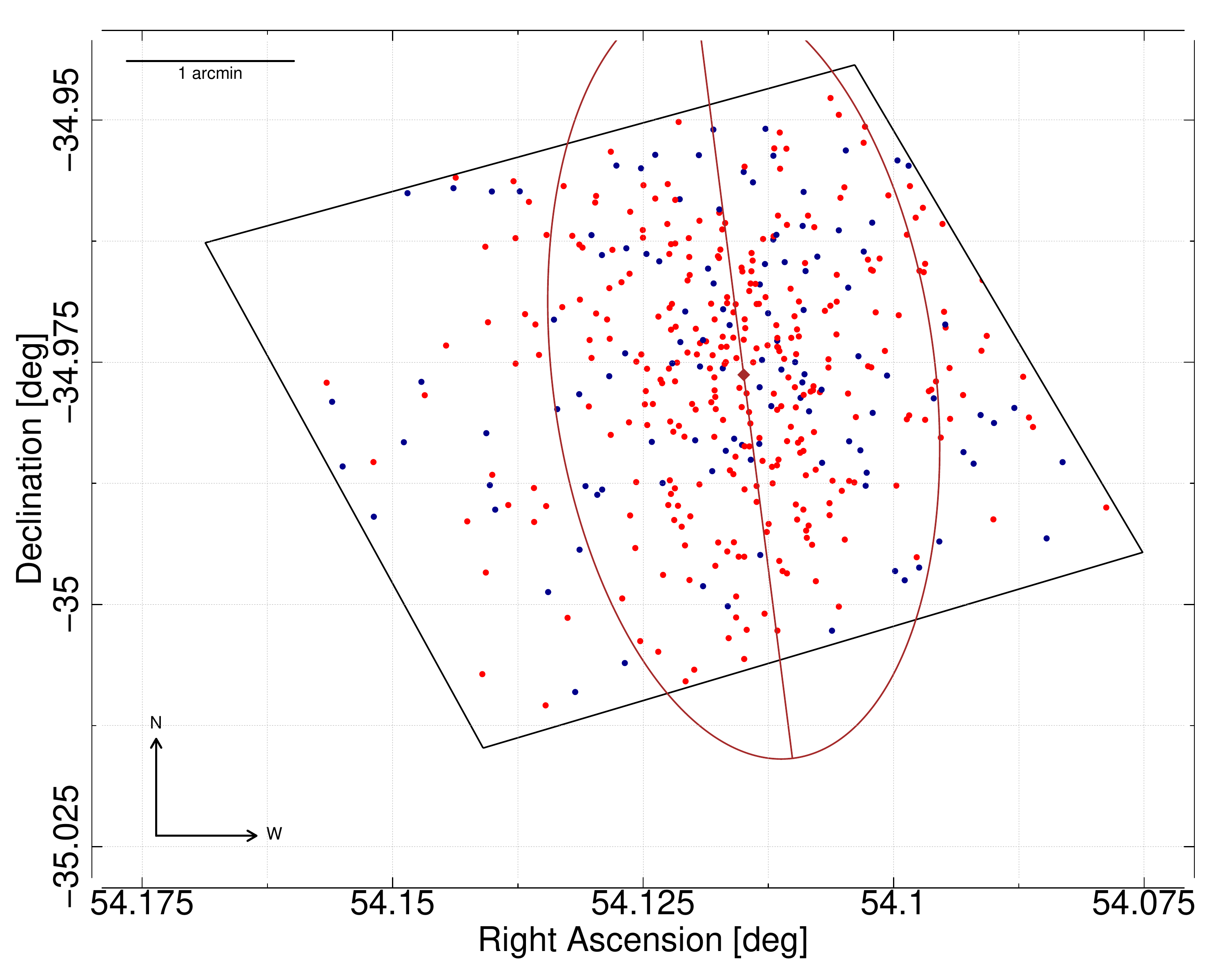}
	\includegraphics[width=0.45\linewidth]{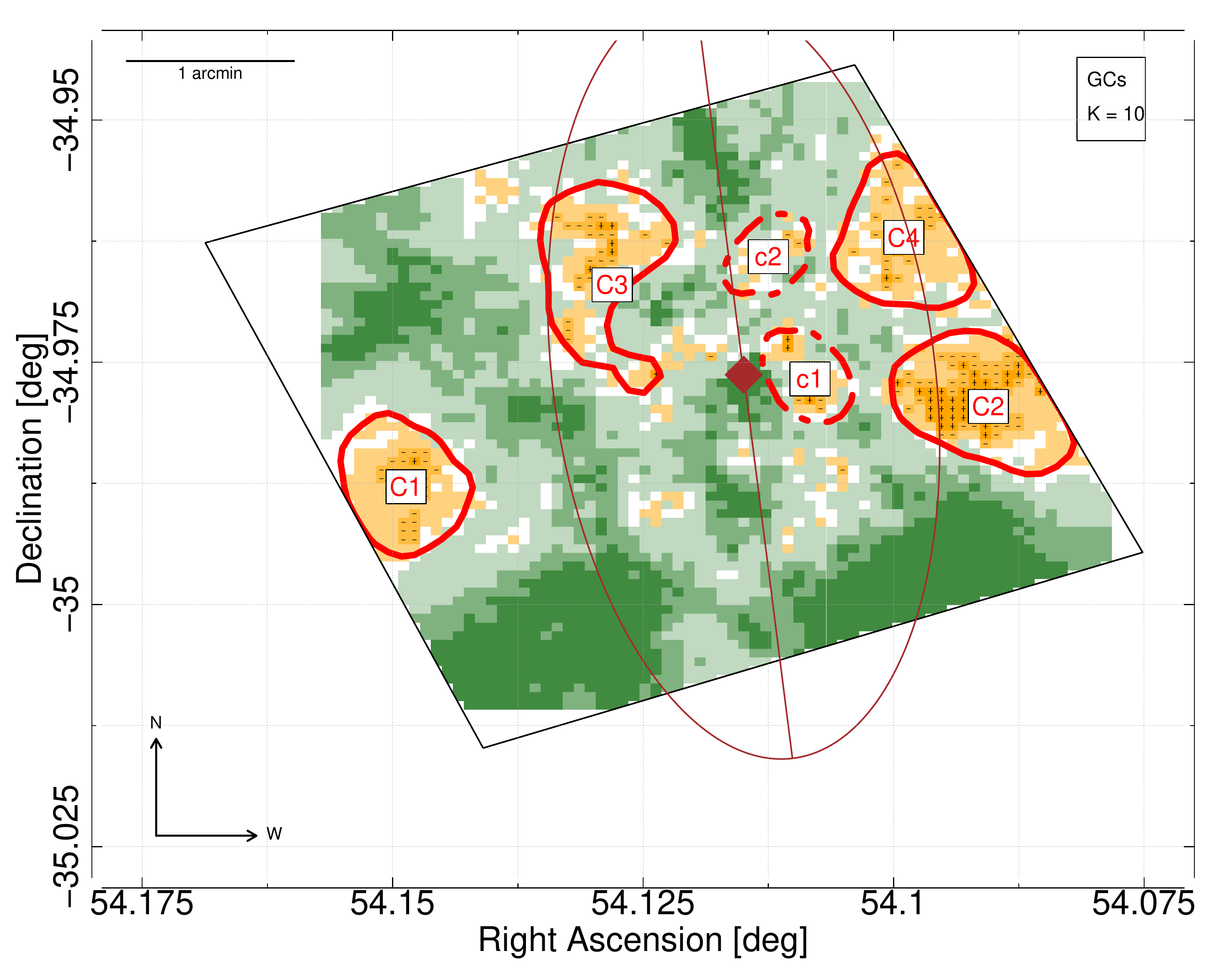}\\
	\includegraphics[width=0.45\linewidth]{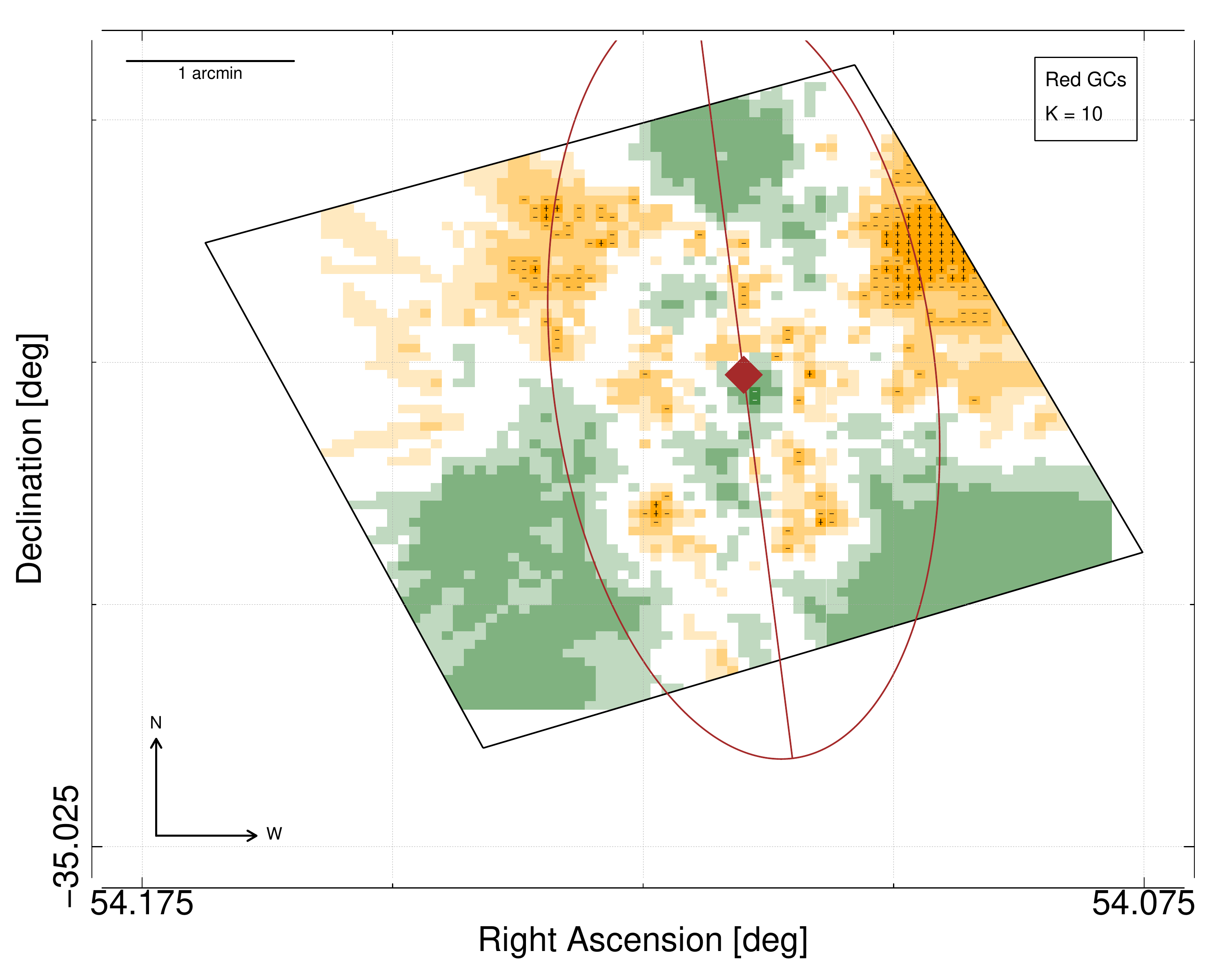}
	\includegraphics[width=0.45\linewidth]{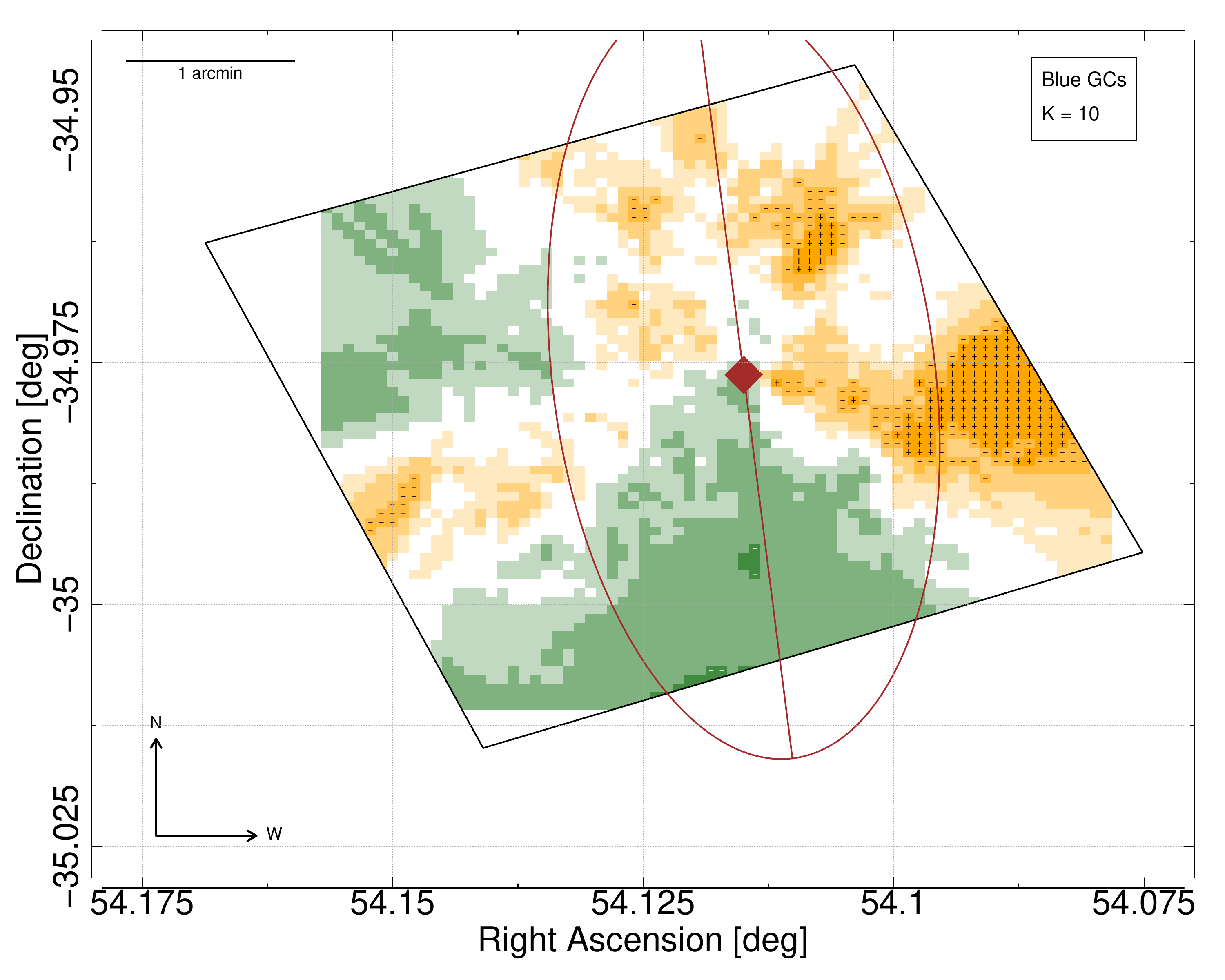}	
    \caption{Spatial distribution of GCs in NGC1380 (see caption of Figure~\ref{fig:ngc1399} for details).}
    \label{fig:ngc1380}
\end{figure}

NGC1380 is an unbarred, lenticular galaxy whose GCS, investigated both from the ground~\citep{kisslerpatig1997}
and using HST data~\citep{chiessantos2007}, features two clearly distinct color classes: the 
blue GCs, which are spherically distributed and mostly located in the outskirts of the galaxy, and the 
red GCs that follow more closely the geometry of the photometric components of the disk and bulge.

Three major overdensities are observed in the residual map generated by all ACSFCS GCs in 
NGC1380~(Figure~\ref{fig:ngc1380}), along the D$_{25}$ elliptical isophote:
C1 is located in the E side of the galaxy, C2 and C4 are located in the W side. While C1 is only
barely visible in the blue residual map, C2 and C4 are clearly detected in both the 
blue and red GCs, although in the red only one overdensity region engulfing both structure is observed. 
The blue residual map suggests that the morphology of C2 and C4 is elongated and aligned along radial 
directions towards the center of the galaxies. C3 is a morphologically complex structure with multiple
peak densities and located at a small galactocentric distance NE of the galaxy center; C3 is only
visible in the red GCs. Two additional small, low significance structures (c1 and c2) are detected
W of the center and could represent the low radius extremities of C2 and C4 that are located
at larger galactocentric distances (the continuity between the structures being more evident in the blue
residual map). The observation of large overdensities structures at relatively large
distances from the center of the galaxy, especially in the spatial distribution of blue GCs, confirms results 
in the literature~\citep{kisslerpatig1997,chiessantos2007} that found that blue GCs dominate the 
NGC1380 GCS at large galactocentric distances. A relatively large region of enhanced positive
residual not detected in the other maps is observed N and NW of the NGC1380 center in the distribution 
of red GCs.

\subsection{NGC1404}
\label{subsec:ngc1404}

\begin{figure}[ht]
    \centering
	\includegraphics[width=0.45\linewidth]{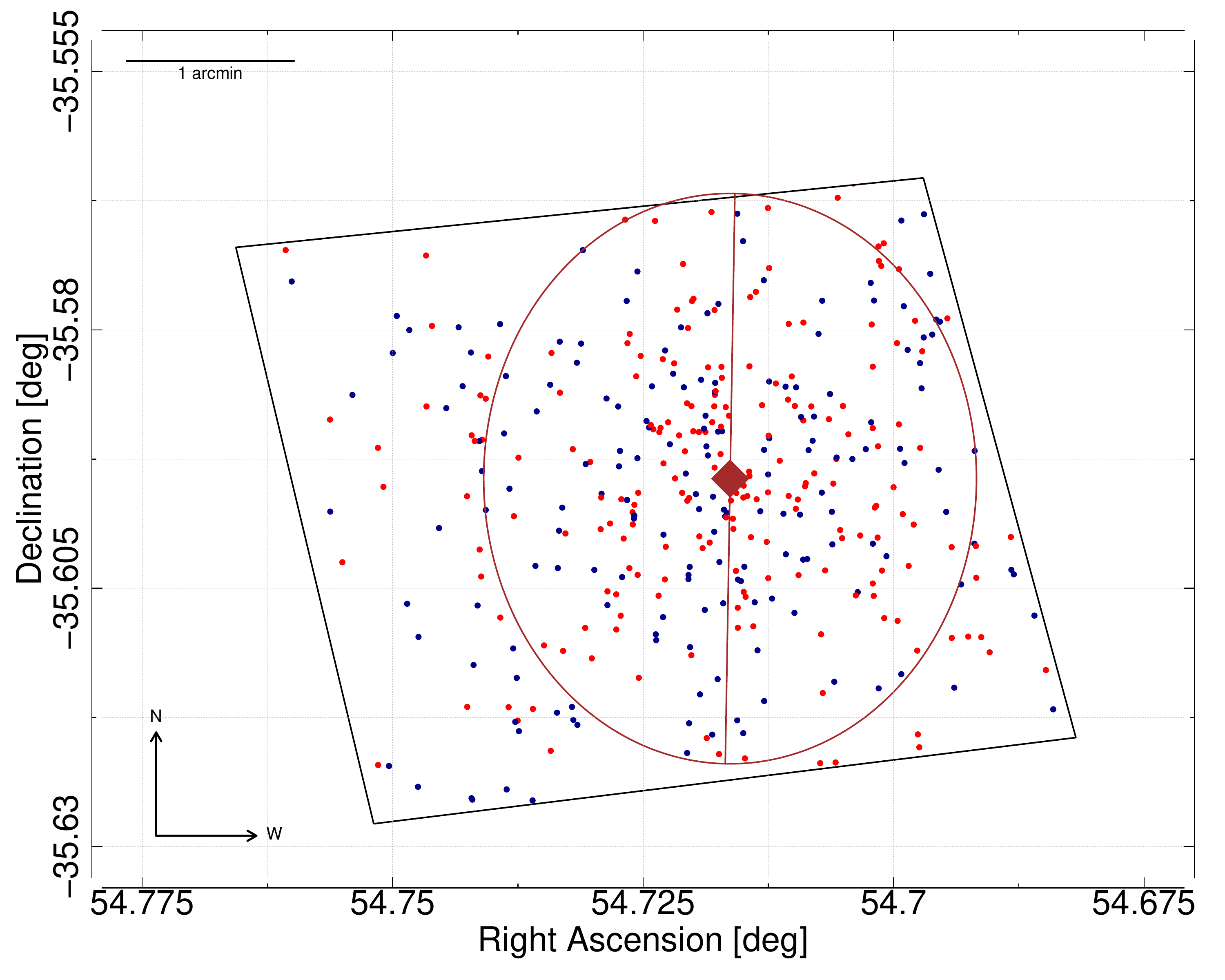}
	\includegraphics[width=0.45\linewidth]{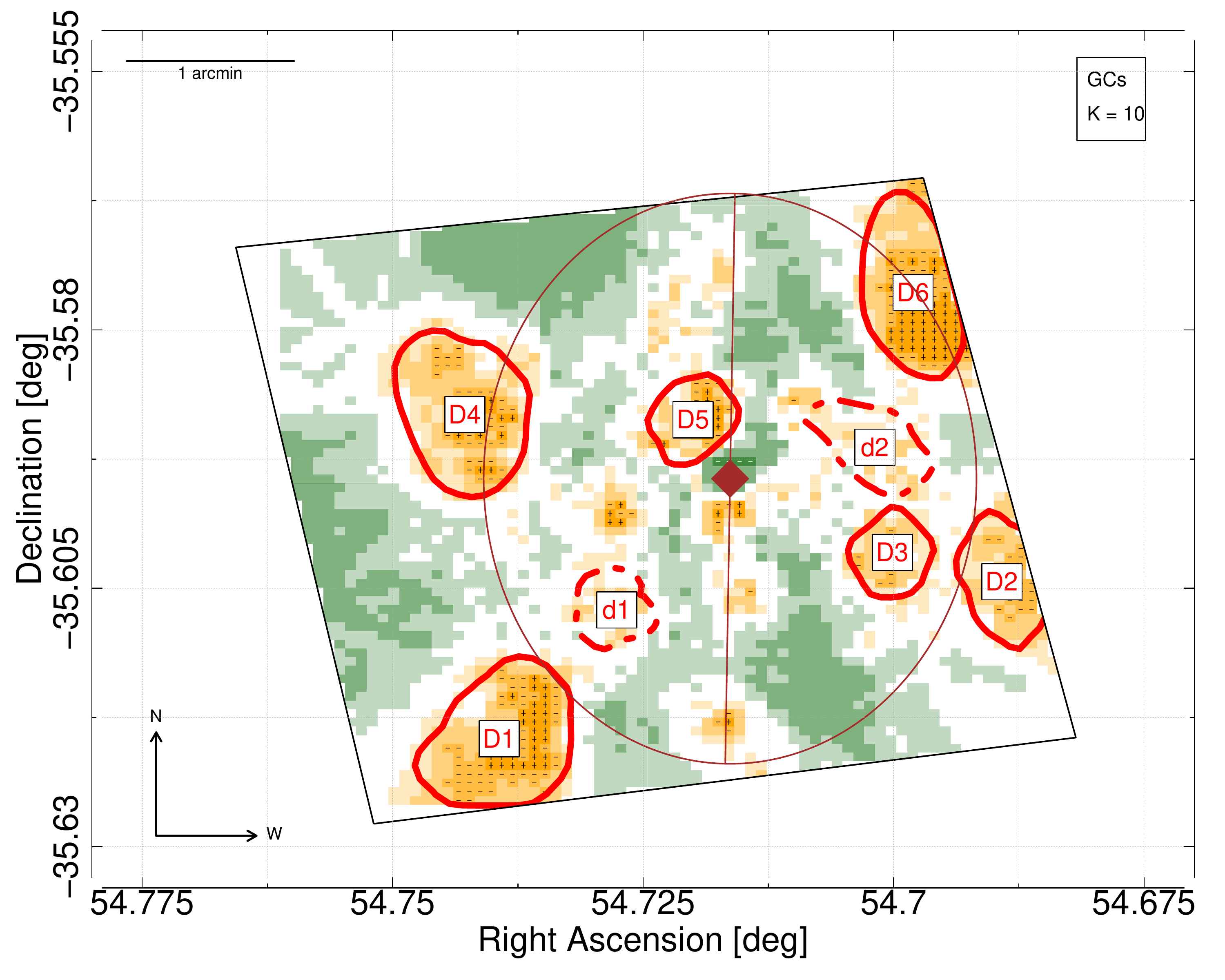}\\
	\includegraphics[width=0.45\linewidth]{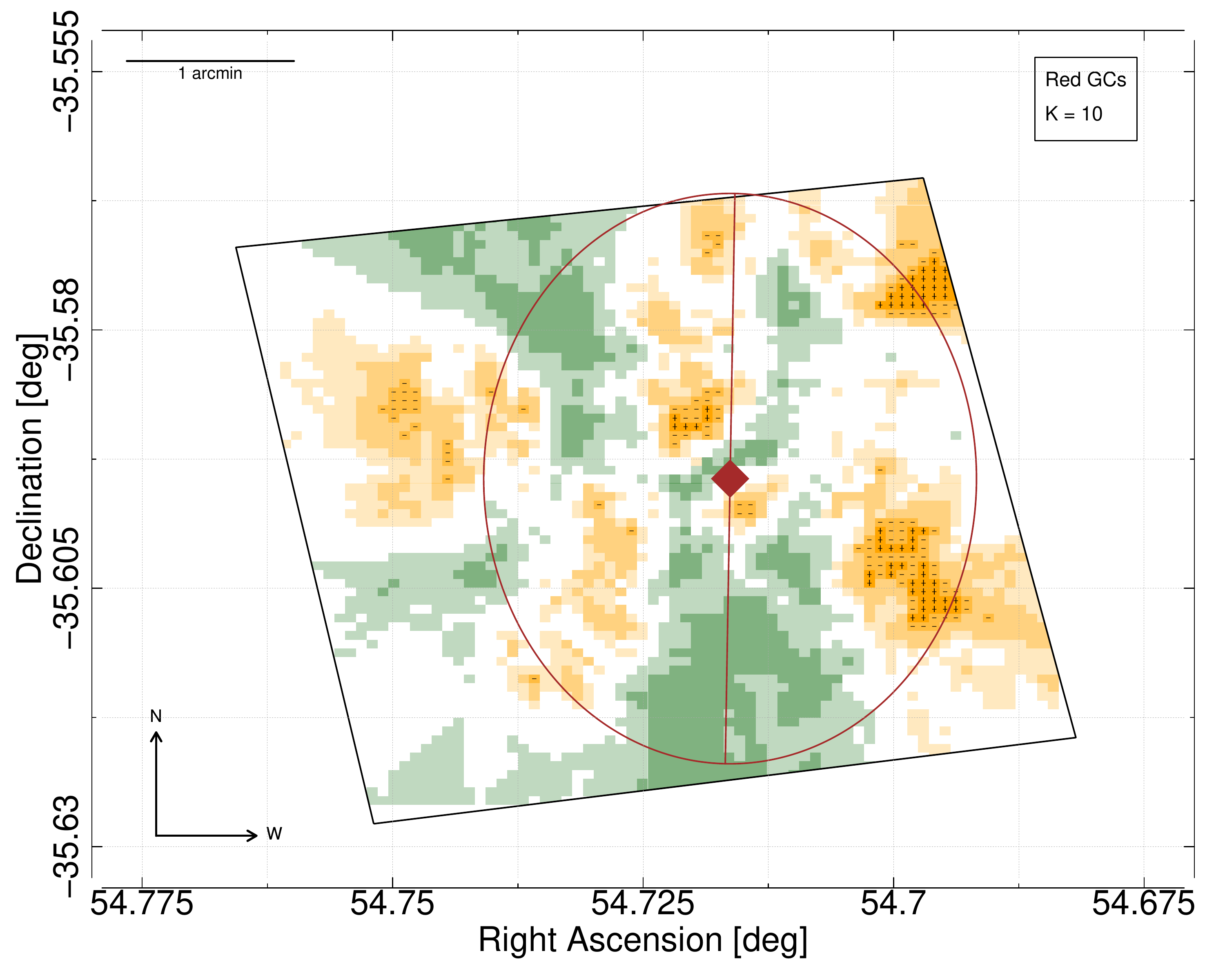}
	\includegraphics[width=0.45\linewidth]{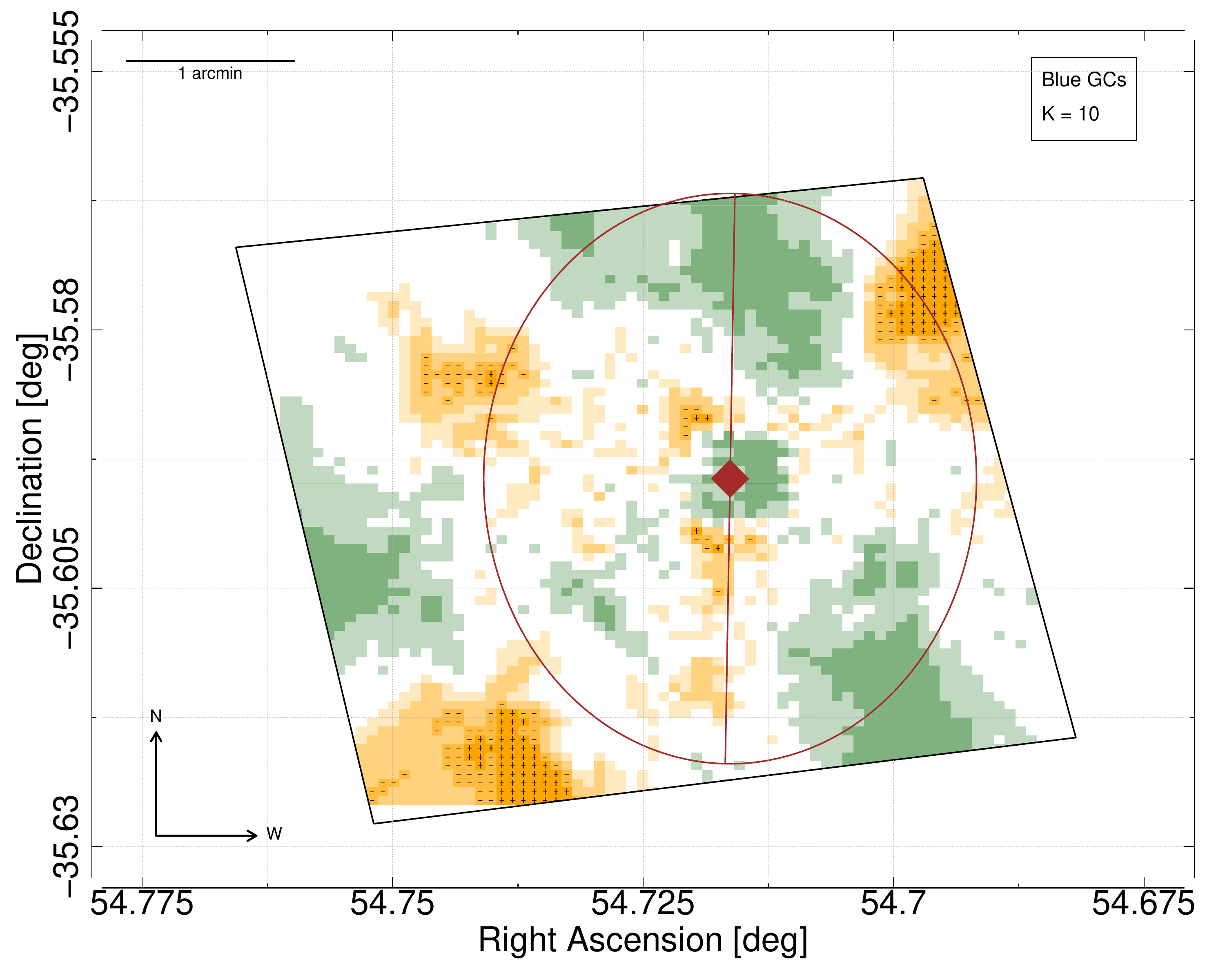}	
    \caption{Spatial distribution of GCs in NGC1404 (see caption of Figure~\ref{fig:ngc1399} for details).}
    \label{fig:ngc1404}
\end{figure}

The GC system of NGC1404, an E2 elliptical galaxy which is the closest neighbor of NGC1399 and 
is located only 9\arcmin~from its very extended GC system, has been investigated in the past using 
observations from the ground~\citep{richtler1992} that have highlighted a relatively compact and 
spherical geometry. More recently,~\cite{bekki2003} used simulations of the NGC1404 GCS 
including the Fornax cluster tidal field to find that the low NGC1404 specific frequency can be 
explained by tidal stripping of NGC1404 GCs caused by NGC1399. They observed that the stripped GCs 
have likely contributed to the 
formation of tidal streams in the cluster core ICGC population whose geometric, metallicity and 
kinematical properties depend on the orbit of their former host galaxy relative to the cluster center 
and to the past global properties of NGC1404 GCS.

Figure~\ref{fig:ngc1404} shows the residual maps of the spatial distributions of ACSFCS GCs
in NGC1404. The general spatial distribution~(top right panel) features 
several large (D1, D2, D4, D6) spatial structures located in proximity of the D$_{25}$ elliptical 
isophote of the host galaxy. D1 and D6, the two 
largest and most significant overdensities, are located in the SE and NW corners of the imaged field respectively, 
while D2 and D4 are approximately located just outside of the D$_{25}$ along a NE to SW direction crossing 
the galaxy center. D6 is located along the direction connecting the center of NGC1404 and NGC1399 and overlaps 
the ``bridge'' observed in the density map of ICGCs in the core of the Fornax cluster by~\cite{dabrusco2016}. D6 is 
also the only structure to be detected with similar size, shape and significance in both the red
and blue GCs residual maps: D2 and D4 are only visible in the red map while D1 is detected only in 
blue residuals. Two additional small but statistically significant, roughly circular structures (D3 and D5) are 
detected at smaller galactocentric distances along the major 
axis of the galaxy N and SW of the galaxy center, respectively. Both can be associated with 
similar structures only in the map of red GCs, although in the blue residuals a moderate circular density peak 
corresponding to the S end of D5 is visible. The two additional, less significant structures d1 and d2 detected 
in the general residual map can only be also observed in the red residual map.

\subsection{NGC1427}
\label{subsec:ngc1427}

\begin{figure}[ht]
    \centering
	\includegraphics[width=0.45\linewidth]{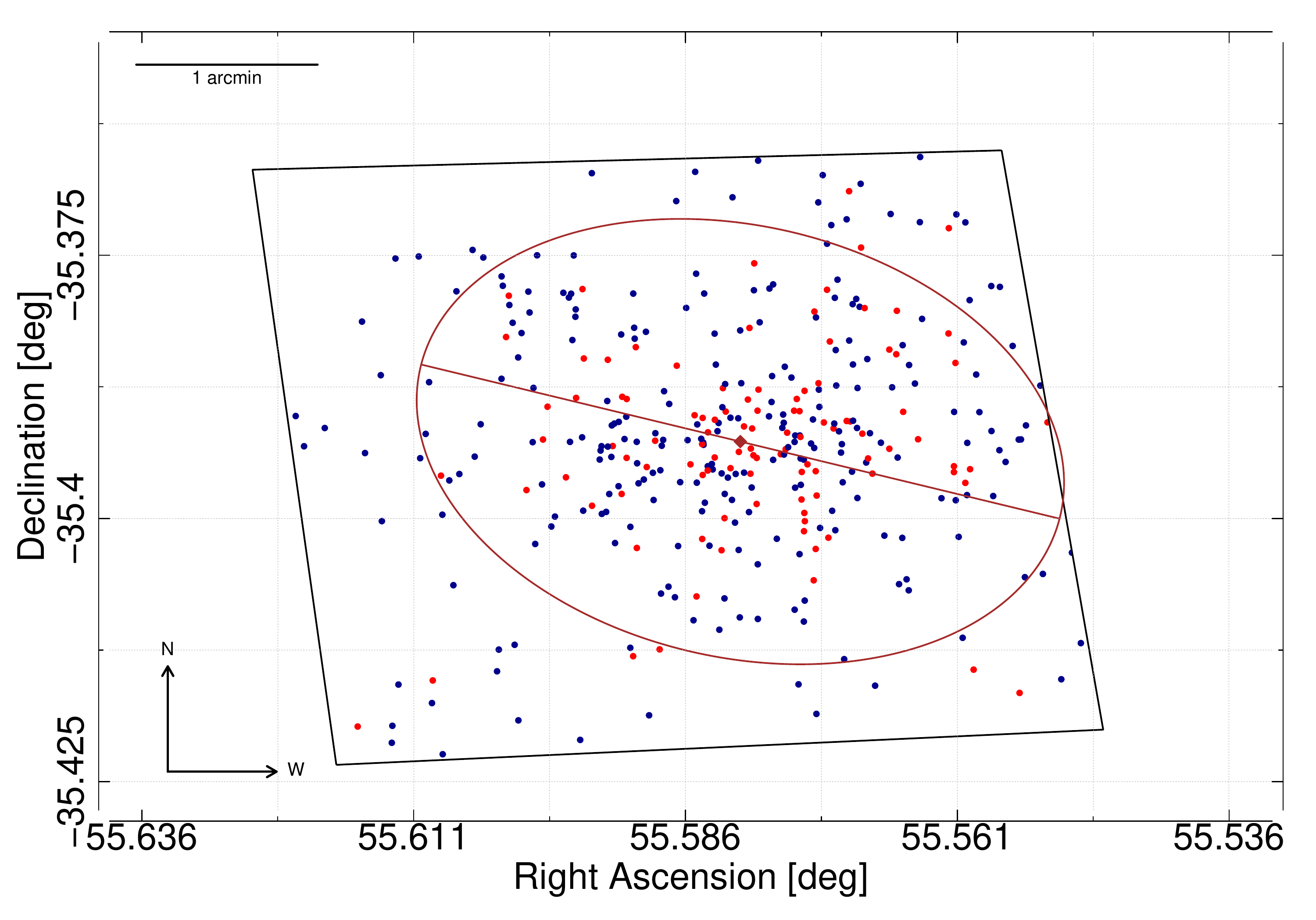}
	\includegraphics[width=0.45\linewidth]{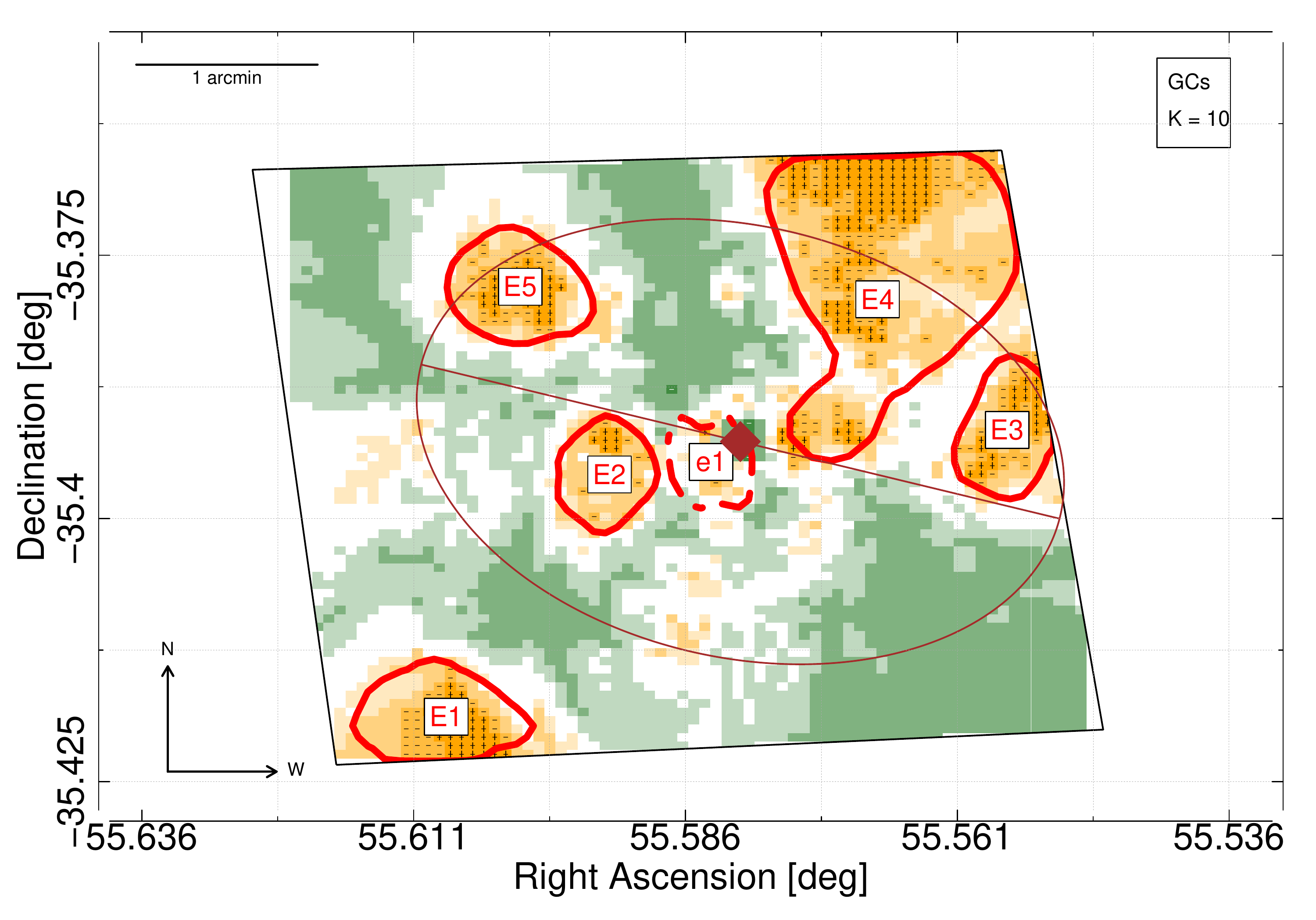}\\
	\includegraphics[width=0.45\linewidth]{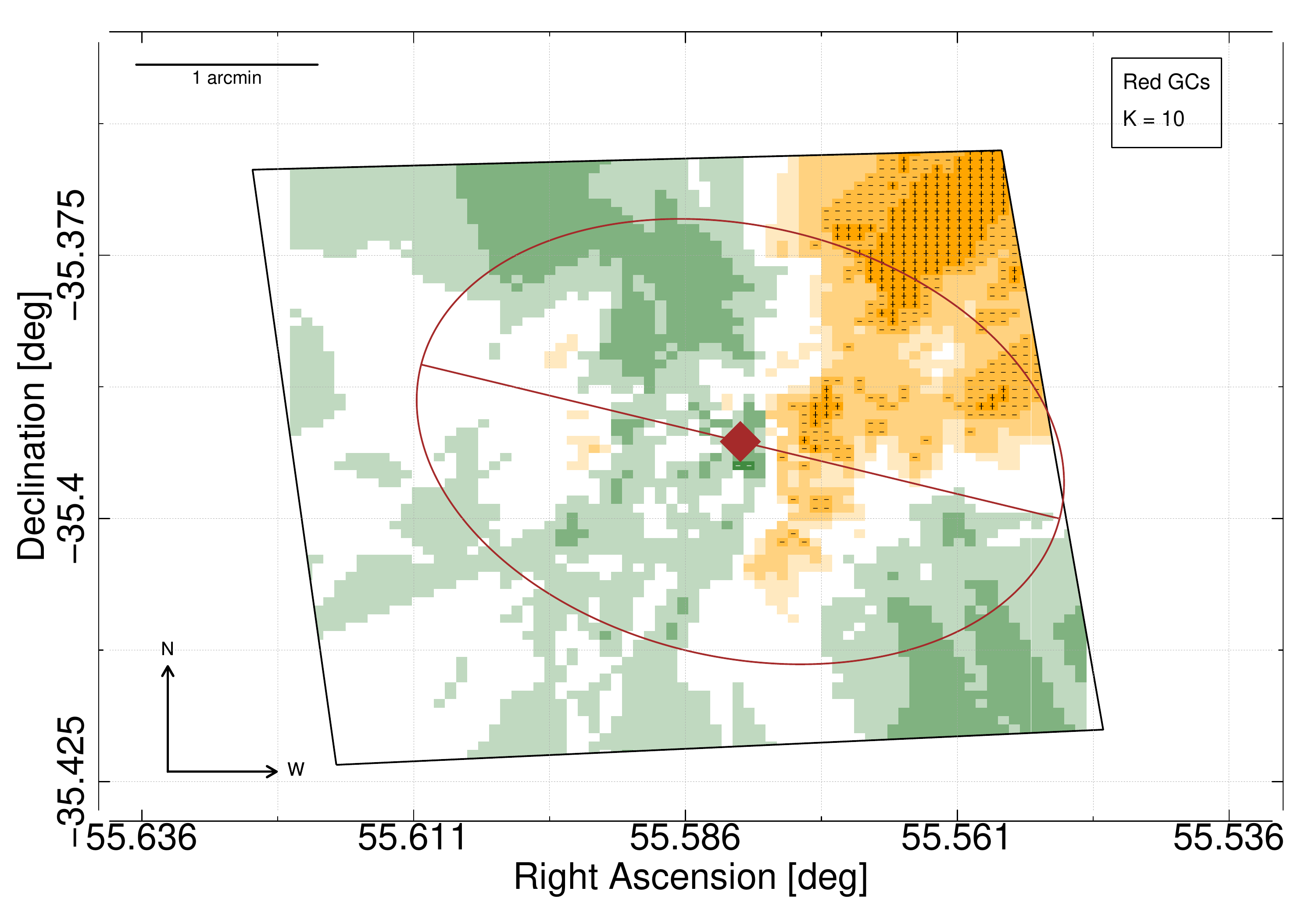}
	\includegraphics[width=0.45\linewidth]{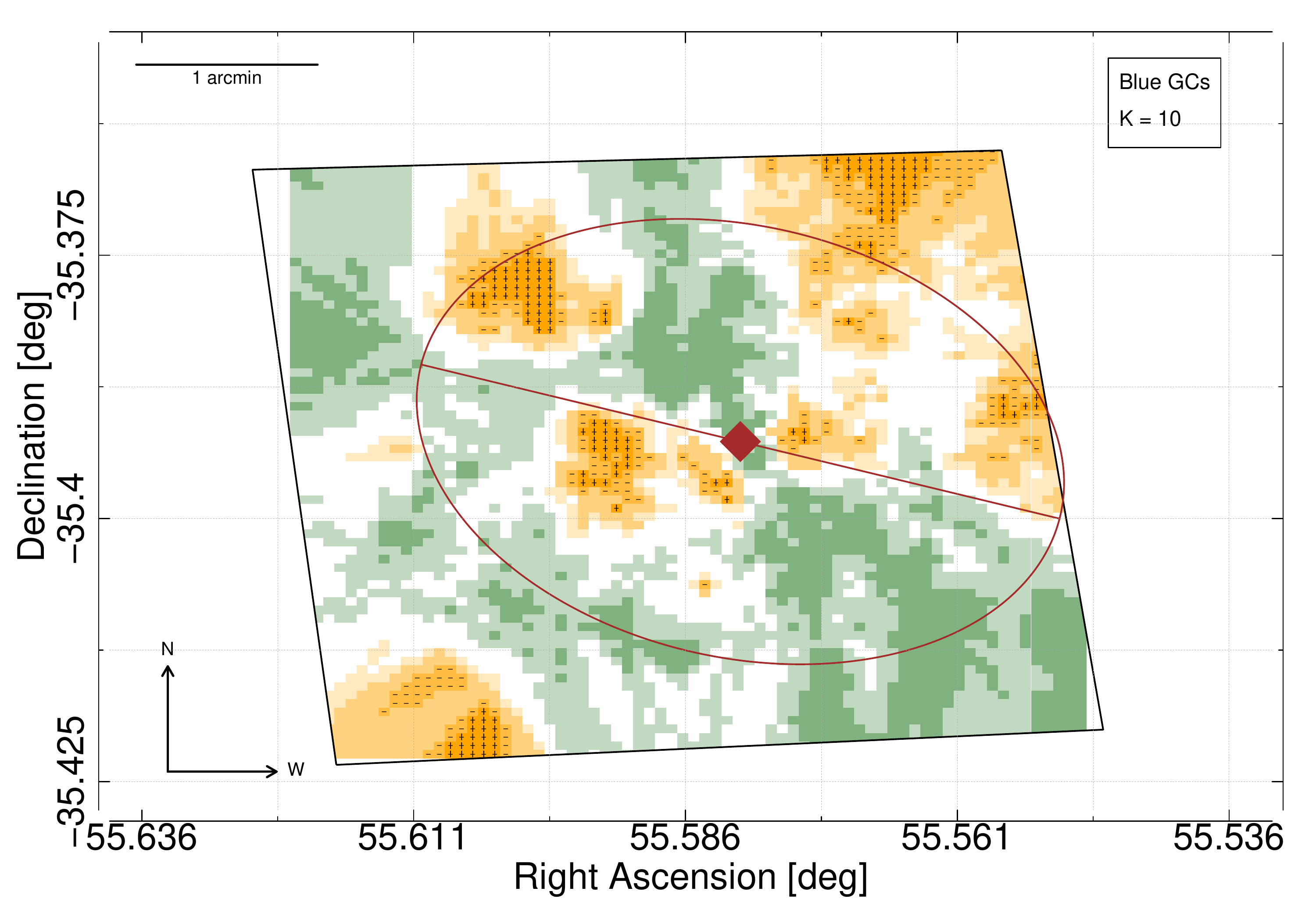}	
    \caption{Spatial distribution of GCs in NGC1427 (see caption of Figure~\ref{fig:ngc1399} for details)}
    \label{fig:ngc1427}
\end{figure}

The~\cite{forte2001} study of the GC system of NGC1427 shown that this low-luminosity elliptical
presents a strongly centrally concentrated red sub-population and a blue component with a much shallower 
radial profile. The spatial distribution of ACSFCS GCs in NGC1427~(Figure~\ref{fig:ngc1427}) is dominated 
by the very large and statistically significant structure E4 that extends radially and displays a complex
morphology with at least two main overdensity regions: the first one, roughly circular, is located W of the 
center of the galaxy, while the second, much larger region lies across the 
D$_{25}$ isophote in the NW quadrant of the galaxy. E4 morphological complexity and extension suggest that it 
might be the result of the superposition of two distinct overdensities. E4 is visible in both the red and blue 
residual maps, although in the blue GCs the two density peaks are clearly separated by a large gap unlike in the
red residuals they are connected. 
Four additional large spatial structures are observed. The roughly circular E3 and E5 are located
respectively close to the N and S intersections of the major axis of the host galaxy with the D$_{25}$ isophote; 
both are also clearly detected in the blue residual maps. E1 is located in the SE corner of imaged field and 
can be clearly observed also in the blue map, while E2 sits E of the galaxy center; both E1 and E2 are only 
detected in the blue residual map. An intermediate structure (e1) is close to the center of the galaxy and 
E2. Since e1 is also only visible in the blue residual map, it may be considered a section of the nearby large 
E2 structure. 

\subsection{NGC1344}
\label{subsec:ngc1344}

\begin{figure}[ht]
    \centering
	\includegraphics[width=0.45\linewidth]{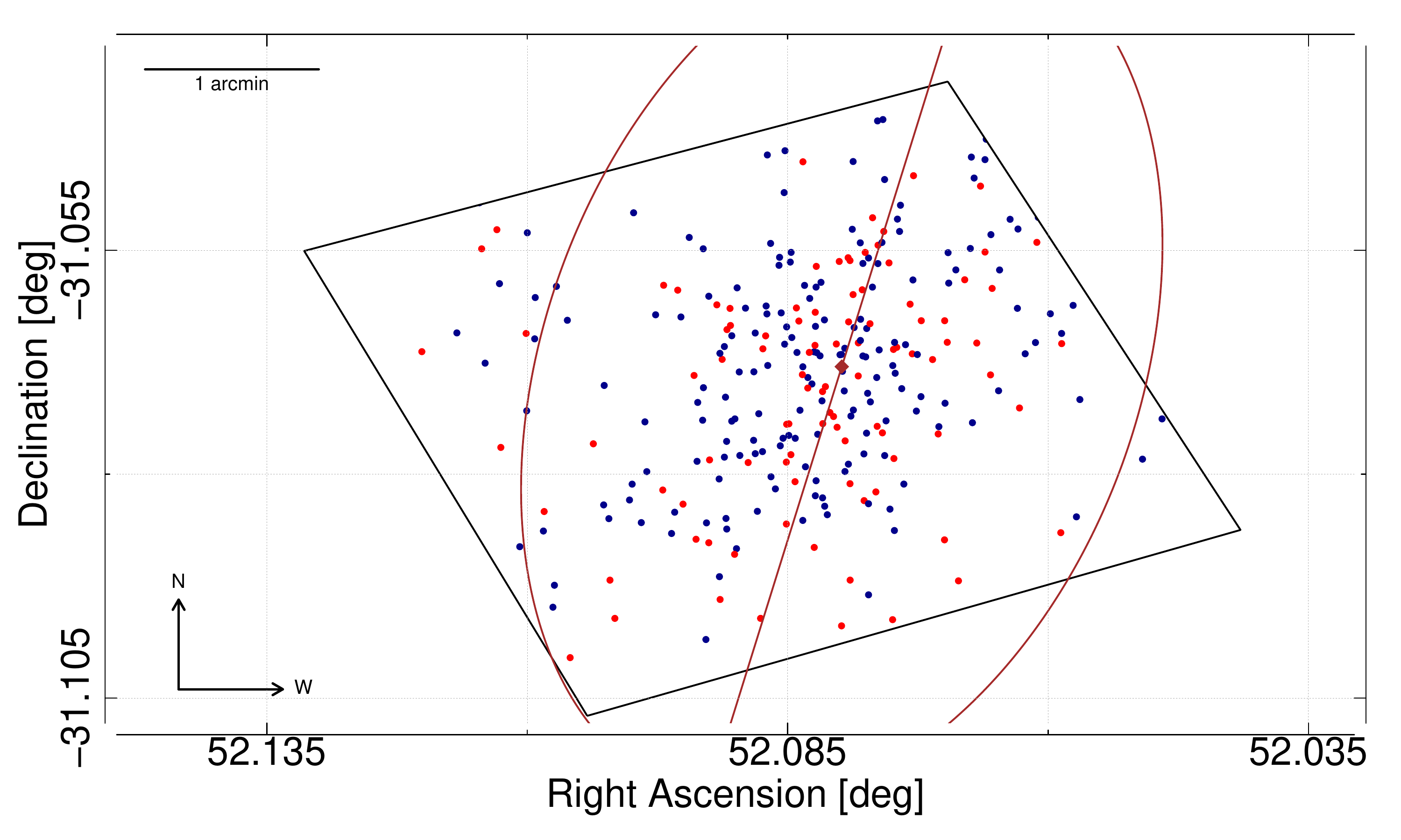}
	\includegraphics[width=0.45\linewidth]{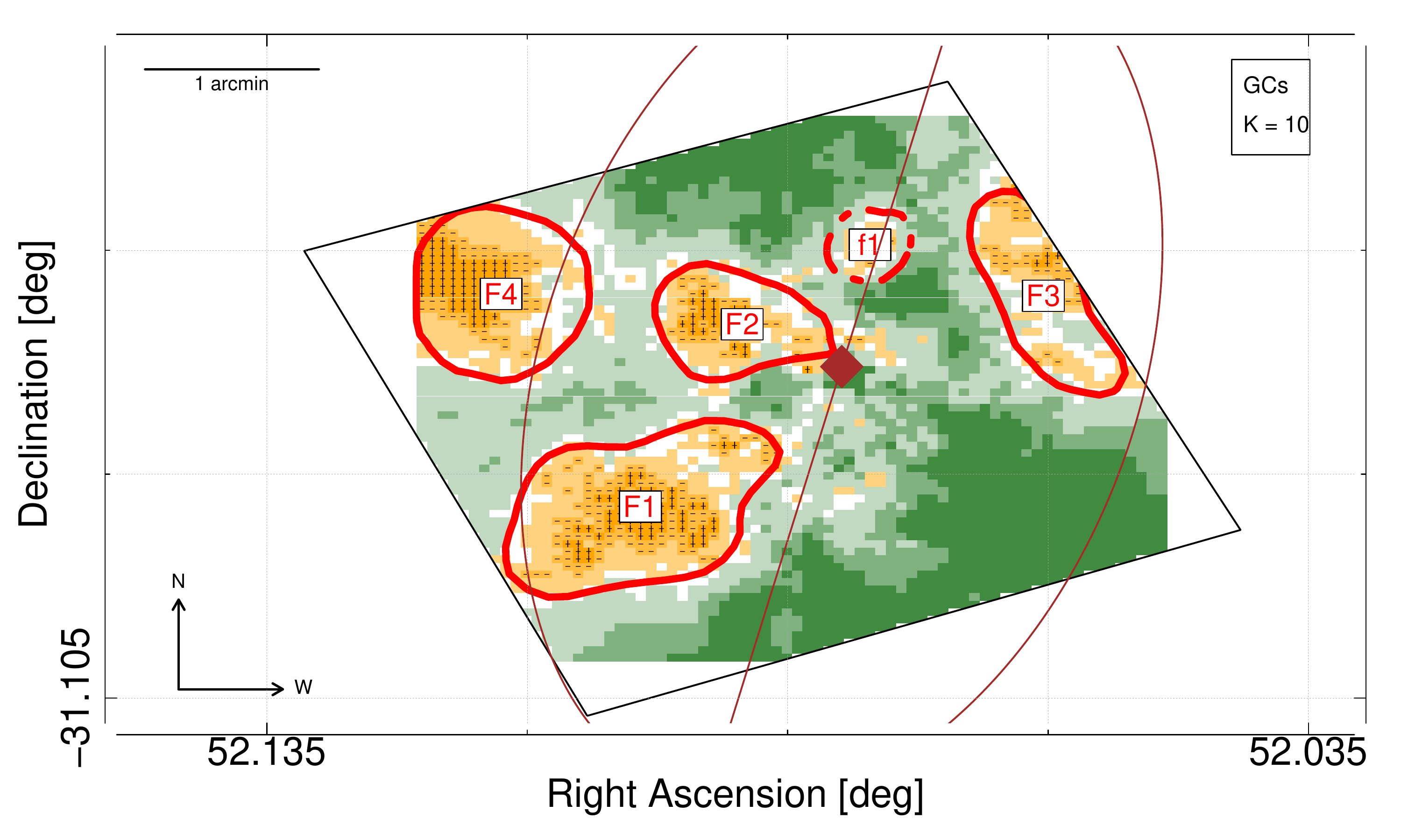}\\
	\includegraphics[width=0.45\linewidth]{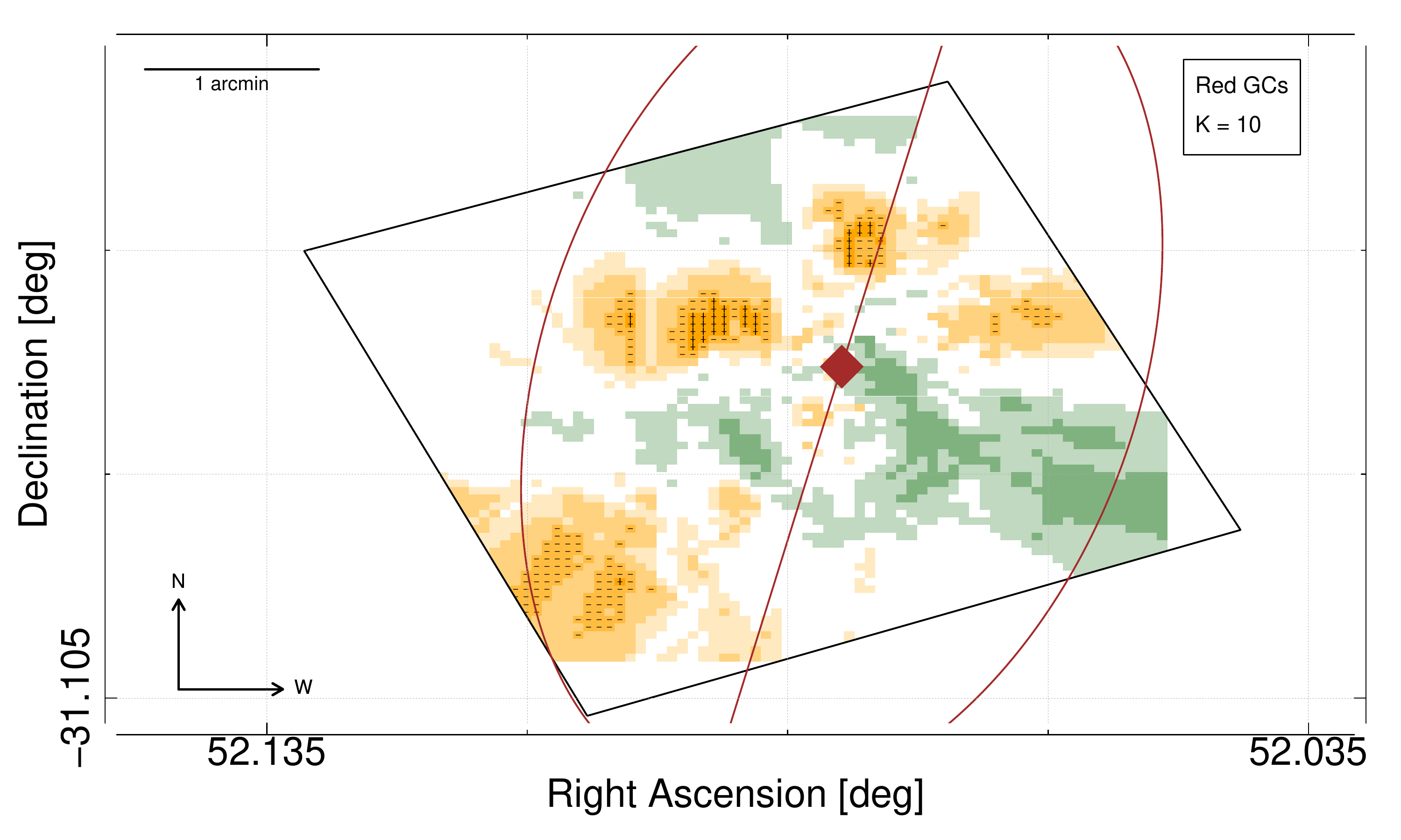}
	\includegraphics[width=0.45\linewidth]{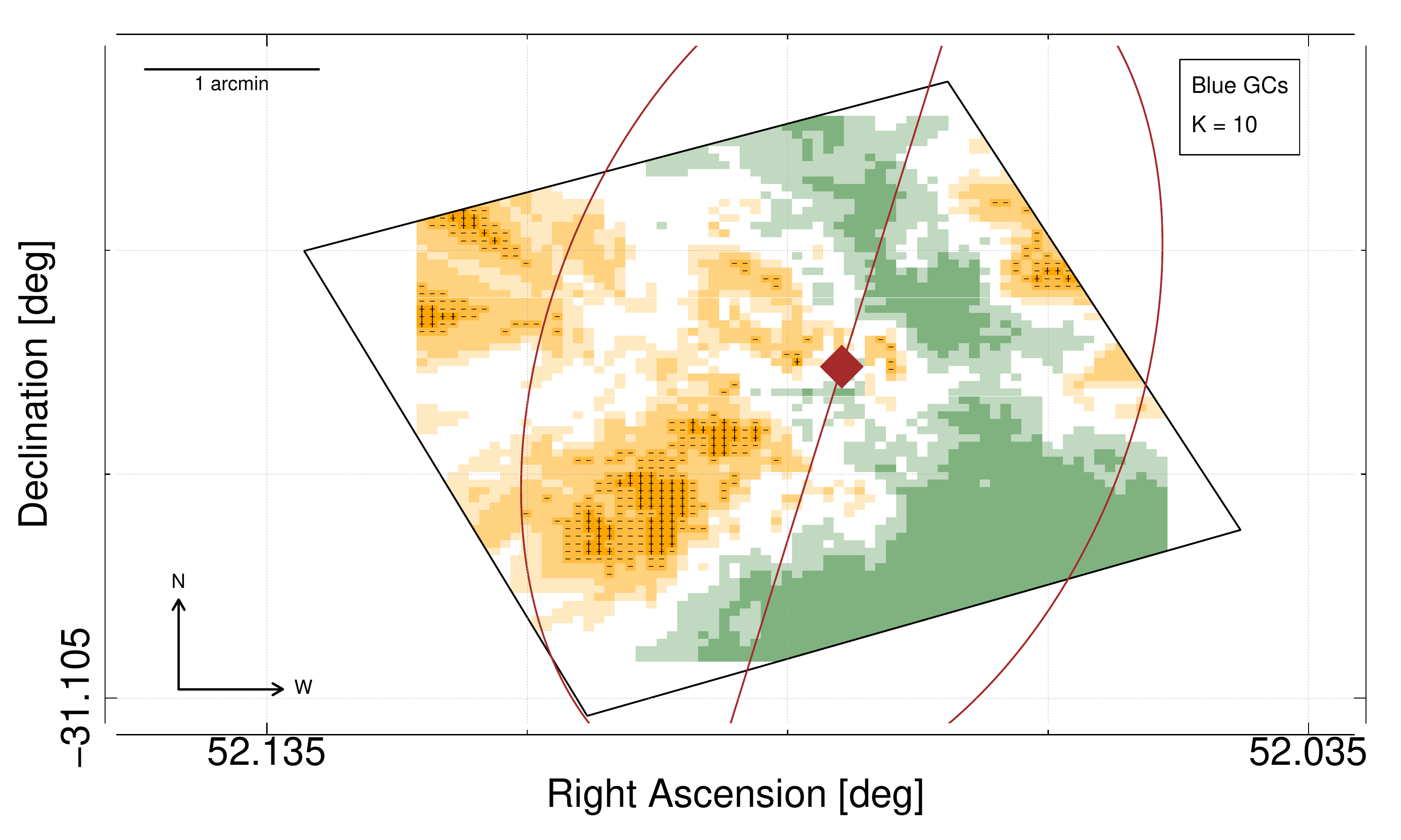}	
    \caption{Spatial distribution of GCs in NGC1344 (see caption of Figure~\ref{fig:ngc1399} for details).}
    \label{fig:ngc1344}
\end{figure}

NGC1344, located in the N, very low-density outskirt of the Fornax cluster and sometimes classified as a
S0 galaxy, is known for low surface brightness features resembling thin shells~\citep{malin1980}, 
that are considered the remnants of relatively recent galactic mergers.~\cite{sikkema2006}
studied, among other shell galaxies, the NGC1344 GCS using large field-of-view, ground based images 
and found that the overall NGC1344 GC population follows closely the photometric geometry of the 
host, but a significant fraction of the red GC subpopulation is also present at large galactocentric 
radii where, typically, blue GCs are more likely observed. 

The residual map generated by all GCs observed in NGC1344~(Figure~\ref{fig:ngc1344}) features an excess of 
GCs in the E side of the galaxy relative to W side~(upper left panel); this asymmetry is also evident 
in the residual maps for all GCs, where three large structures (F1, F2 and F4) are detected in the E half 
of the galaxy, while only one structure (F3) is located in the W section. F1 and F2, both visible in the 
red and blue GC residual maps, have elongated shapes and are located along radial directions spanning a 
large range of galactocentric distances. The 
roughly circular, large structure F4 lies outside of the D$_{25}$ isophote and is not detected in the 
red map. F3, which displays at least two separate density peaks, straddles the border of the 
observed field and extends radially in the NE to SW direction; it is clearly detected in the blue residual
map. The intermediate, small structure f1 sits on the N section of the major axis and is 
only detected in the red residual map although with a much larger statistical significance that in general GC
distribution. 

\subsection{NGC1387}
\label{subsec:ngc1387}

\begin{figure}[ht]
    \centering
 	\includegraphics[width=0.45\linewidth]{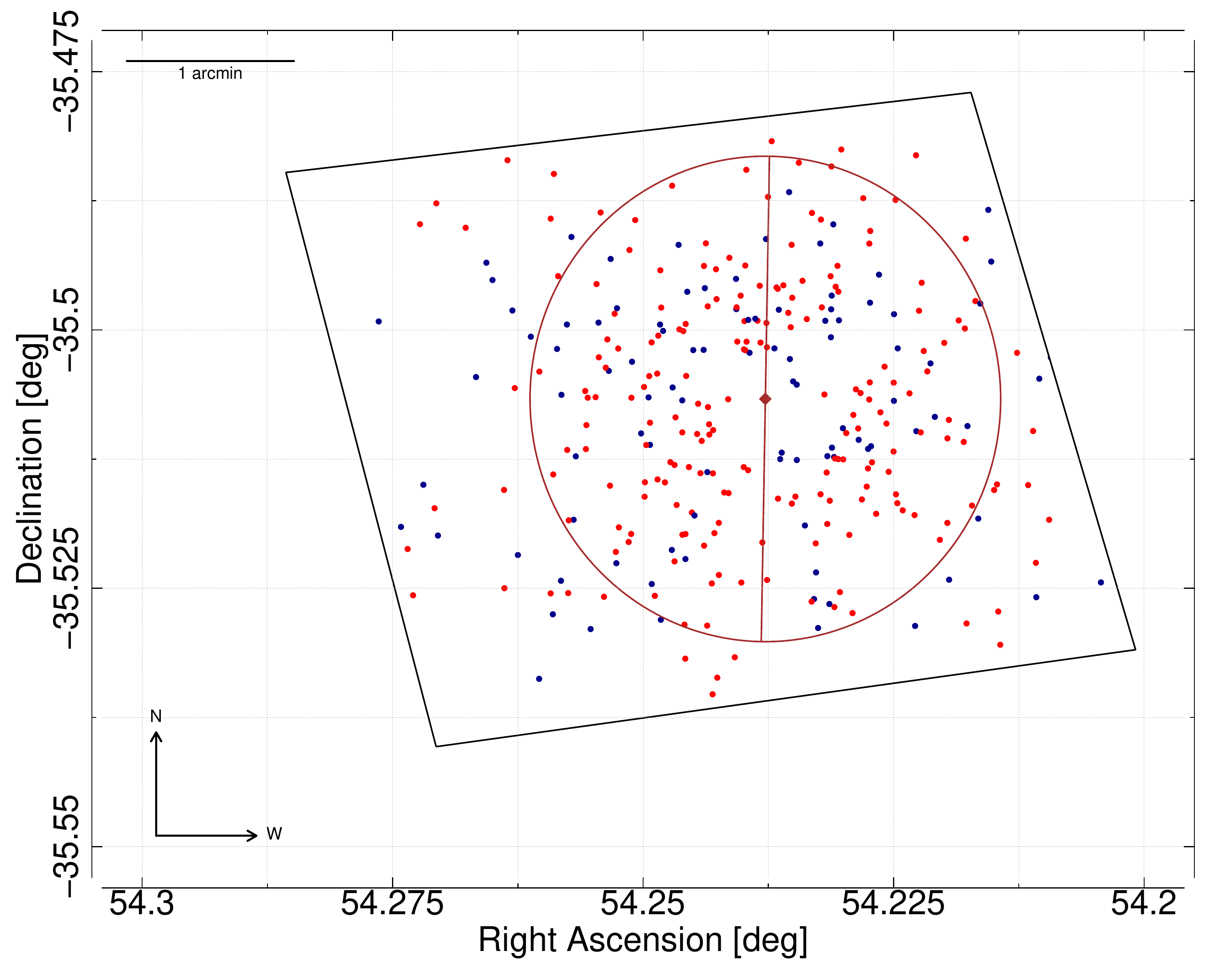}
 	\includegraphics[width=0.45\linewidth]{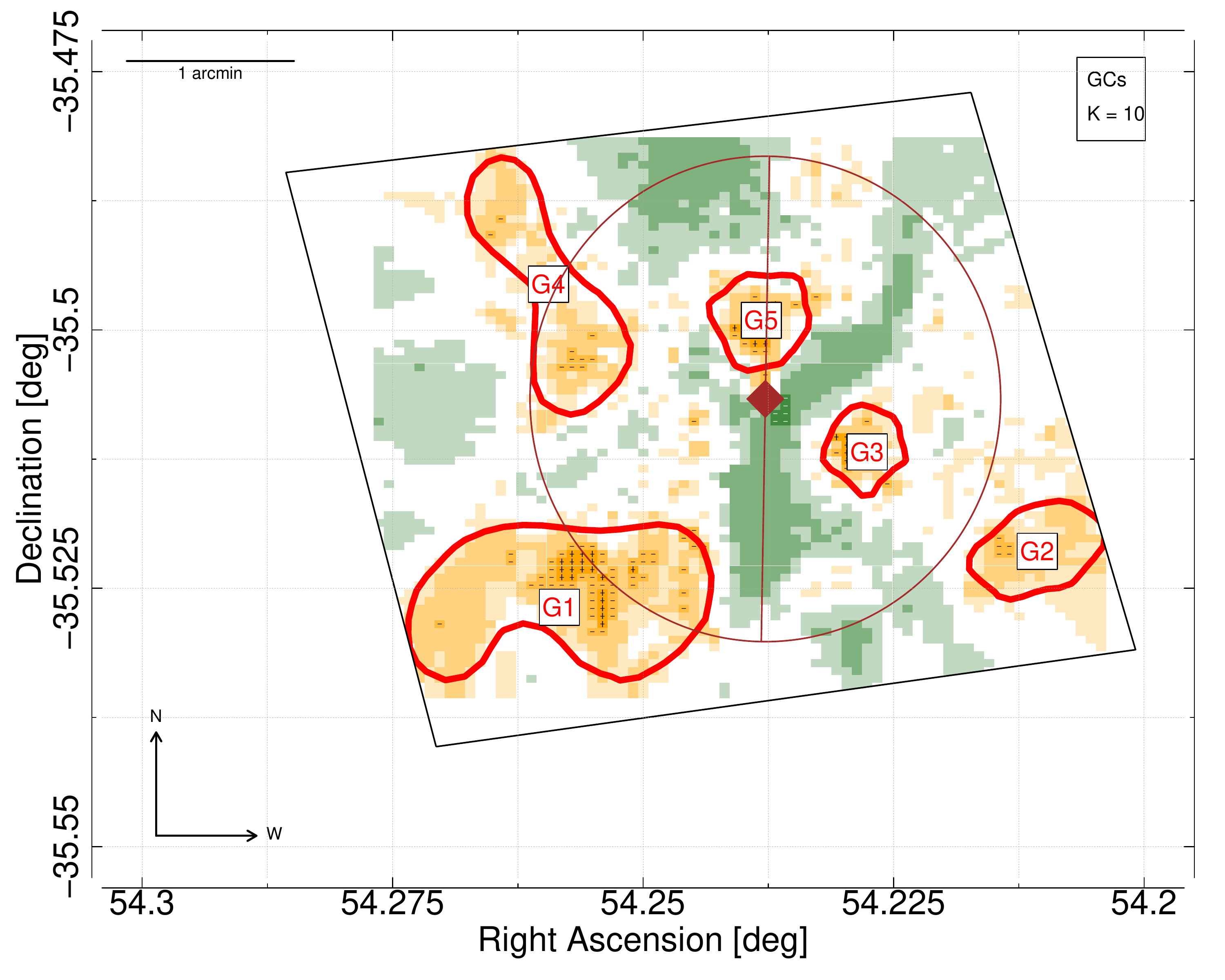}\\
 	\includegraphics[width=0.45\linewidth]{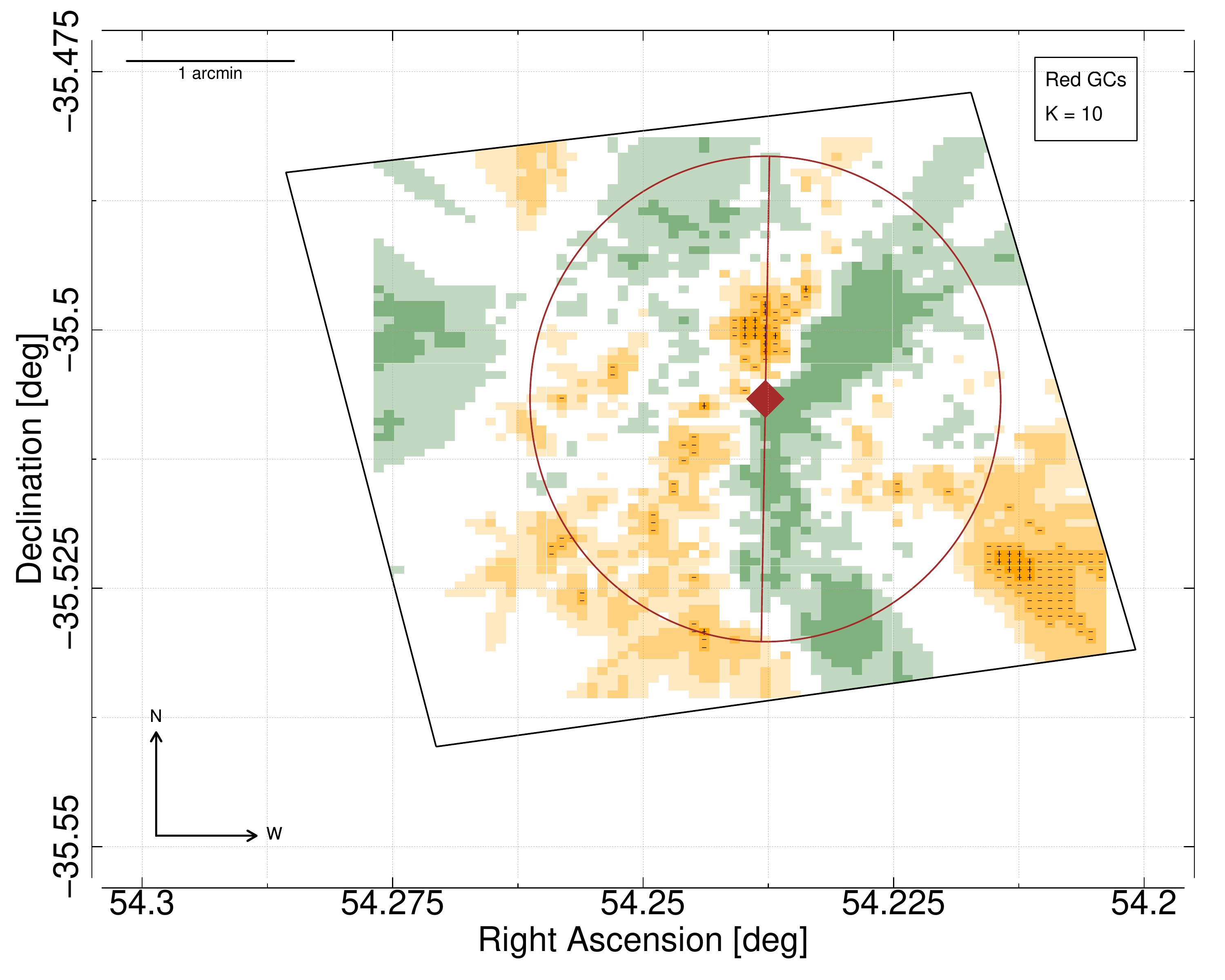}
	\includegraphics[width=0.45\linewidth]{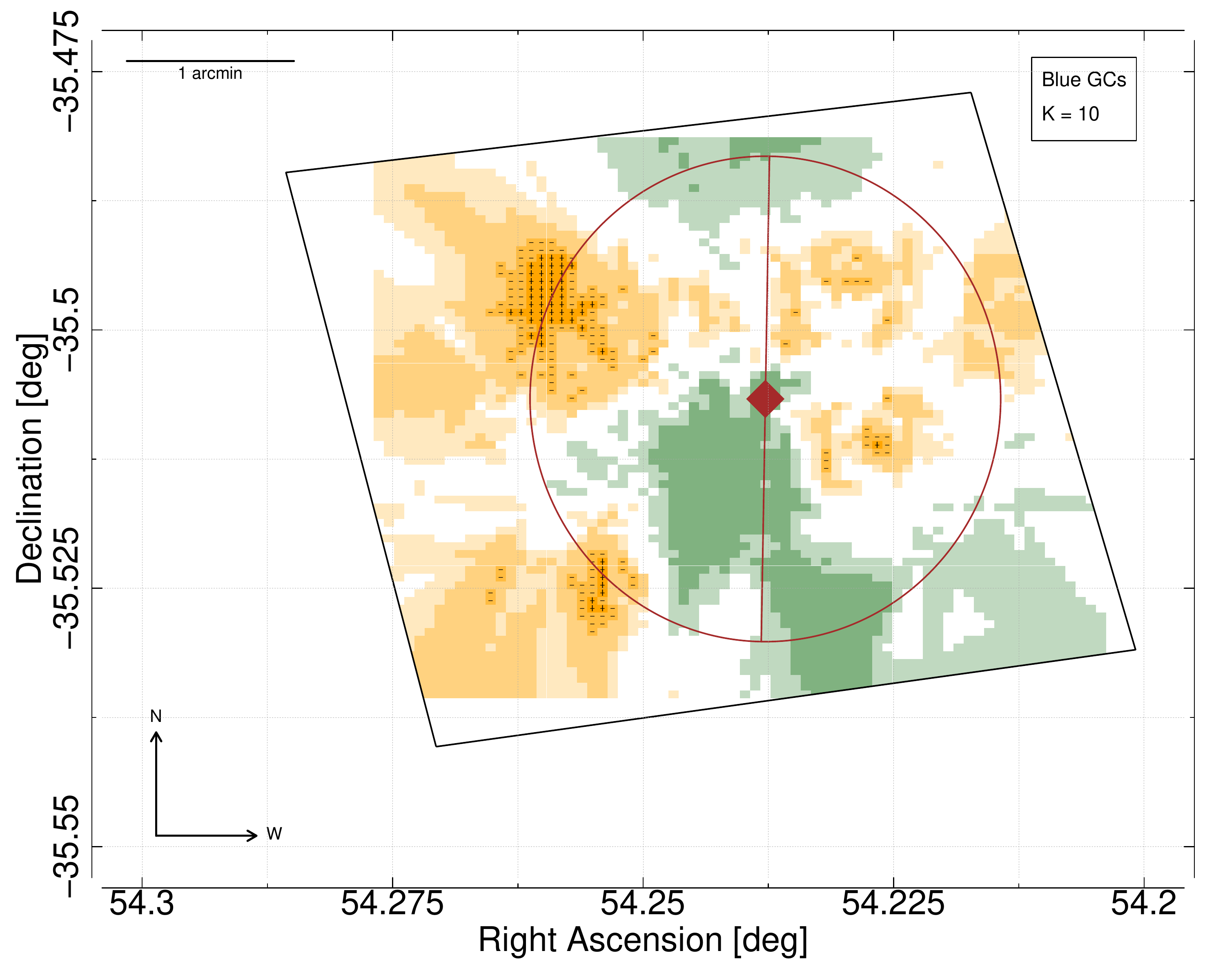}
    \caption{Spatial distribution of GCs in NGC1387~(see caption of Figure~\ref{fig:ngc1399} for details).}
    \label{fig:ngc1387}
\end{figure}

The GCS of NGC1387 is characterized by an almost spherical spatial distribution 
and the lack of significant azimuthal anisotropy~\citep{bassino2006}; the GC color subclasses display 
different radial profiles, with red GCs more centrally concentrated and of a shallower but more radially
more extended blue GCs profile.~\cite{bassino2006} suggested that the 
current properties of the NGC1387 GCS have been determined by the hierarchical merging evolution
history of the host~\citep{beasley2002} and the tidal interaction with the nearby 
giant elliptical NGC1399 which has stripped NGC1387 of the outermost GCs. 

Figure~\ref{fig:ngc1387} shows the residual maps based on the spatial distribution of the ACSFCS GCs 
observed in NGC1387. In the upper right panel, the map for all GCs features three extended structures 
straddling or located just outside the D$_{25}$ isophote of the host galaxy (G1, G2 and G4). 
G1 and G4 have complex, elongated morphology suggestive 
of the presence of substructures, partially confirmed by the different shapes of these structures in the residual 
maps for different color GCs classes, respectively. In the red map, G4 is very weak while G1 is associated with an 
overdensity region with lower statistical significance and smaller size, that is mostly located along the D$_{25}$ 
isophote; both structures are clearly visible in the residual map generated by blue GCs instead. G2 is only detected 
in the red residual map. Two additional high significance, roughly circular overdensities at small galactocentric 
distances can be found N (G5) and E (G3) of the center of the galaxy, with similar properties in the residual maps of both
red and blue GCs. Structures G1 and G4, clearly visible in the residual map of blue ACSFCS GCs, 
are located along the direction connecting NGC1387 to the center of NGC1399 and overlap the well known 
blue GCs ``bridge'' in the spatial distribution of Fornax cluster core ICGCs reported 
by~\cite{dabrusco2016,kim2013,bassino2006} (see Section~\ref{subsec:spatialstructures_position}).

\subsection{NGC1374}
\label{subsec:ngc1374}

\begin{figure}[ht]
    \centering
	\includegraphics[width=0.45\linewidth]{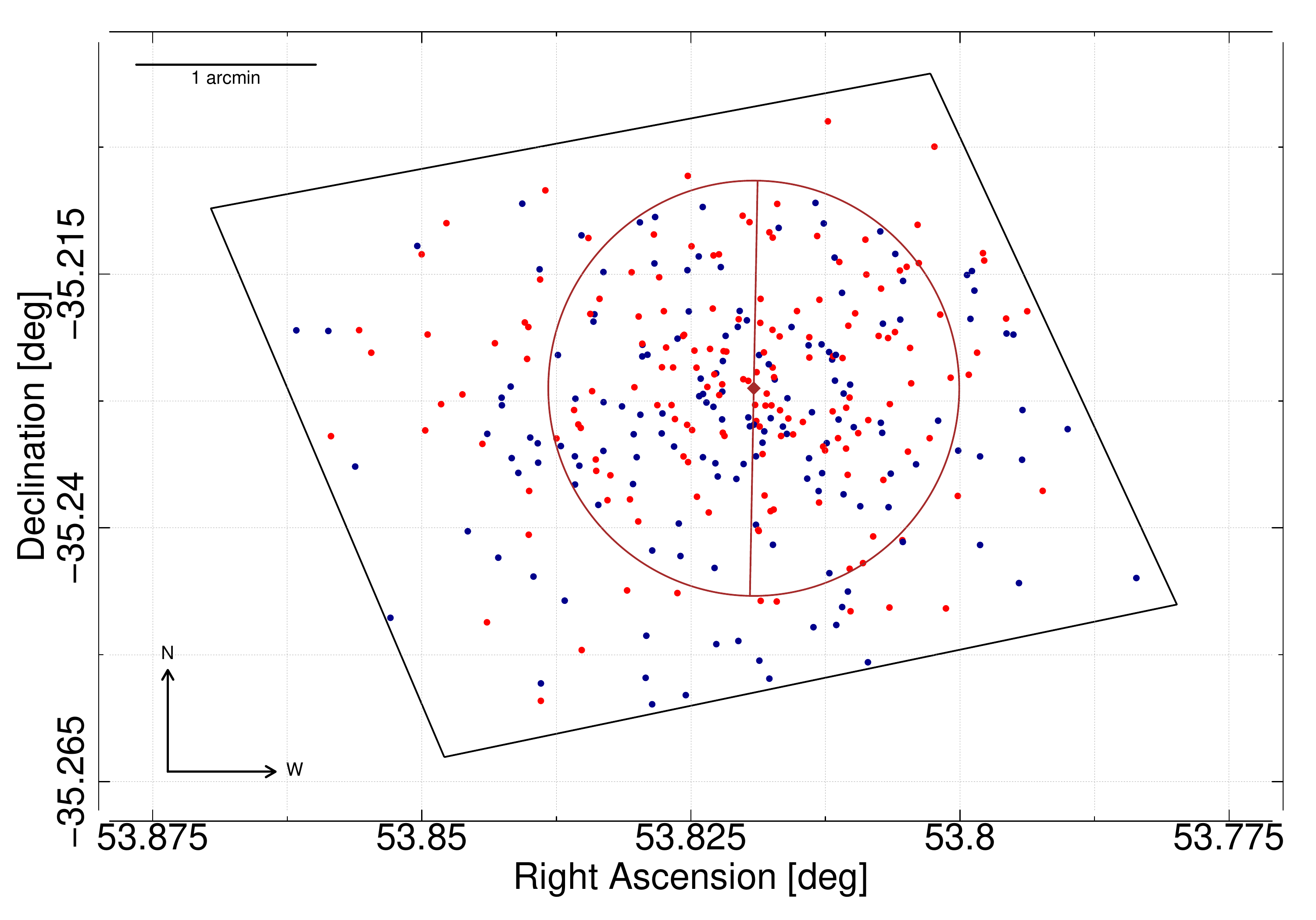}
	\includegraphics[width=0.45\linewidth]{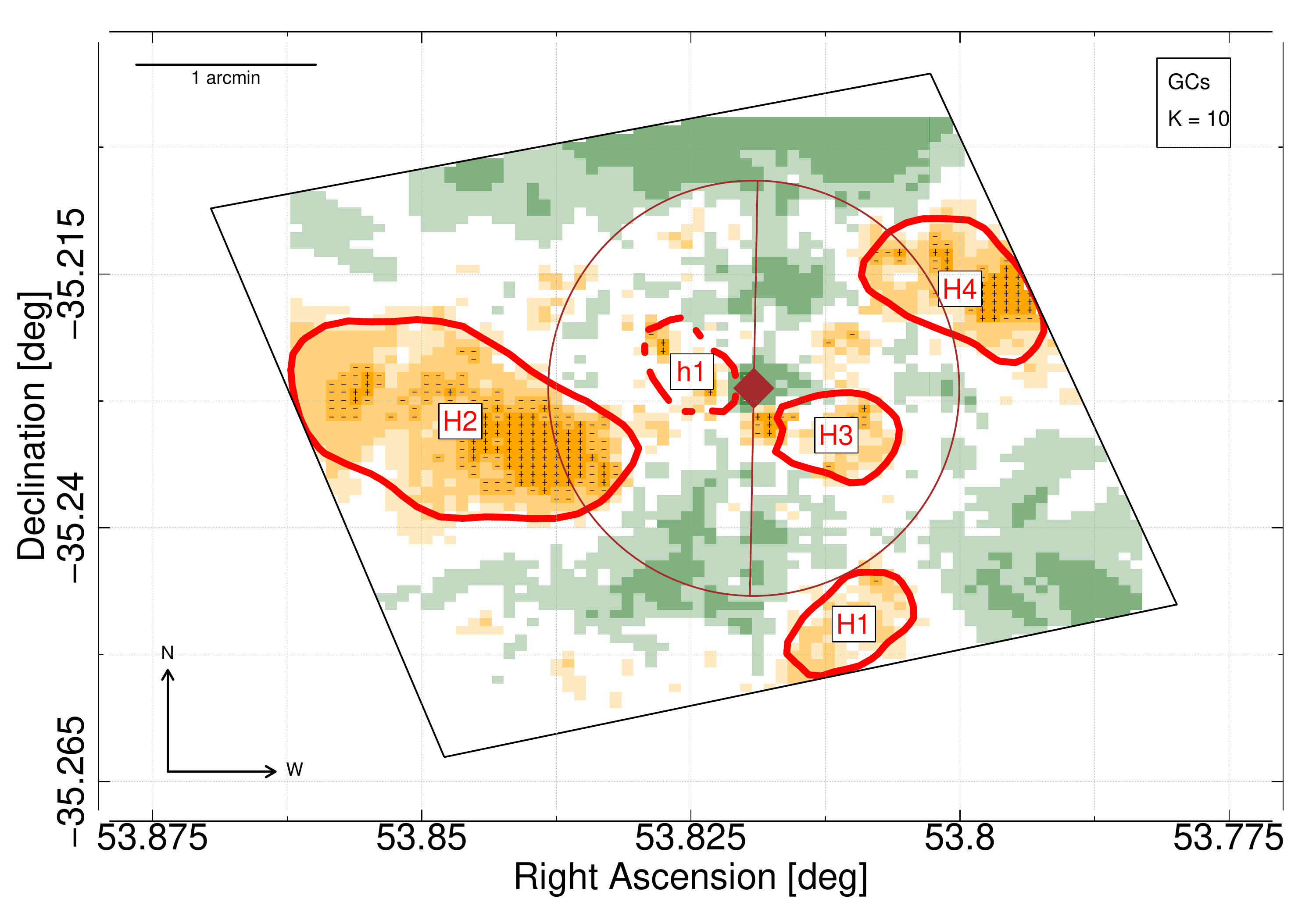}\\
	\includegraphics[width=0.45\linewidth]{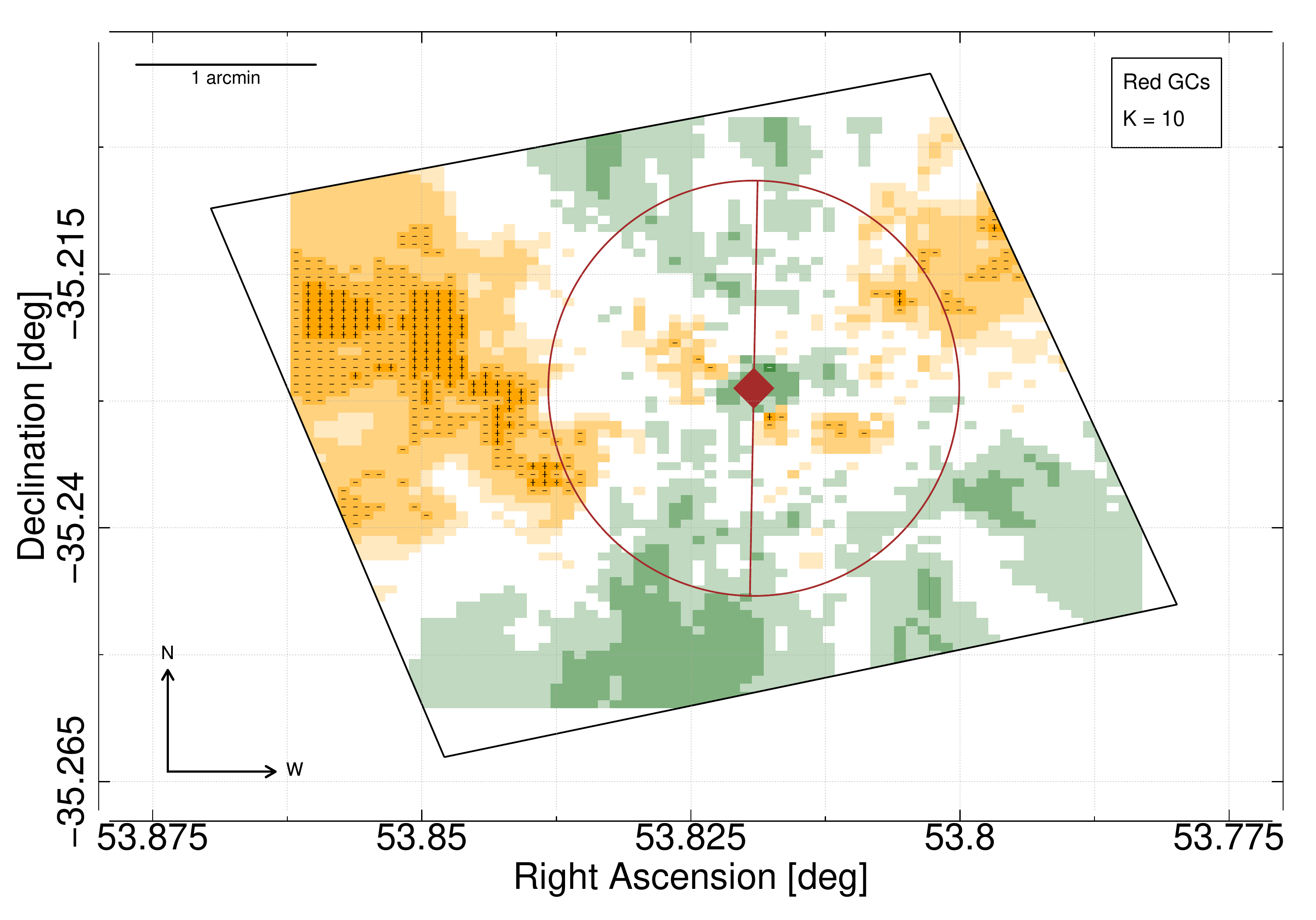}
	\includegraphics[width=0.45\linewidth]{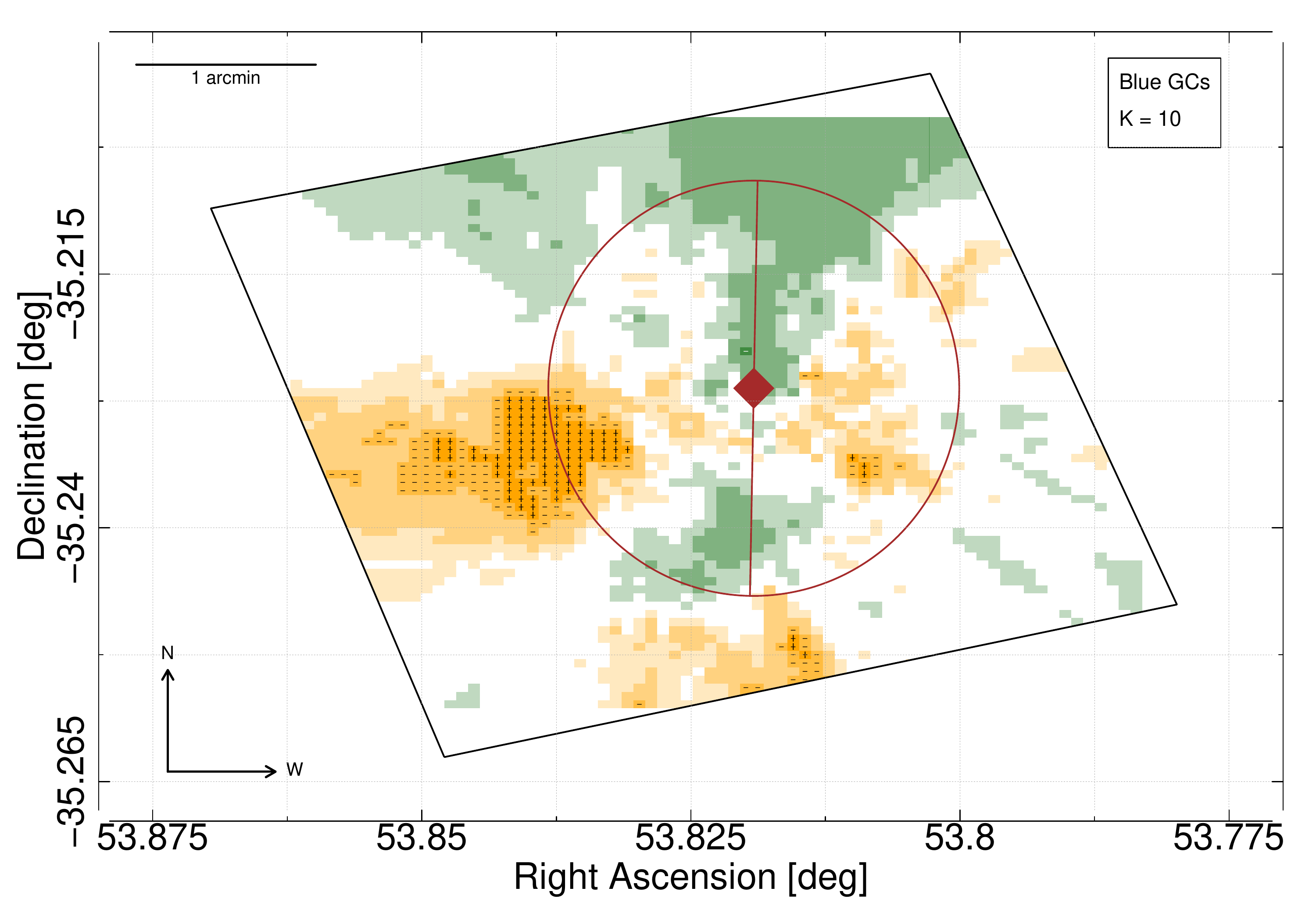}	
    \caption{Spatial distribution of GCs in NGC1374 .}
    \label{fig:ngc1374}
\end{figure}

The GC cluster system of NGC1374 has been investigated, using wide-field imaging data, by~\cite{bassino2006}.
They observed an evident color bimodality and reported that the radial profiles of blue and red subclasses
at the largest probed galactocentric distances are not consistent with the background source density probably 
because of contributions from the GC systems of companion galaxies NGC1375 and NGC1373. 

The spatial distribution of all ACSFCS GCs in NGC1374 (upper panel in Figure~\ref{fig:ngc1374})
features H1, H2 and H4, three very significant structures located close to the D$_{25}$ elliptical 
isophotes of the host galaxy in the S, E and W sides of the observed field respectively. These structures 
have elongated shapes with multiple overdensity peaks suggestive of spatial substructures. 
H2 points in an almost radial direction reaching well beyond the D$_{25}$ ellipse, while H1 and H4 
lie almost tangentially to the same isophote. The two additional structures H3 and h1 are located 
at small galactocentric distances SW and NE of the galaxy center, respectively. 
The red and blue residual maps~(lower panels in Figure~\ref{fig:ngc1374}) display significant differences: 
in the red GCs, H2, H4 and H3 are clearly detected (although the shapes of H2 and H3 differ from the 
general residual map and H3 has a much lower statistical significance), while in the blue GCs only H1, 
H2 and H3 are firmly detected. 

\subsection{NGC1351}
\label{subsec:ngc1351}

\begin{figure}[ht]
    \centering
	\includegraphics[width=0.45\linewidth]{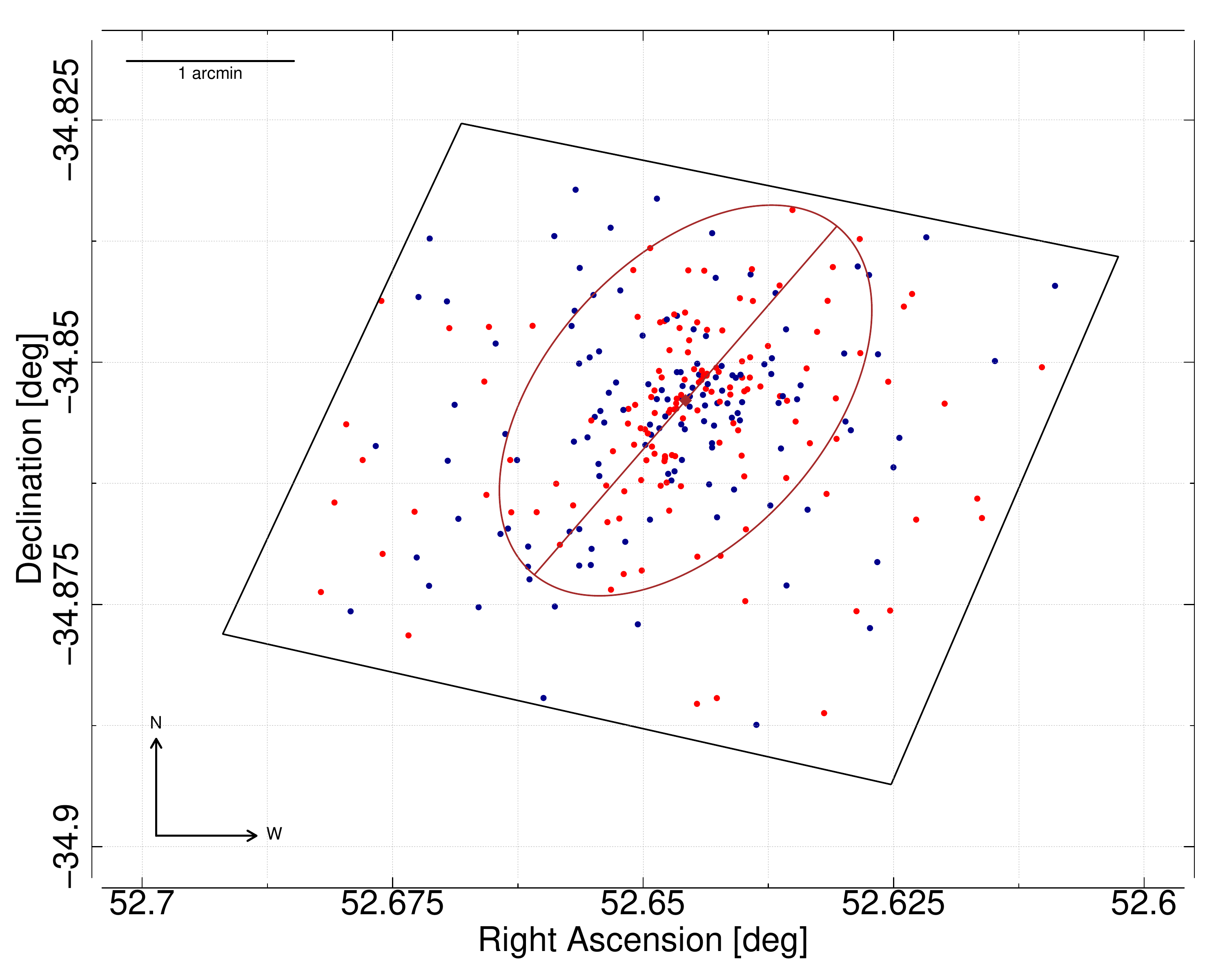}
	\includegraphics[width=0.45\linewidth]{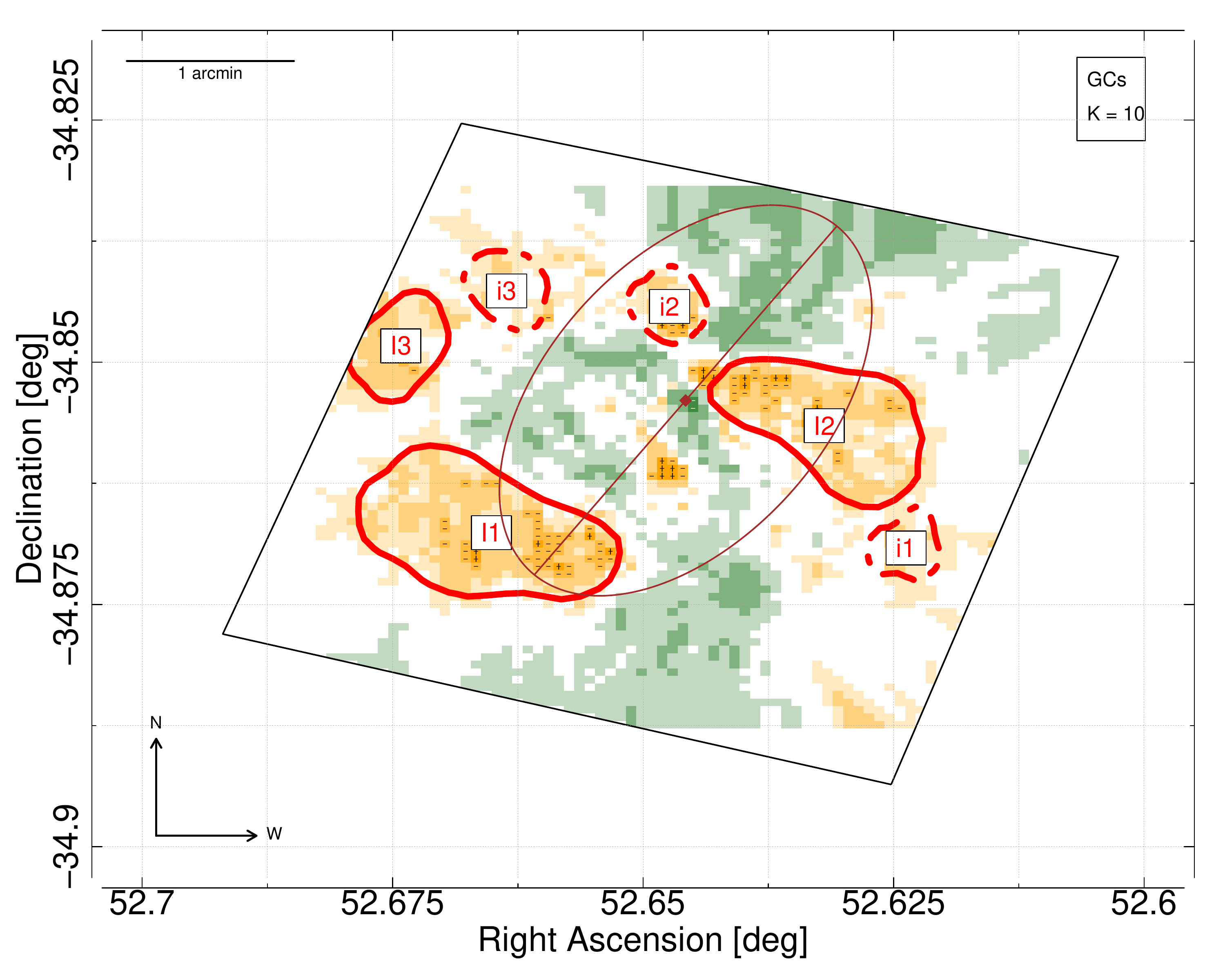}\\
	\includegraphics[width=0.45\linewidth]{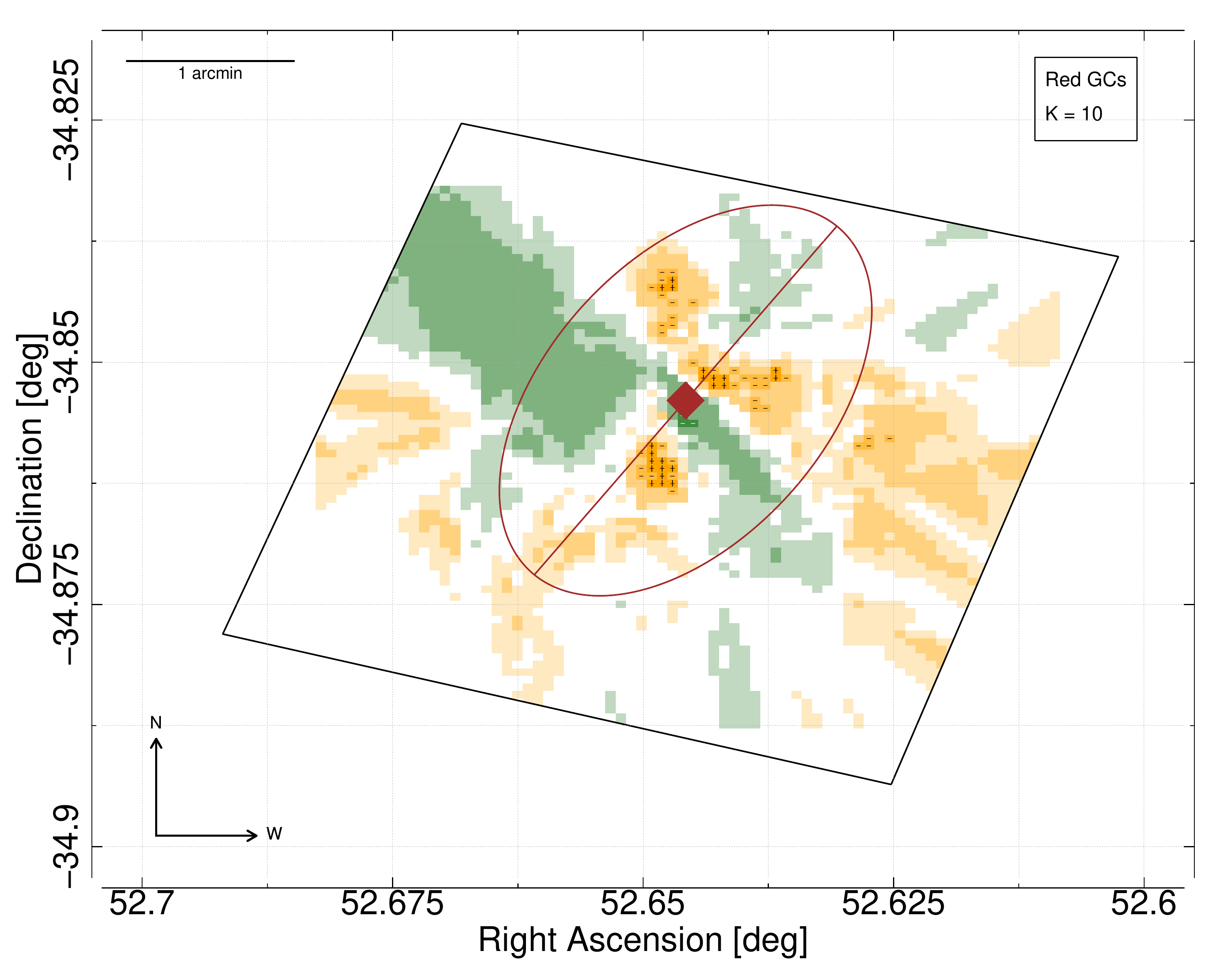}
	\includegraphics[width=0.45\linewidth]{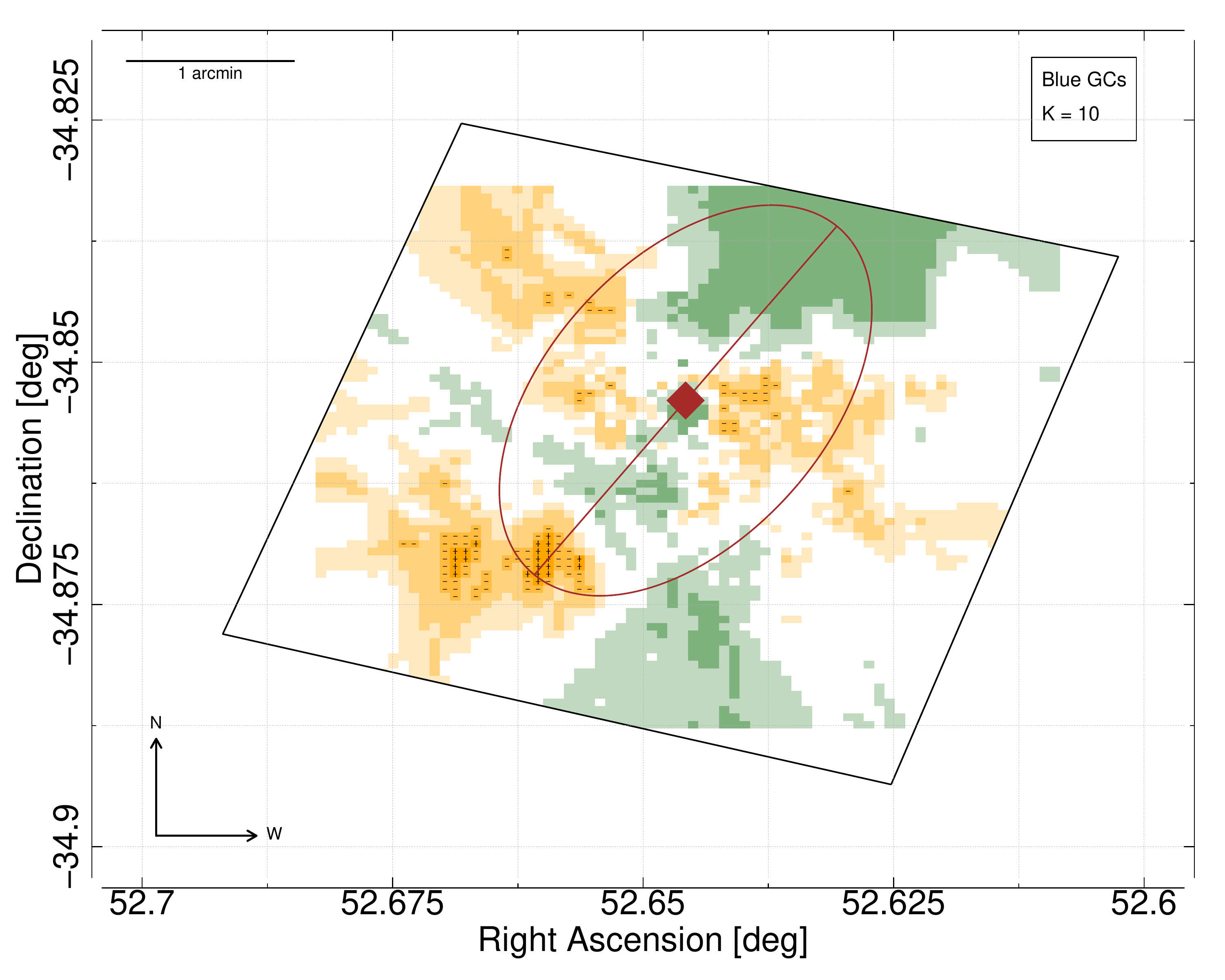}	
    \caption{Spatial distribution of GCs in NGC1351 (see caption of Figure~\ref{fig:ngc1399} for details).}
    \label{fig:ngc1351}
\end{figure}

The globular clusters system of NGC1351 features several large spatial structures~(Figure~\ref{fig:ngc1351}). 
The very statistically significant overdensities I1 and I2 display similarly 
elongated morphology: I1 lies along an almost tangential direction to the D$_{25}$ elliptical isophote 
of the host galaxy at the S end of the major axis, unlike I2 that originates in the center
and radially extends to the W outskirts of the observed field. Both structures are also visible
in the blue GCs residual map, while I1 is not detected in the red residual map. 
Four additional density structures (I3 and the smaller i1, i2 and i3) have roughly circular shapes and
are located at different galactocentric distances in the E and W sides of the galaxy. 
i1, i2 and i3 are observed in the blue GCs map, while I3 is only weakly detected in the red GCs 
residual map. In the distribution of red GCs, i2 and the E edge of structure I2 are apparently 
connected, while another significant overdensity peak, not associated with any structure in the 
residual map generated by all GCs, is located S of the galaxy center. In the blue residual map, 
a geometrically incoherent region of positive residuals can be seen in the regions occupied by i2, i3 and i1.

\subsection{NGC1336}
\label{subsec:ngc1336}

\begin{figure}
    \centering
	\includegraphics[width=0.45\linewidth]{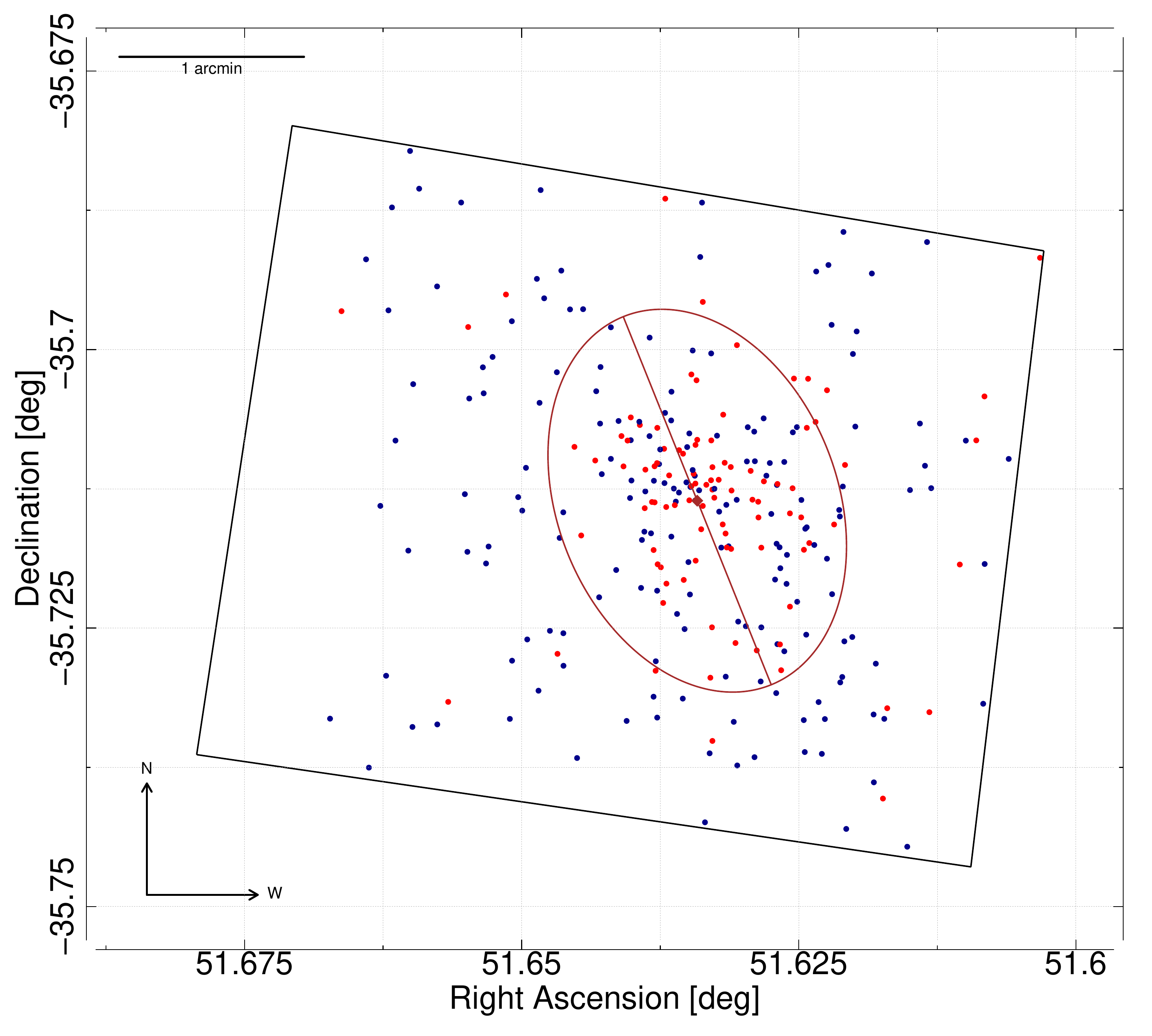}
	\includegraphics[width=0.45\linewidth]{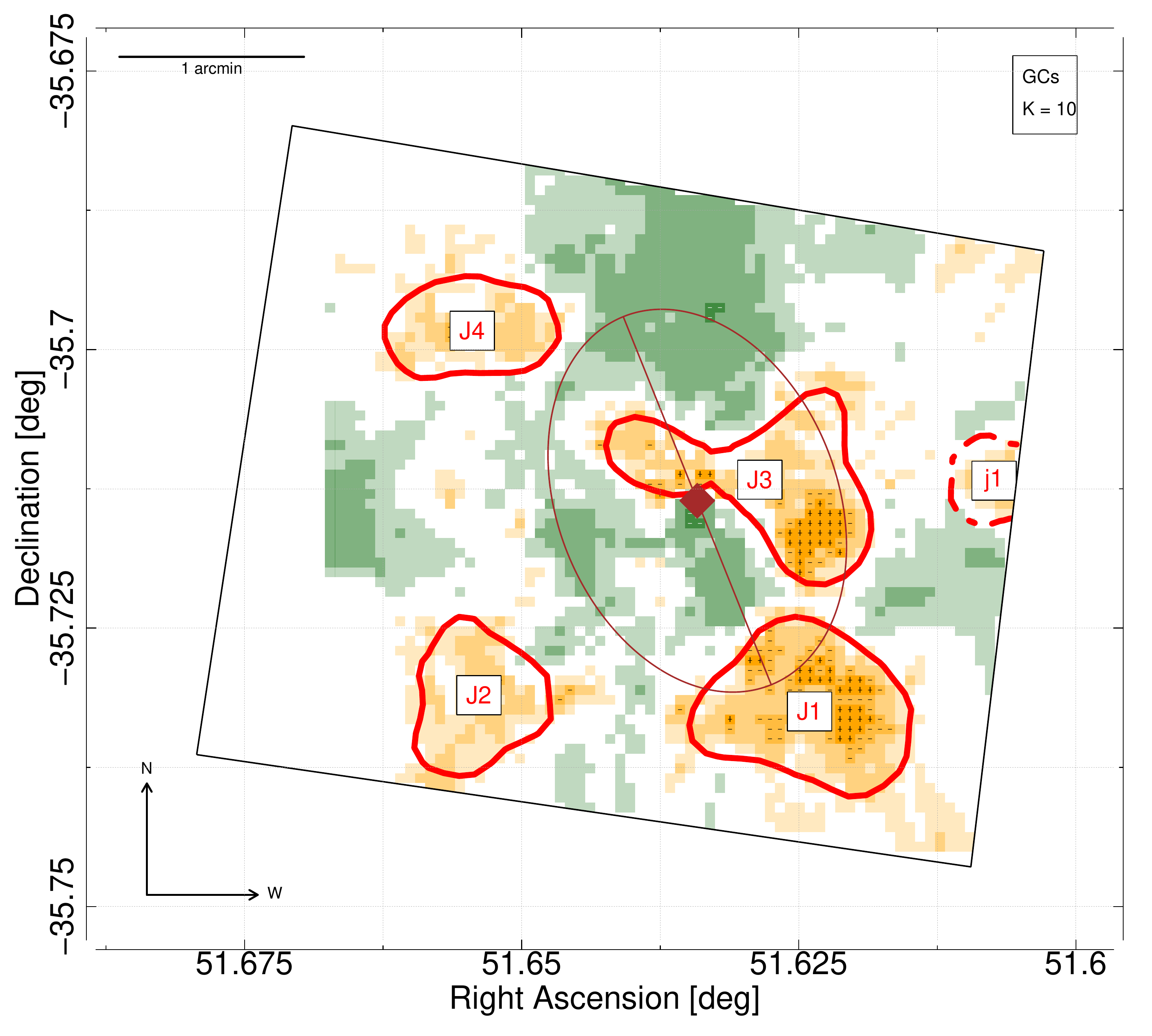}\\
	\includegraphics[width=0.45\linewidth]{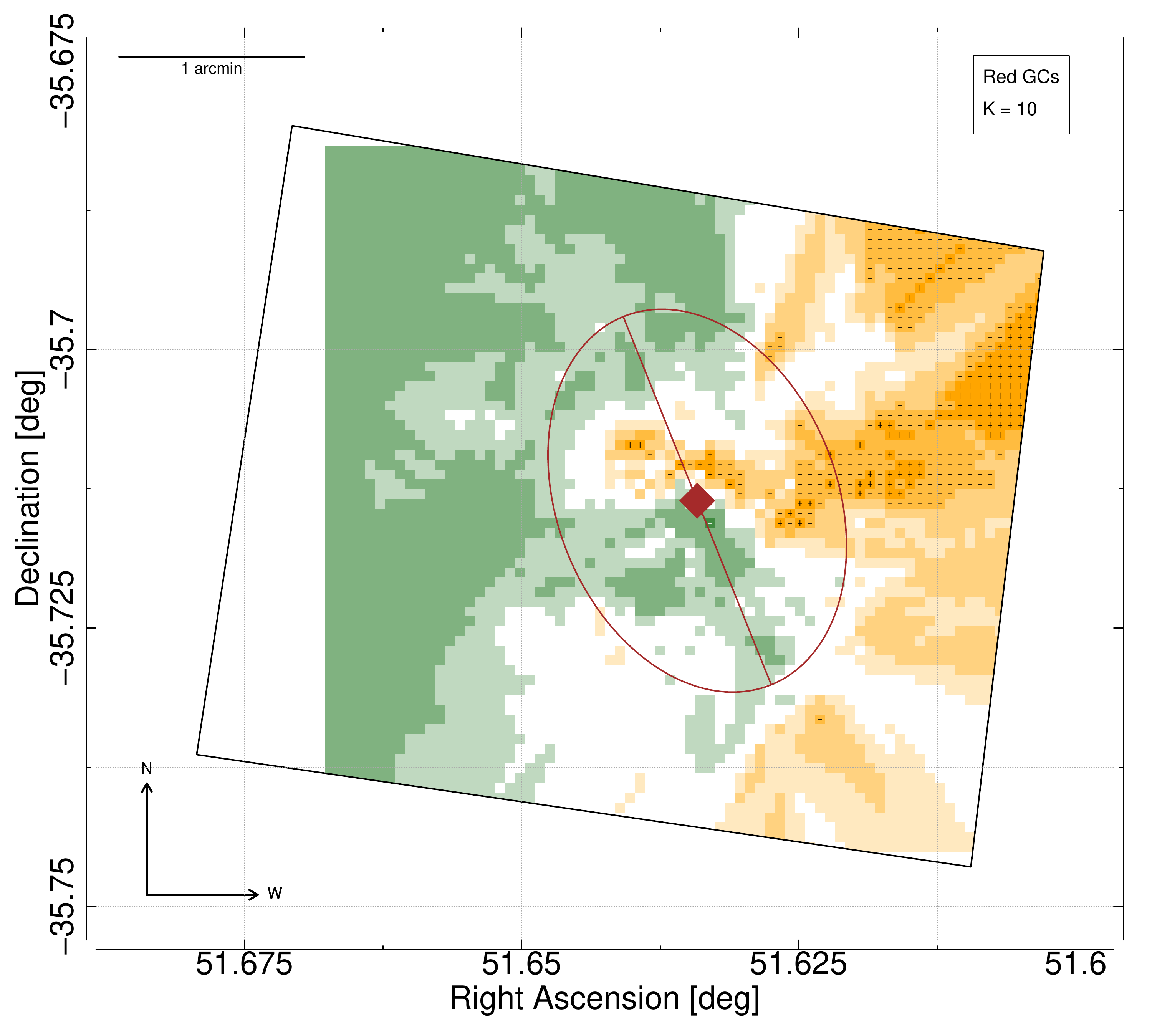}
	\includegraphics[width=0.45\linewidth]{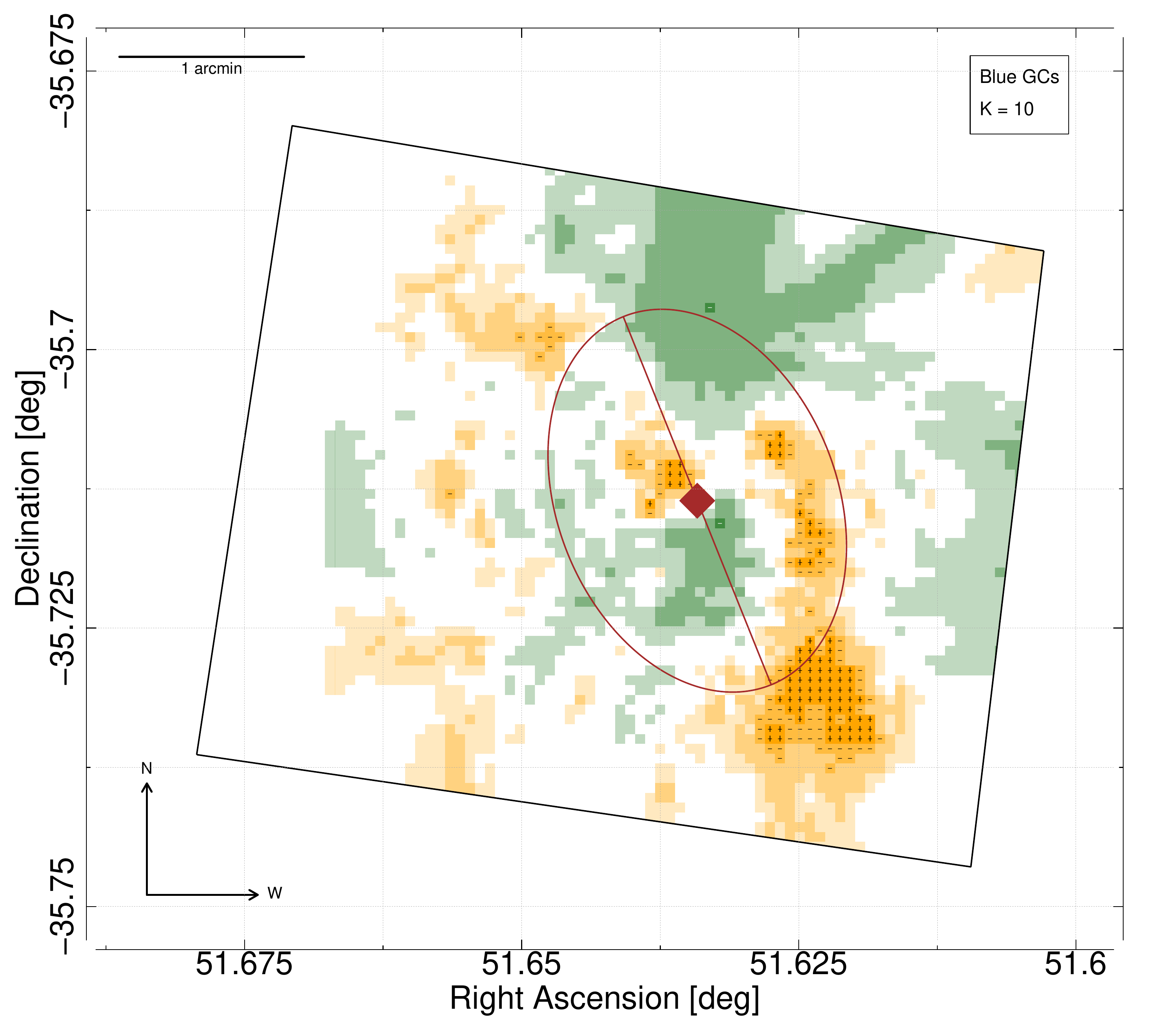}	
    \caption{Spatial distribution of GCs in NGC1336 (see caption of Figure~\ref{fig:ngc1399} for details)}
    \label{fig:ngc1336}
\end{figure}

Recent studies have found that NGC1336 possesses two kinematically decoupled components that
indicate that this galaxy has undergone a major merger~\citep{fahrion2019}. Moreover, the GCS of 
NGC1336 has a very high specific frequency~\citep{liu2019}, second only to NGC1399 in Fornax, 
that can be explained by lack of significant GCs disruption favored by the isolation 
from the other members of the Fornax clusters, possibly suggesting that NGC1336 is infalling 
in the cluster and has not interacted with other members of cluster yet. 

The spatial distribution of GCs in NGC1336~(Figure~\ref{fig:ngc1336}) is dominated by two large 
structures (J1 and J3) that, together, occupy a significant fraction of the area of the host galaxy 
within the D$_{25}$ elliptical isophote. J1 is located at the S intersection
of the major axis with the D$_{25}$ and extends towards large galactocentric distances. J3 is a  
geometrically complex, hook-shaped structure containing multiple overdensity peaks that spans a large
range of radii and is likely the result of the chance projection of multiple structures. 
J3 is detected in both the residual maps of red and blue GCs, while J1 is detected only in 
blue map where it is connected to J3. The remaining structures (J2, J4 and j1) have all low statistical
significance and can only be seen in the blue GCs.

\section{Properties of the GC spatial structures as a function of $K$}
\label{sec:appendix2}

The parameter $K$ of the KNN method determines the number of ``neighbors'' used to 
estimate the density maps of both observed and simulated GC distributions and, in turn, 
the residual maps~(See 
Section~\ref{sec:method} for details). For this reason, its choice affects the properties 
of the spatial structures detected in the GCs spatial distributions. As discussed in 
Section~\ref{sec:spatialdistribution}, the practical choice of the $K$ value used to study 
the spatial structures in the GC distribution depends on the scales to be 
investigated and the desired type of characterization of the structures observed. 
In general, large values of $K$ produce residual maps with larger spatial structures 
while lower values produce maps highlighting smaller spatial features that can be 
difficult to distinguish from noise and whose properties are poorly constrained. 

\begin{figure*}[ht]
    \centering
	\includegraphics[width=0.19\linewidth]{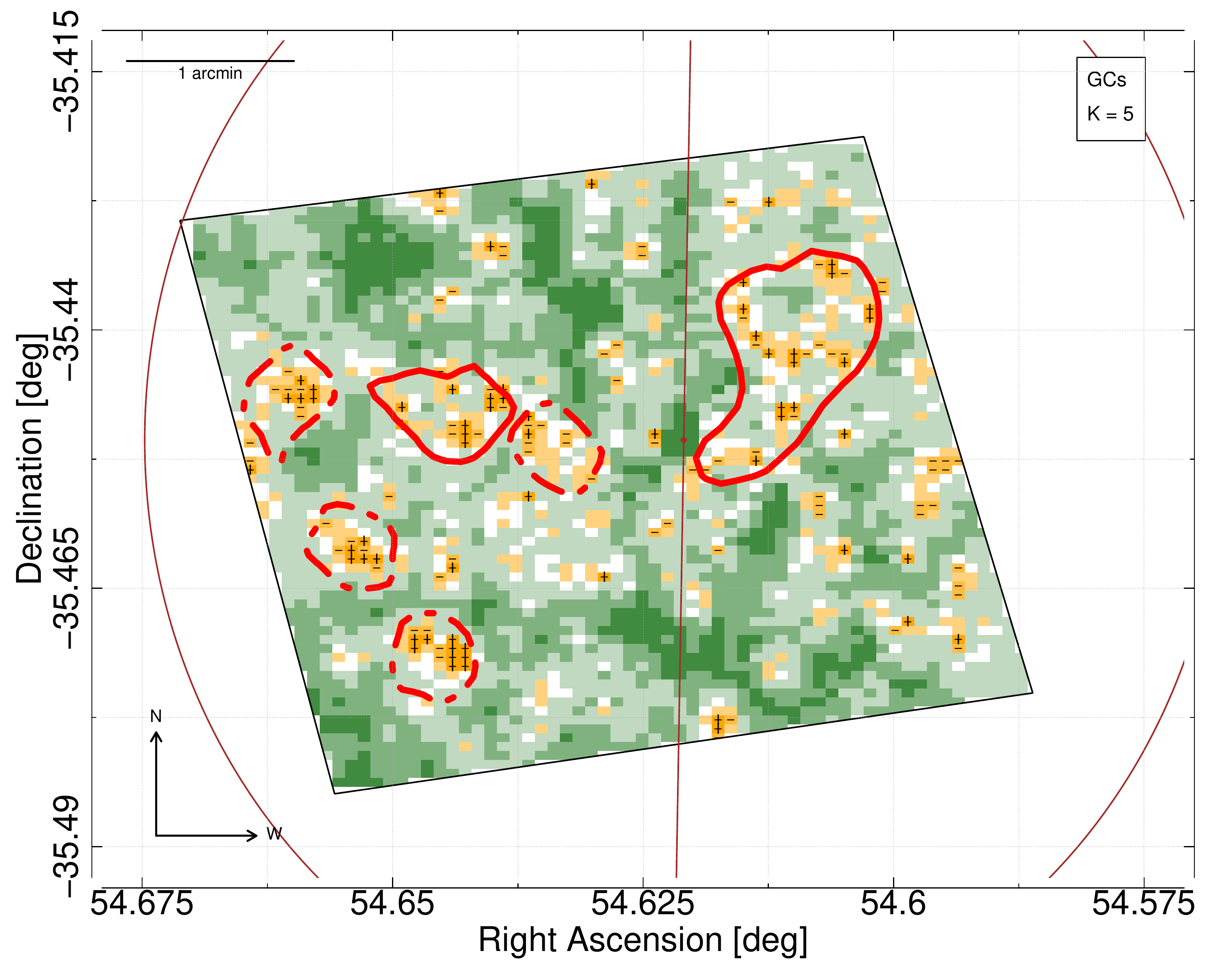}
	\includegraphics[width=0.19\linewidth]{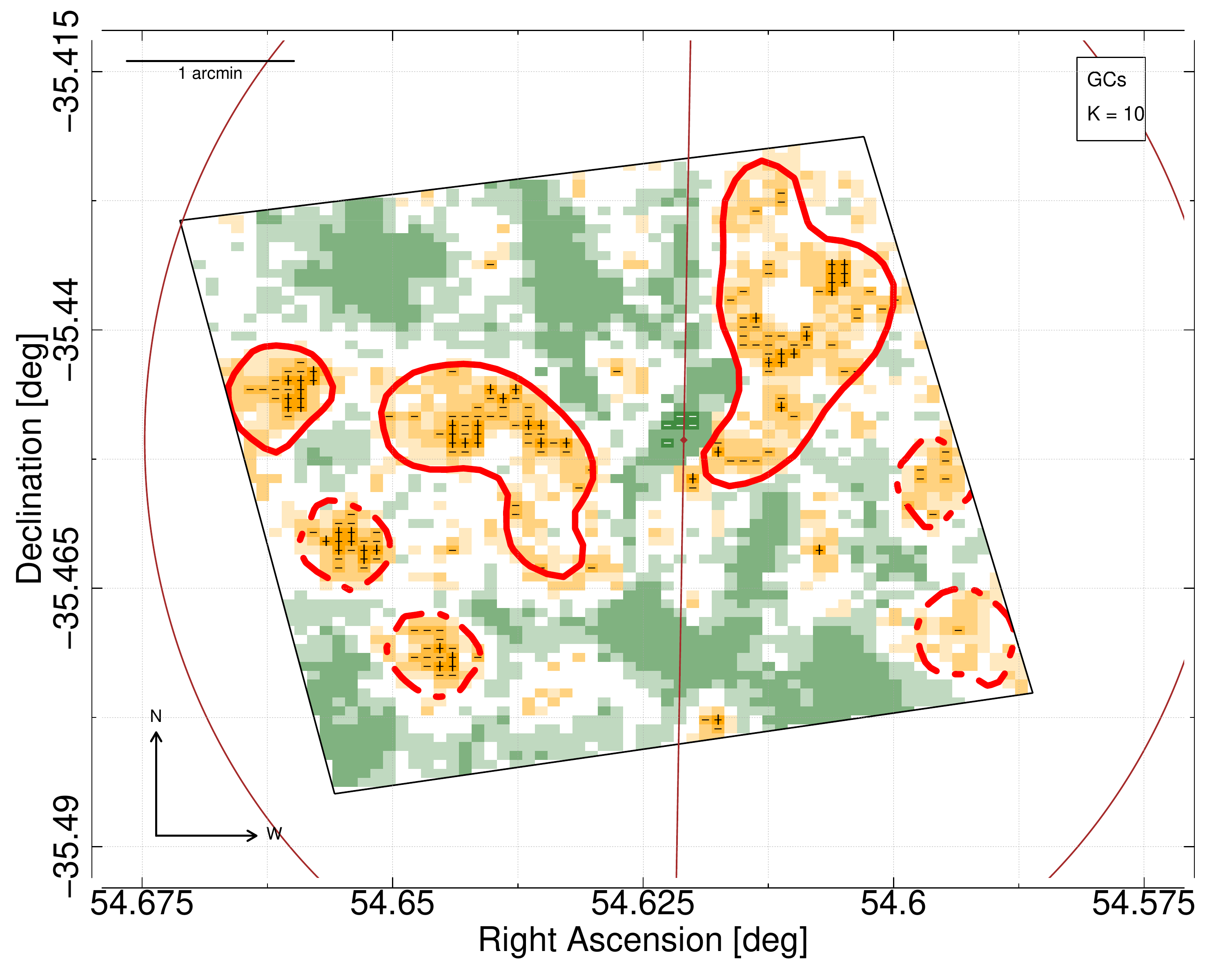}
	\includegraphics[width=0.19\linewidth]{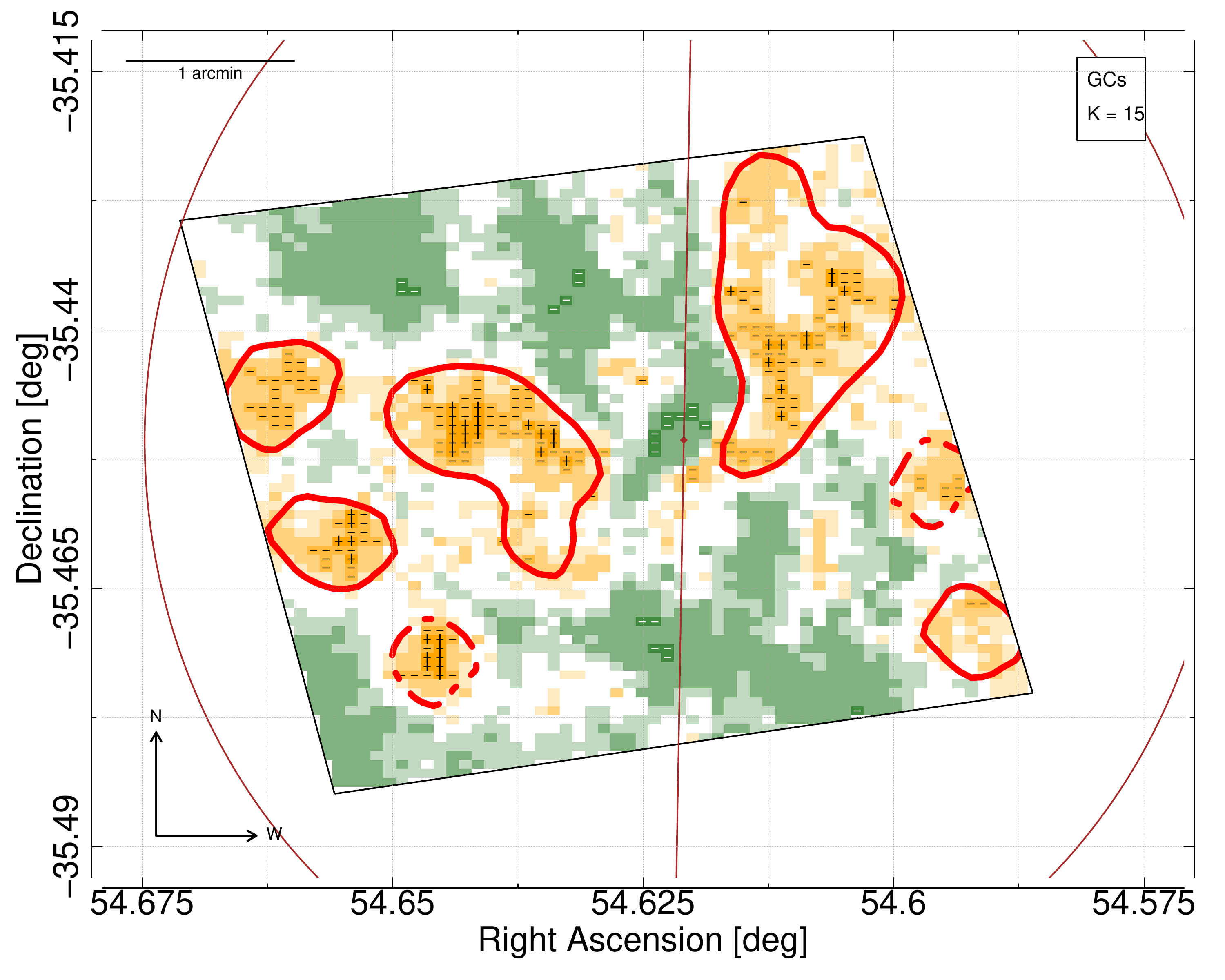}
	\includegraphics[width=0.19\linewidth]{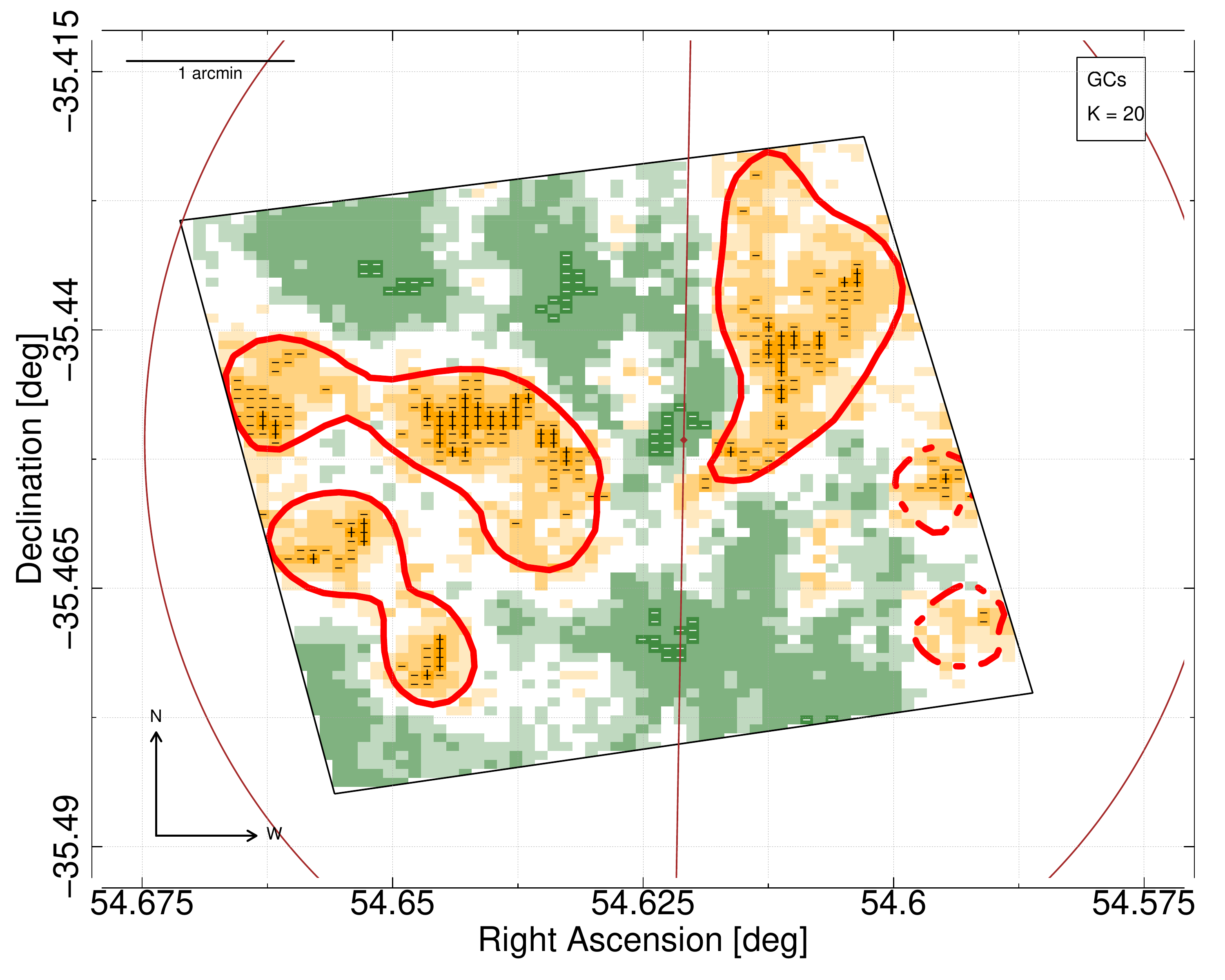}
	\includegraphics[width=0.19\linewidth]{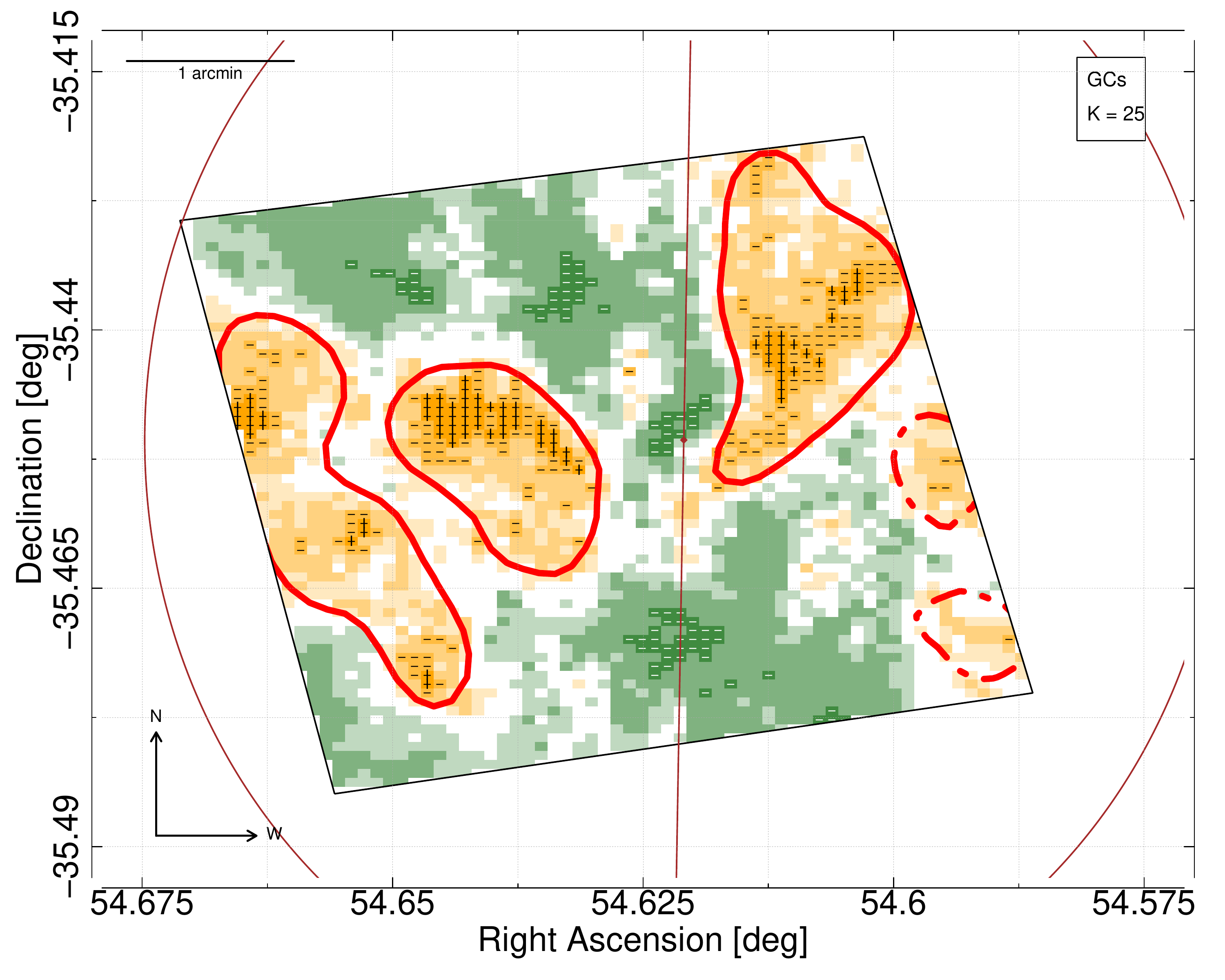}	\\
	\includegraphics[width=0.19\linewidth]{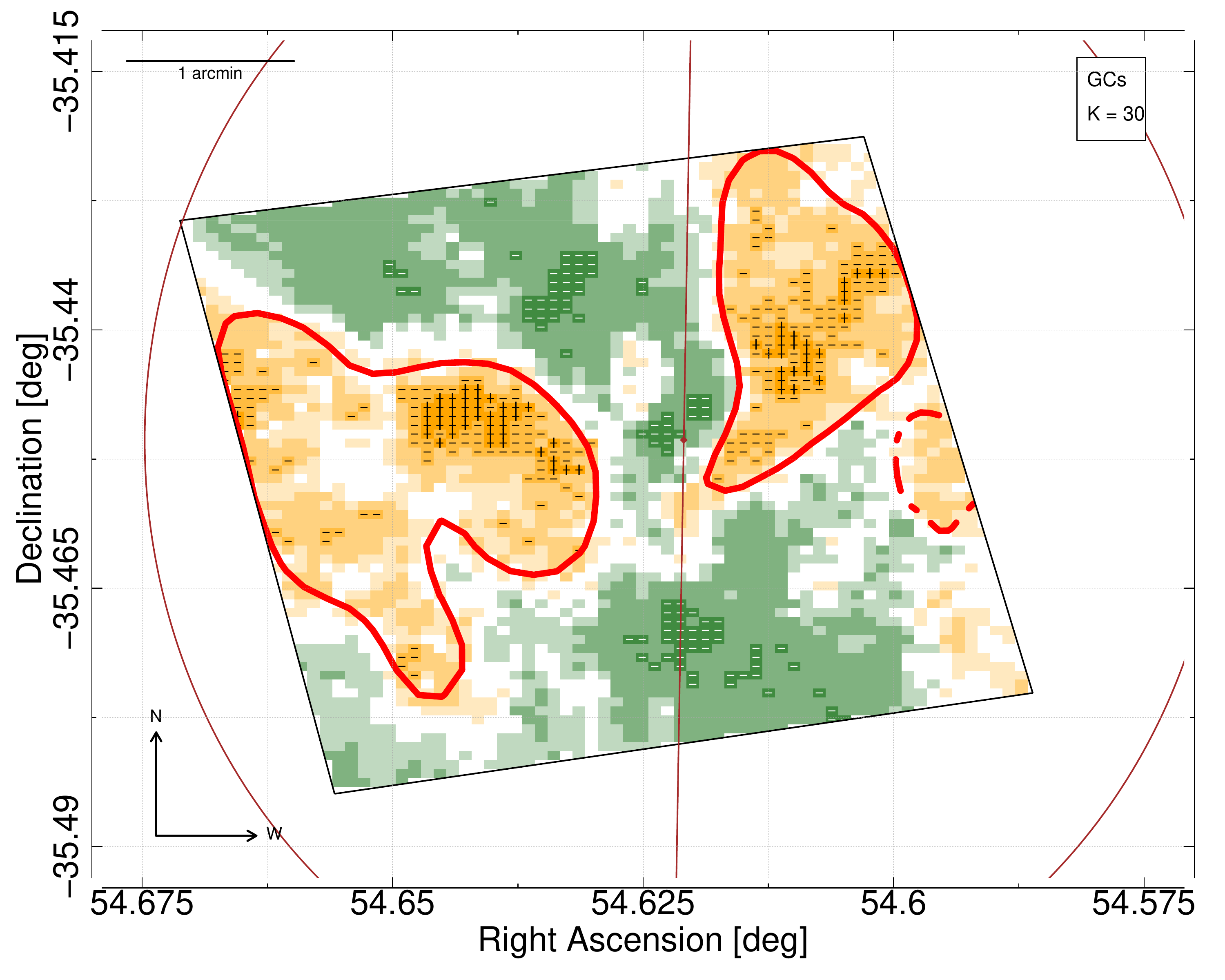}
	\includegraphics[width=0.19\linewidth]{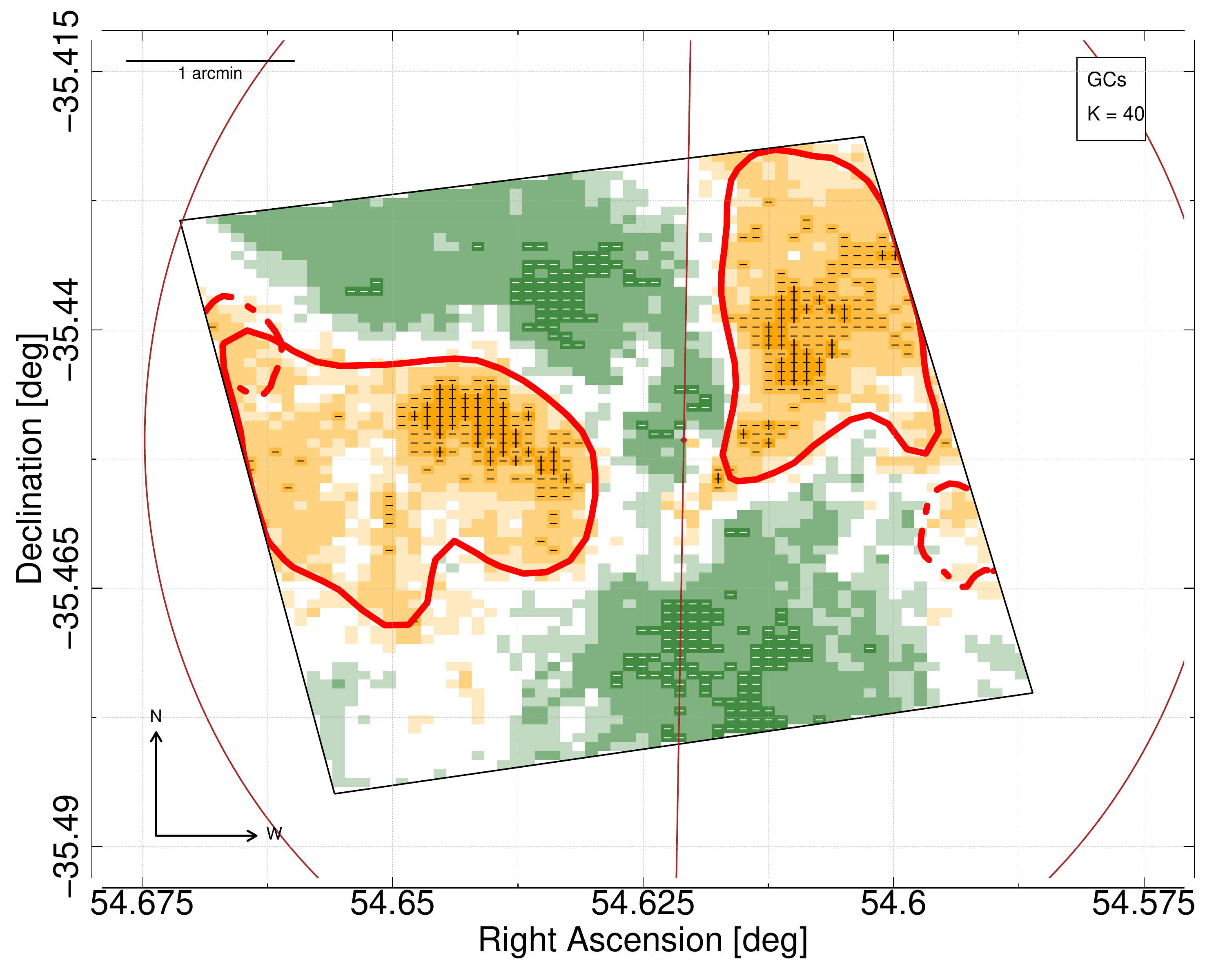}
	\includegraphics[width=0.19\linewidth]{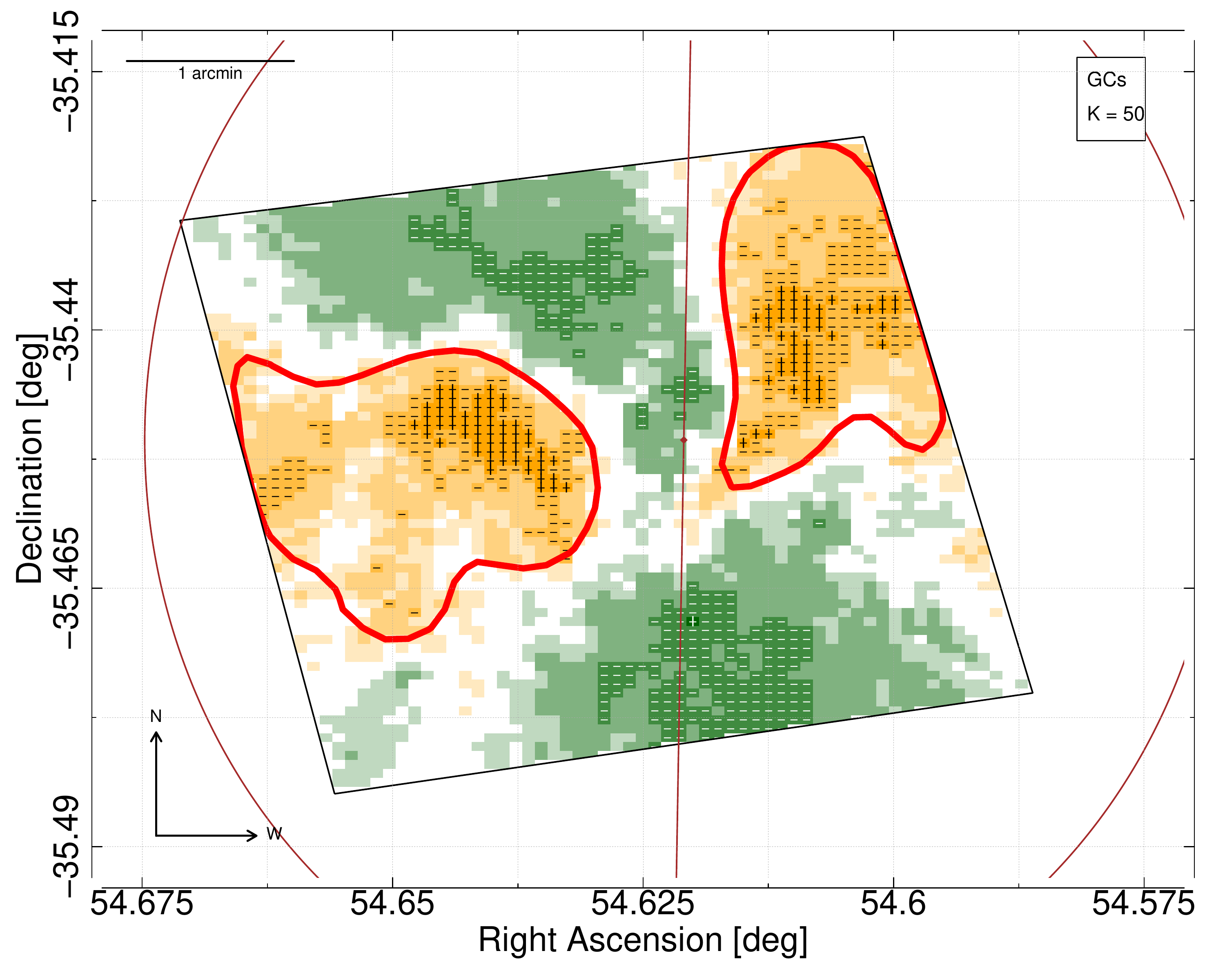}
	\includegraphics[width=0.19\linewidth]{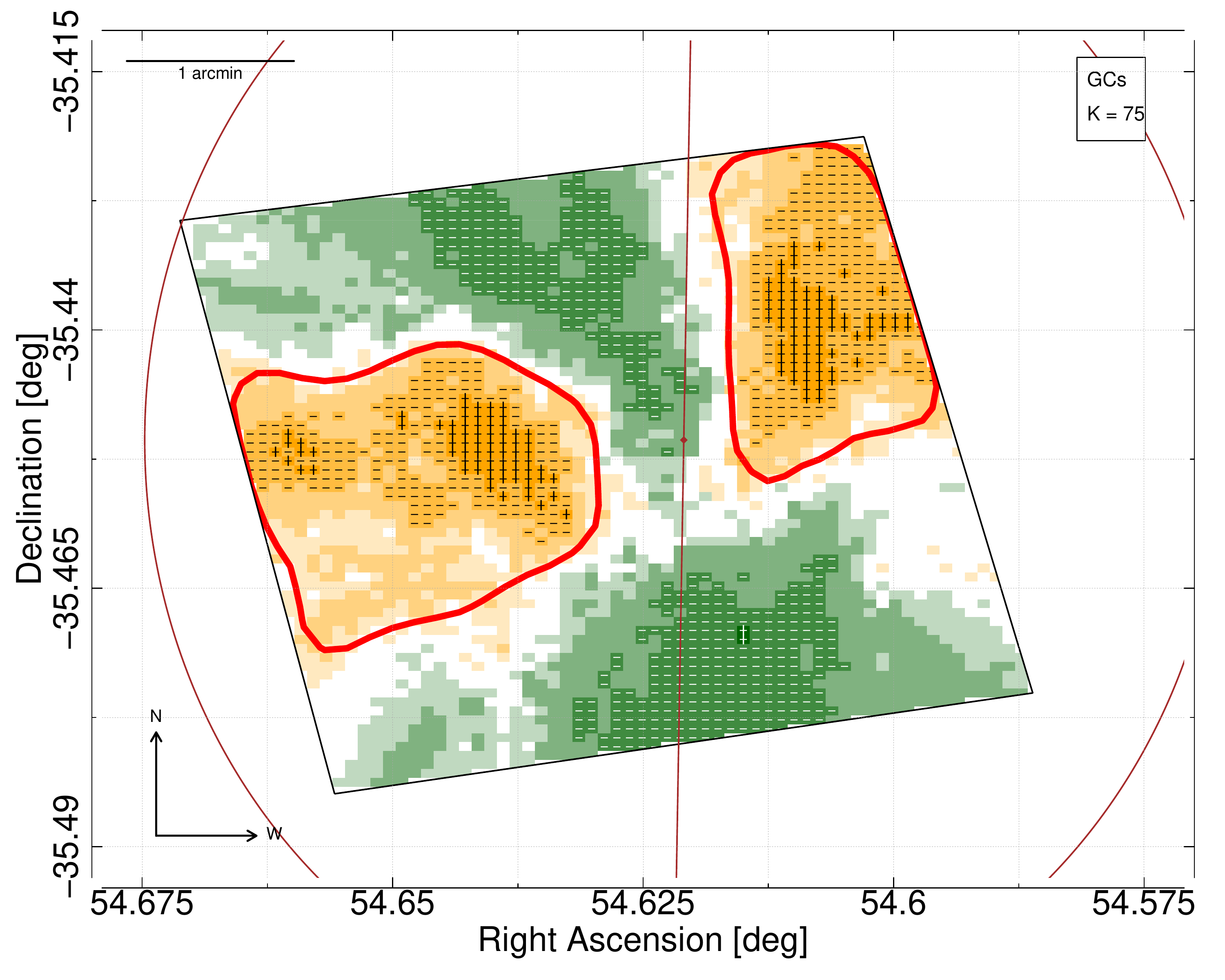}
	\includegraphics[width=0.19\linewidth]{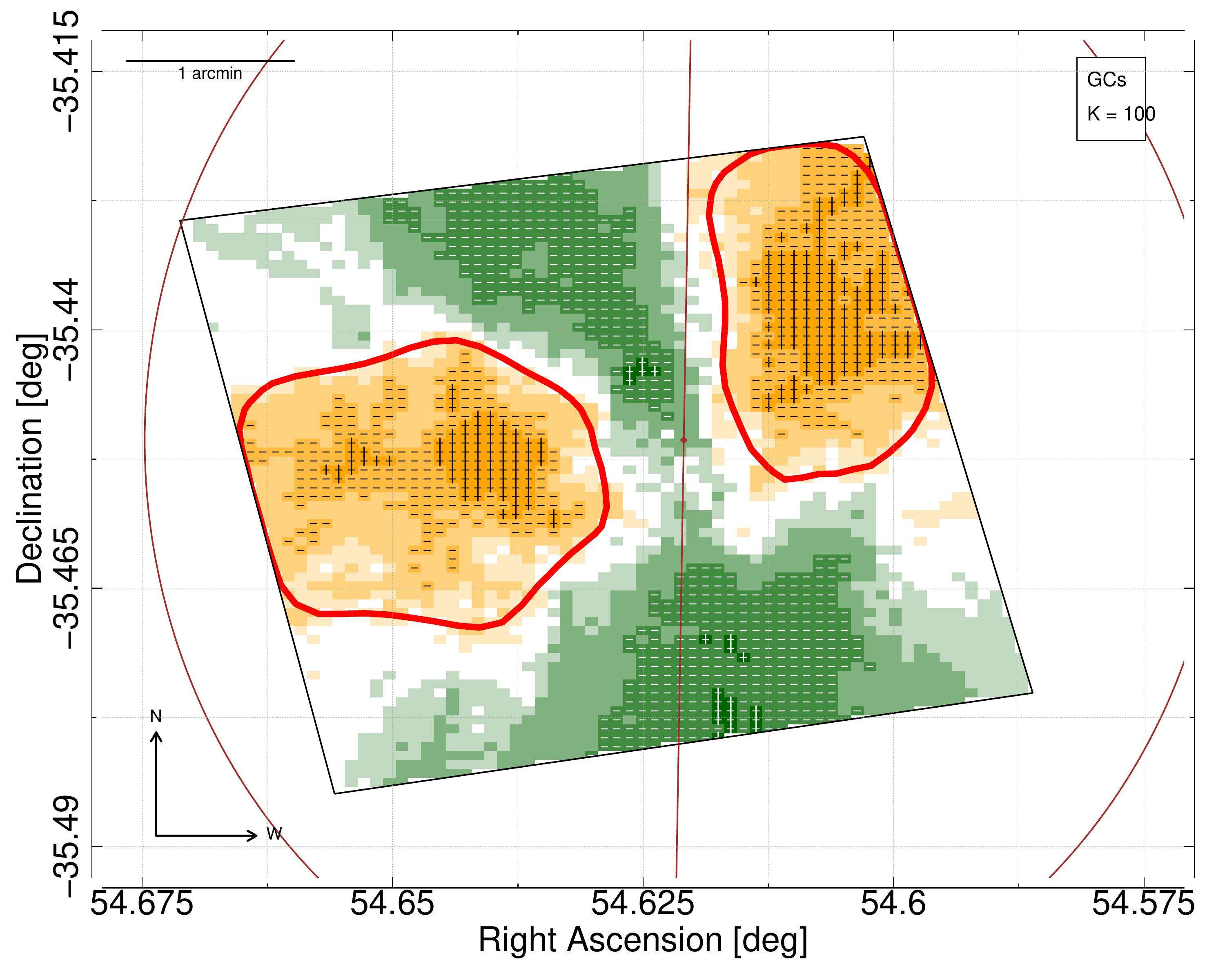}\\
	\includegraphics[width=0.16\linewidth]{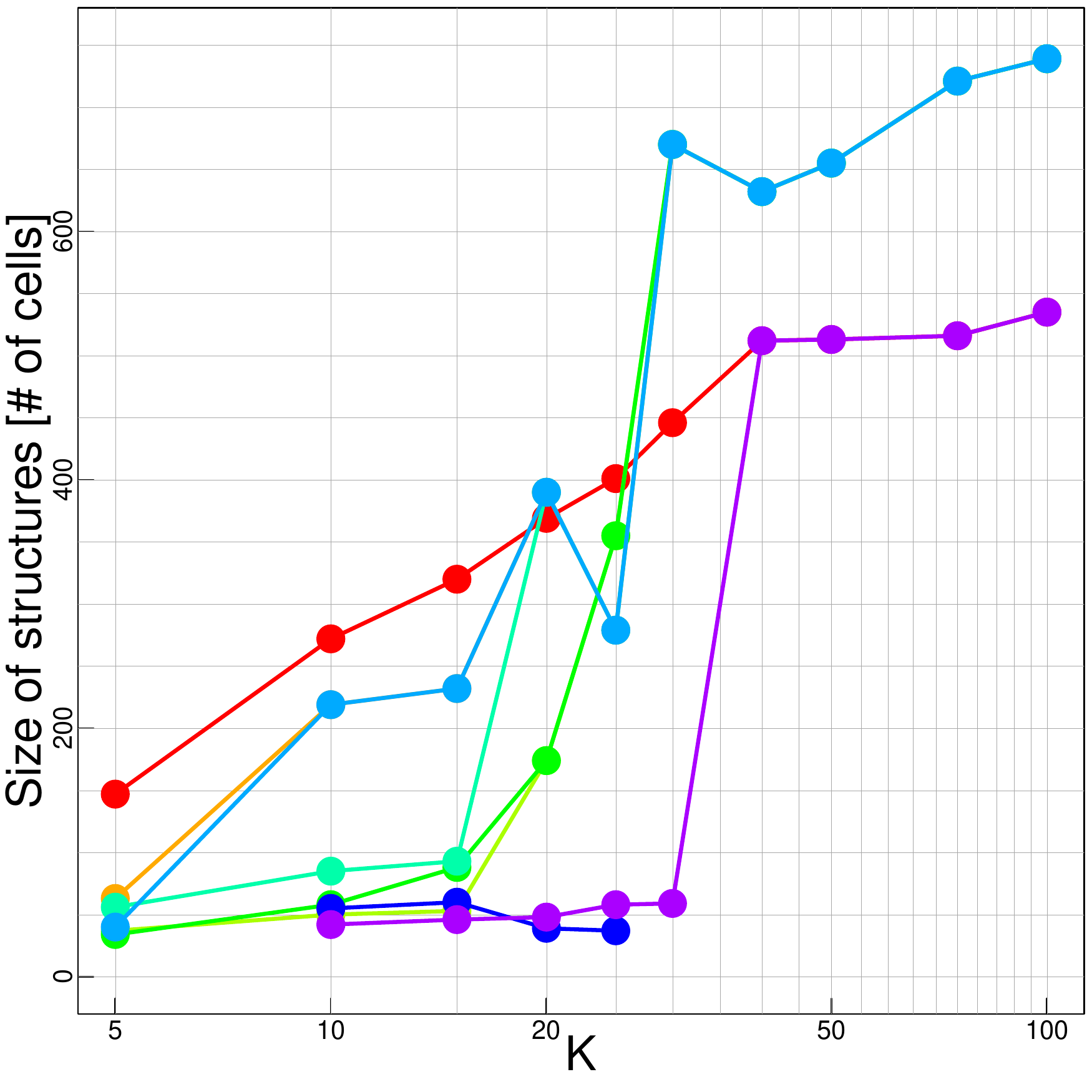}
	\includegraphics[width=0.16\linewidth]{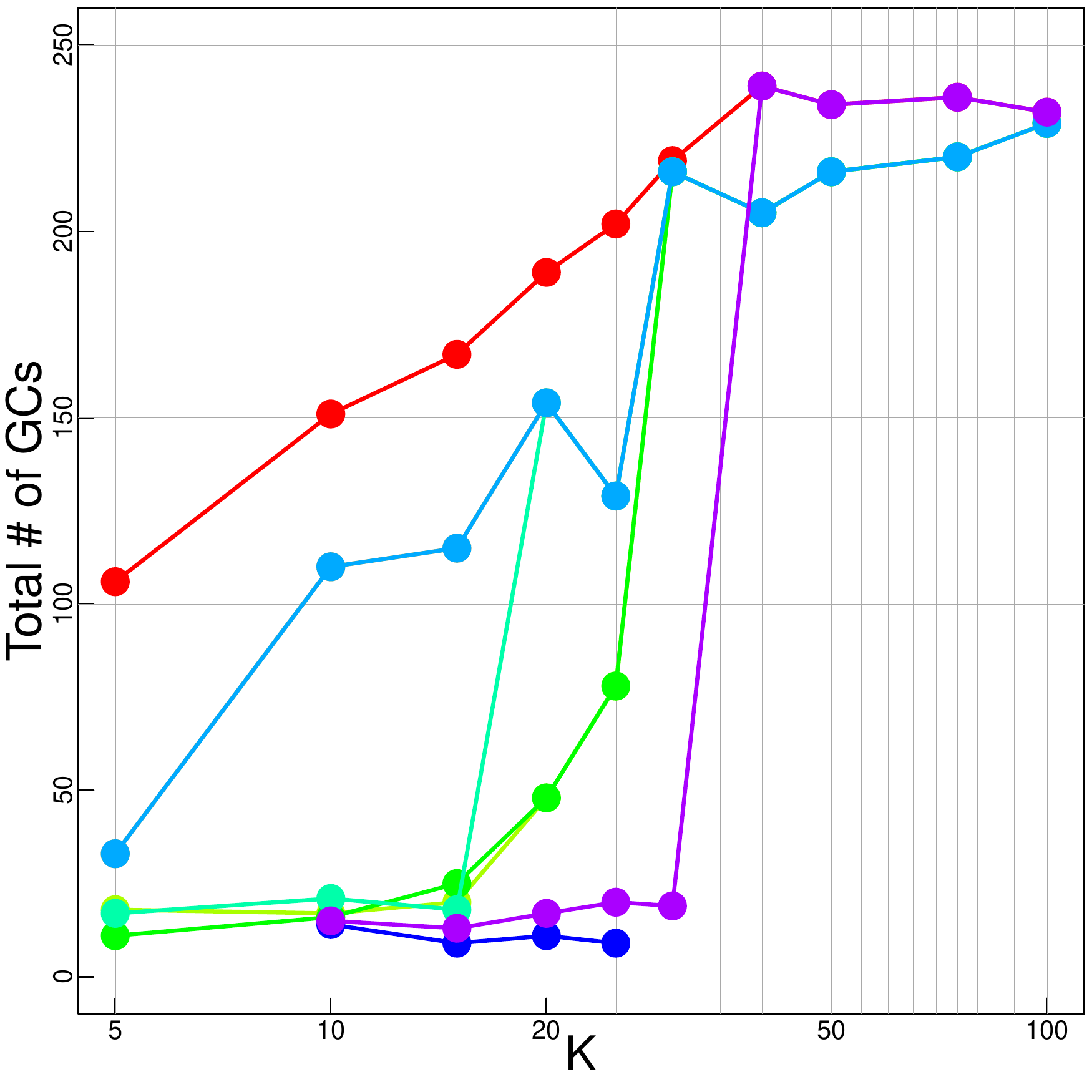}
	\includegraphics[width=0.16\linewidth]{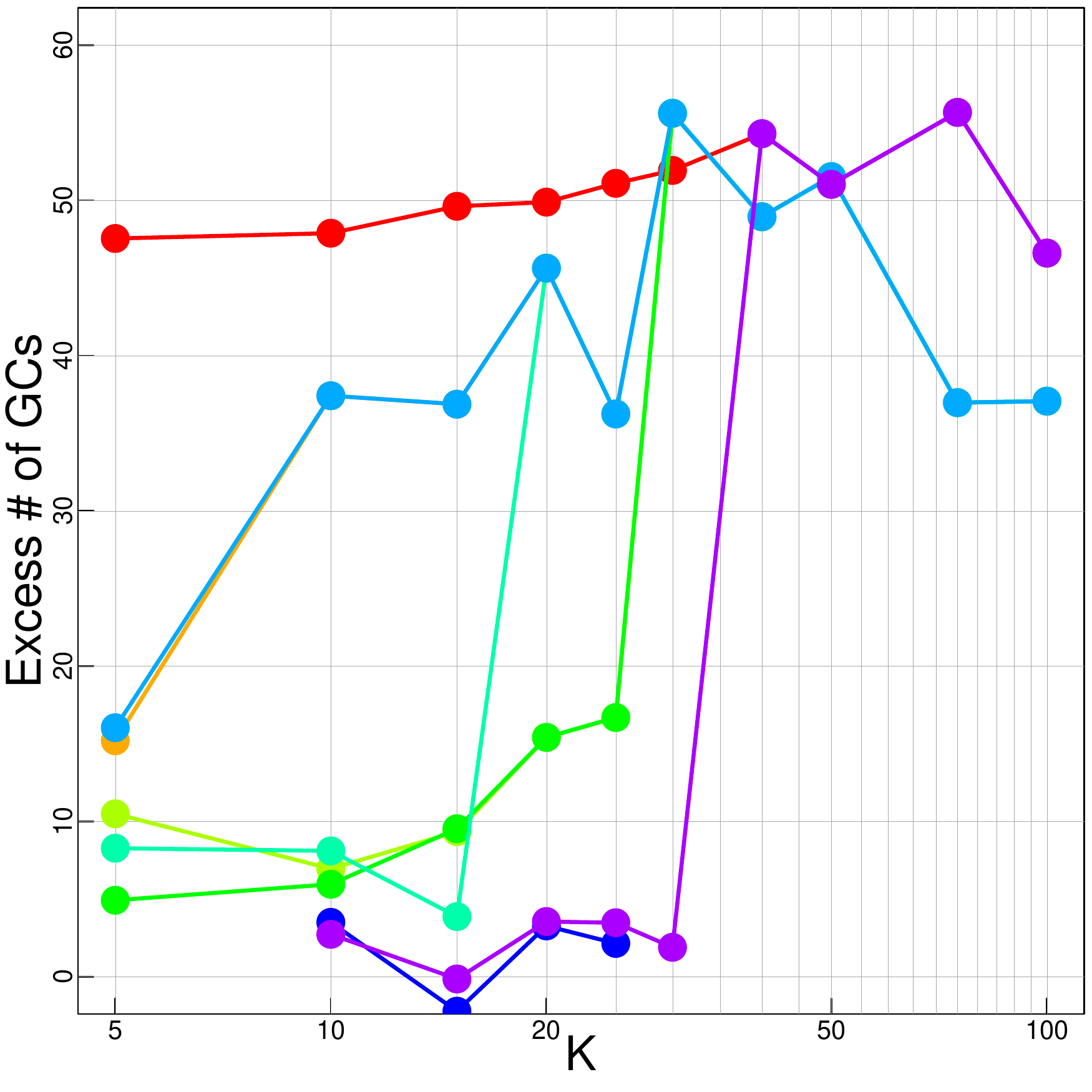}
	\includegraphics[width=0.16\linewidth]{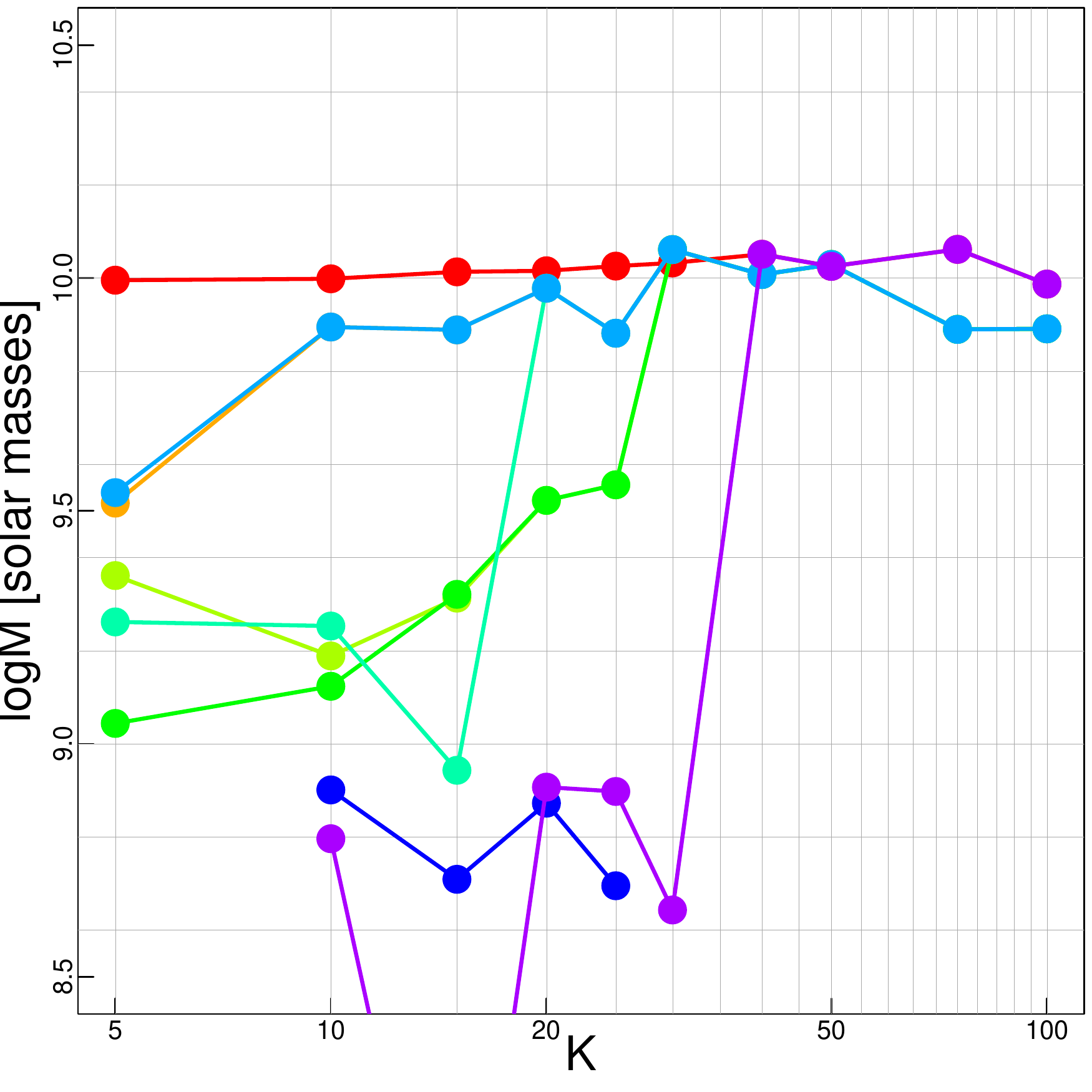}
	\includegraphics[width=0.16\linewidth]{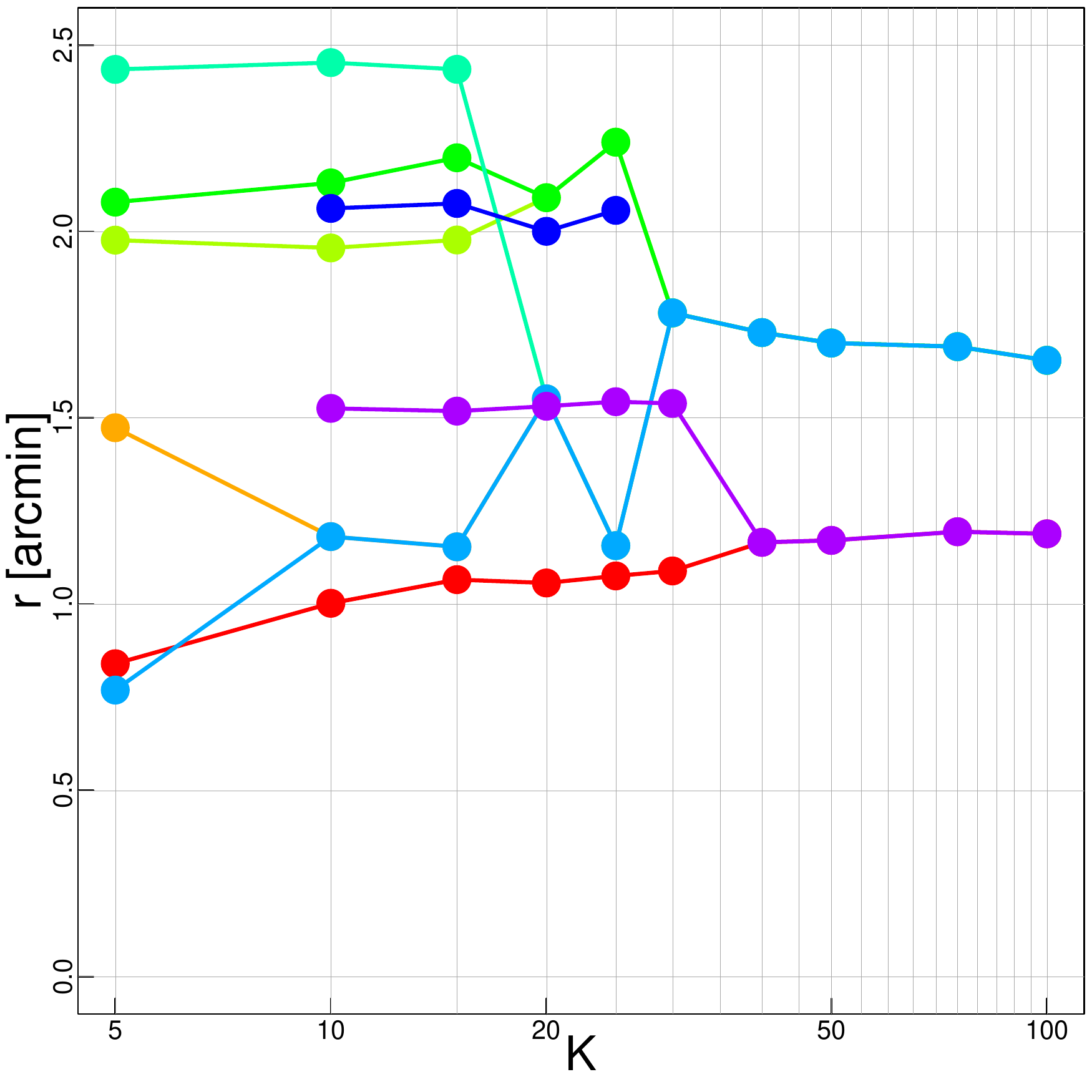}
	\includegraphics[width=0.16\linewidth]{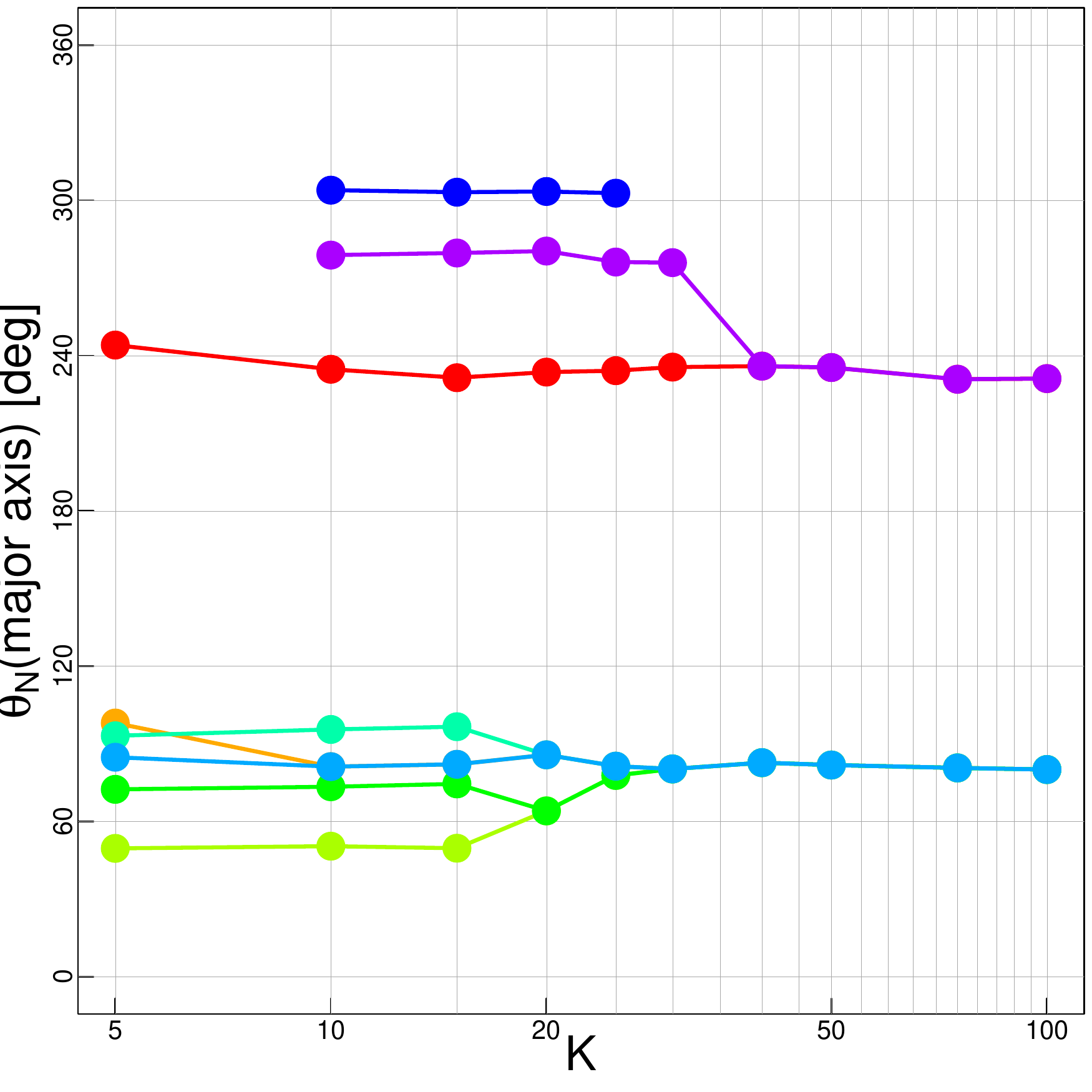}\\	
    \caption{Upper and mid panels: residual maps of the spatial distribution of GCs in NGC1399 for 
    for $K\!=\!\{5,10,15,20,25,30,40,50,75,100\}$. Red solid and dotted lines show 
    the boundaries of large and intermediate structures detected in each residual 
    map. Lower panel, from left to right: evolution of sizes, total and excess numbers of GCs, 
    mass of the progenitor, average galactocentric distance and average azimuthal position as
    a function of $K$ for the all the intermediate and large structures detected in the spatial 
    distribution of NGC1399.}
 	\label{fig:ngc1399_kevolution}
 \end{figure*}

Figure~\ref{fig:ngc1399_kevolution} shows 
the dependence of the residual maps and the properties of the residual structures in the 
case of NGC1399: the upper and mid rows display the residual maps 
for $K\!=\!\{5,10,15,20,25,30,40,50,75,100\}$ obtained from all GCs, with the approximate 
contours of large and intermediate structures~(defined as explained in Section~\ref{sec:method})
shown as dotted and solid lines respectively. The panels in the lower row display (from left
to right) the size, total and excess numbers of GCs, mass of the progenitor, average 
galactocentric distance and average azimuthal position of the structures as a function of 
$K$. In general, the size, total and excess numbers of GCs of structures grow with larger
$K$ either gradually (when the same structure includes nearby cells of the 
residual maps) or abruptly as two
or more structures detected at a given value of $K$ merge at a larger $K$. Structures can 
also disappear for growing $K$'s without merging into other existing structures, as it occurs for 
the small structures located in the NW corner of the NGC1399 residual maps obtained 
for $K\leq 25$. The evolution of
the dynamical mass inferred for the precursors of the GC structures follows the trend observed 
in the excess number of GCs, as per Equation~\ref{eq:dynamicalmass}. Both the mass-weighted radial
and azimuthal positions of the structures as a function of $K$ quickly stabilize as the major
structures start to dominate the residual maps. The dominating structures usually converge
towards either amorphous, roughly circular morphology (ADs) as smaller structures 
with more complex shapes either merge into the major structures or disappear, or form 
geometrically complex, potentially composite structures (Hybrids).

\begin{sidewaysfigure}[ht]
 	\includegraphics[width=\linewidth]{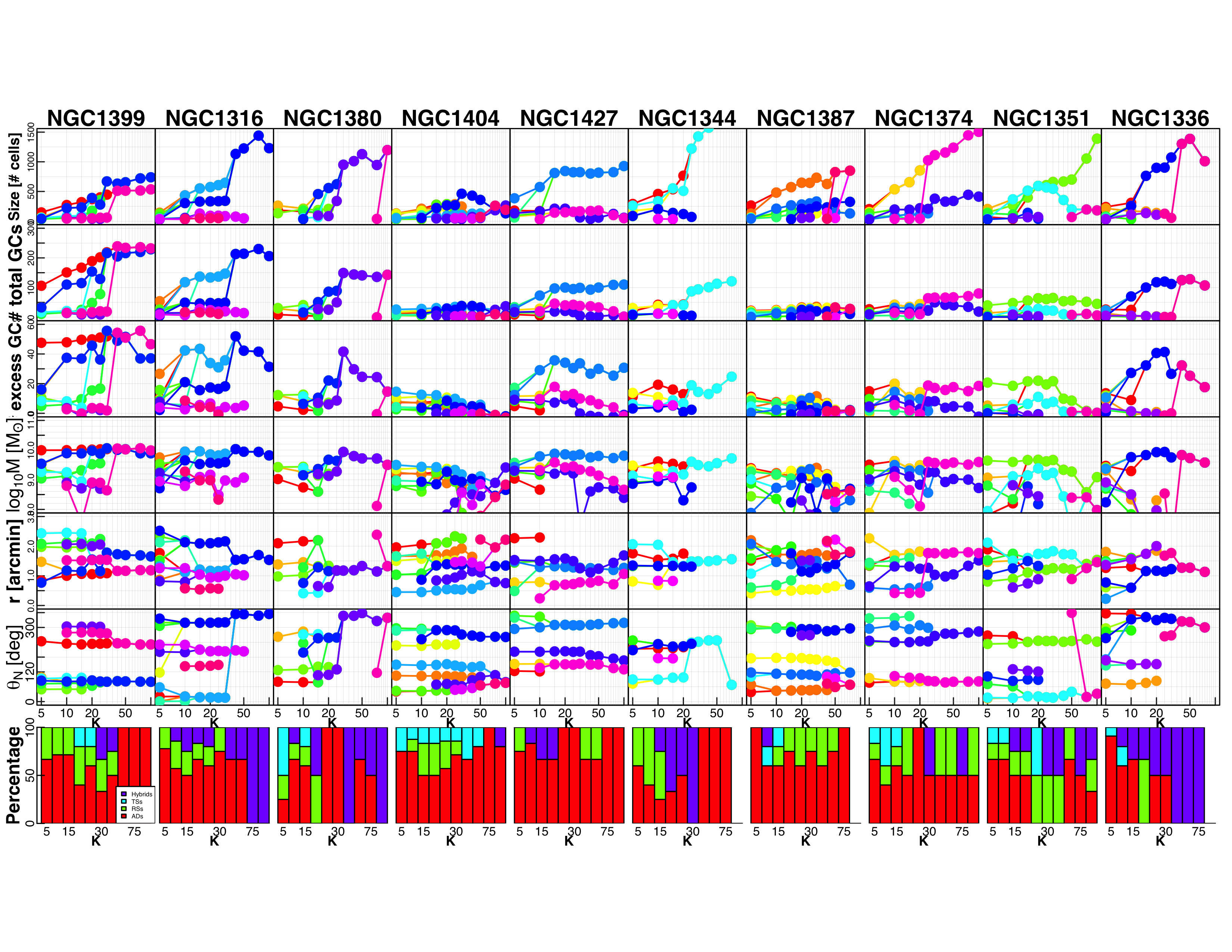}
    \caption{From top to bottom row: size, total and excess numbers of
    GCs, mass of the progenitor, average galactocentric distance, average azimuthal 
    positions and distribution of morphological classes as a function of $K$ 
    ($K\!=\!\{5,10,15,20,25,30,40,50,75,100\}$) for all GC structures detected in the
    galaxies studied in this paper. Each bin in the distribution of morphological class
    is associated with a different $K$ value investigated.} 
 	\label{fig:properties_kevolution}
 \end{sidewaysfigure}

Figure~\ref{fig:properties_kevolution} shows the size, total and excess numbers of 
CGs, the dynamical mass of the progenitors, radial and azimuthal positions and 
morphology for all the 
structures detected in the residual maps of the GC distributions of all galaxies 
investigated in this paper for values of $K\!=\!\{5,10,15,20,25,30,40,50,75,100\}$. 
The same general trends described for NGC1399 structures above are observed for the 
other galaxies. 

\end{document}